\newif\ifjournalversion \journalversionfalse

\newif\ifcrimson \crimsontrue  
\newif\ifreview  \reviewfalse    
\newif\ifcolorlinks \colorlinkstrue 
\newif\ifprintversion \printversionfalse 
\newif\iffinetypo \finetypotrue 
\newif\ifversionid \versionidtrue 
\newif\ifacresetsection \acresetsectiontrue 
\ifjournalversion
  \finetypofalse \colorlinksfalse \versionidfalse \printversionfalse \reviewfalse 
\fi

\ifjournalversion\PassOptionsToClass{12pt}{article}\else\PassOptionsToClass{10pt}{article}\fi
\documentclass[a4paper,twoside]{article}

\newcommand{\papertitle}{Roadmap: Image reconstruction in emission tomography for the era of fast Monte Carlo
  simulation}

\ifprintversion\PassOptionsToPackage{bindingoffset=6mm}{geometry}\fi
\usepackage[a4paper,textwidth=153mm,textheight=250mm,top=24mm,headsep=10mm,hcentering,headheight=14pt]{geometry}
\usepackage[T1]{fontenc}
\usepackage{lmodern}
\ifcrimson
  \iffinetypo\usepackage[osf]{crimson}\else\usepackage{crimson}\fi 
  \RequirePackage{iftex}
  \iffinetypo\ifpdftex
    \makeatletter
    \begingroup\fontfamily{\rmdefault}\selectfont\fontfamily{Crimson-TLF}\selectfont\endgroup 
    \edef\typo@fam{\rmdefault}
    \@ifundefined{T1/\typo@fam/b/n}{\DeclareFontShape{T1}{\typo@fam}{b}{n}{<-> Crimson-Bold-tlf-t1}{}}{}
    \@ifundefined{T1/\typo@fam/bx/n}{\DeclareFontShape{T1}{\typo@fam}{bx}{n}{<-> Crimson-Bold-tlf-t1}{}}{}
    \makeatother
  \fi\fi
  \usepackage[scale=0.81,semibold]{librefranklin} 
\fi
\usepackage{silence}
\usepackage{microtype} 
\iffinetypo\usepackage[british]{babel}\else\usepackage[english]{babel}\fi 
\usepackage{csquotes}
\ifjournalversion
  \usepackage[backend=biber, style=authoryear-comp, sorting=nyt, language=english, doi=true, isbn=false, url=false,
    eprint=false, maxbibnames=6, minbibnames=3, maxcitenames=2, mincitenames=1, giveninits=true, uniquename=false,
    uniquelist=false, dashed=false]{biblatex}
  \DefineBibliographyStrings{english}{andothers={et al}} 
  \AtBeginDocument{\let\cite\parencite} 
\else
  \usepackage[backend=biber, style=numeric-comp, sorting=none, language=english, doi=true, isbn=false, url=false,
    eprint=false, maxbibnames=6, minbibnames=3, maxcitenames=2, mincitenames=1, giveninits=true,
    backref=true, backrefstyle=three]{biblatex} 
  \DefineBibliographyStrings{english}{backrefpage={cited on p\adddot}, backrefpages={cited on pp\adddot}}
\fi

\usepackage{amsmath}
\usepackage{amsthm}
\usepackage{amssymb}
\usepackage{caption}
\usepackage[printonlyused]{acronym} 
\makeatletter
\ifdefined\AC@verridelabel\renewcommand*\AC@verridelabel[1]{}\fi 
\ifdefined\@verridelabel\renewcommand*\@verridelabel[1]{}\fi     
\makeatother
\usepackage[dvipsnames]{xcolor}
\usepackage{graphicx}
\graphicspath{{images/}}

\usepackage{hyperref}
\usepackage{pdftexcmds} 
\ifcolorlinks
  \definecolor{linkblue}{RGB}{31,78,121}
  \definecolor{citeclaret}{RGB}{128,24,48}
  \hypersetup{colorlinks=true, linkcolor=linkblue, urlcolor=linkblue, citecolor=citeclaret}
\else
  \hypersetup{hidelinks}
\fi
\colorlet{acrocolor}{black} 
\makeatletter
\AtBeginDocument{\@ifpackageloaded{hyperref}{%
  \def\AC@hyperlink#1#2{{\hypersetup{linkcolor=acrocolor}\hyperlink{#1}{#2}}}%
  \def\AC@hyperref[#1]#2{#2}}{}}
\makeatother
\usepackage{orcidlink}
\usepackage[noabbrev,nameinlink]{cleveref}
\hypersetup{
  pdftitle={\papertitle},
  pdfauthor={G. Schramm et al.},
  pdfkeywords={Monte Carlo simulation, PET, SPECT, image reconstruction, forward model, scatter, positron range},
  bookmarksnumbered
}

\usepackage{booktabs}
\usepackage{tabularx}
\usepackage{multicol} 
\usepackage{flafter} 
\usepackage{tikz}
\usepackage{pgfplots} 
\pgfplotsset{compat=1.17}
\definecolor{plotblue}{RGB}{31,119,180}
\definecolor{plotorange}{RGB}{214,110,20}
\definecolor{plotgreen}{RGB}{44,140,60}
\definecolor{plotred}{RGB}{179,26,31}
\providecommand{\plotnums}{} 
\pgfplotsset{
  roadmap/.style={
    scale only axis,
    axis line style={black!60, line width=0.4pt},
    every tick/.style={black!60, line width=0.4pt},
    xtick pos=bottom, ytick pos=left, major tick length=2.5pt, tick align=outside,
    xminorticks=false, yminorticks=false,
    xmajorgrids, ymajorgrids, grid style={black!28, dotted, line width=0.4pt},
    tick label style={font=\footnotesize\plotnums, black!80},
    label style={font=\small},
    xticklabel={\pgfmathprintnumber[1000 sep={}, assume math mode=true]{\tick}},
    legend style={draw=none, fill=white, fill opacity=0.9, text opacity=1, font=\footnotesize,
      inner sep=1pt, row sep=-1pt},
    legend cell align=left,
    every axis plot/.append style={line width=0.5pt},
    clip mode=individual,
  },
  roadmap log y/.style={log basis y=10,
    yticklabel={10\textsuperscript{\pgfmathparse{int(round(\tick))}\pgfmathresult}}},
}
\tikzset{series label/.style={font=\footnotesize, anchor=west, align=left, inner sep=0pt},
  panel/.style={anchor=north west, font=\small\bfseries, inner sep=4pt}}
\usepackage{bm} 
\newcommand{\isotope}[2][]{\textsuperscript{\plotnums #1}#2}

\usetikzlibrary{arrows.meta, shapes}

\defbibheading{bibintoc}[\refname]{\section*{#1}\addcontentsline{toc}{section}{#1}\markright{#1}}

\newif\ifauthorline
\newcommand{\authorline}[1]{%
  \par\ifauthorline\vskip-\medskipamount\vskip\smallskipamount\fi
  {\noindent\small\itshape #1\par}\nobreak\vskip\medskipamount
  \global\authorlinetrue
  \everypar{{\setbox0\lastbox}\global\authorlinefalse\everypar{}}%
}
\newcommand{\sectionauthor}[2][]{%
  \authorline{\if\relax\detokenize{#1}\relax\else #1: \fi #2}%
  \addtocontents{toc}{\protect\tocauthor{\if\relax\detokenize{#1}\relax\else #1: \fi #2}}%
}
\newcommand{\tocauthor}[1]{{\leftskip=2.3em\noindent\small\itshape #1\par}}

\makeatletter
\renewcommand\section{\@startsection{section}{1}{\z@}{-3.5ex \@plus -1ex \@minus -.2ex}{2.3ex \@plus.2ex}%
  {\normalfont\Large\bfseries\sffamily}}
\renewcommand\subsection{\@startsection{subsection}{2}{\z@}{-3.25ex\@plus -1ex \@minus -.2ex}{1.5ex \@plus .2ex}%
  {\normalfont\large\bfseries\sffamily}}
\renewcommand\subsubsection{\@startsection{subsubsection}{3}{\z@}{-3.25ex\@plus -1ex \@minus -.2ex}{1.5ex \@plus .2ex}%
  {\normalfont\normalsize\bfseries\sffamily}}
\renewcommand\paragraph{\@startsection{paragraph}{4}{\z@}{3.25ex \@plus1ex \@minus.2ex}{-1em}%
  {\normalfont\normalsize\bfseries}}
\makeatother

\numberwithin{equation}{section}
\counterwithin{figure}{section}
\counterwithin{table}{section}

\usepackage{fancyhdr}
\newcommand{\runningtitle}{\papertitle} 
\makeatletter
\DeclareRobustCommand{\rh@nolabel}{\let\label\@gobble} 
\renewcommand{\sectionmark}[1]{\markright{\thesection\quad{\rh@nolabel #1}}}
\renewcommand{\subsectionmark}[1]{}
\let\@mkboth\@gobbletwo 
\makeatother
\fancypagestyle{plain}{\fancyhf{}\fancyfoot[C]{\small\thepage}}
\makeatletter
\newcommand{\newsectionpage}{%
  \clearpage
  \ifprintversion\if@twoside\ifodd\c@page\else\hbox{}\thispagestyle{empty}\newpage\fi\fi\fi
  \thispagestyle{plain}%
  \ifacresetsection\acresetall\fi} 
\makeatother

\DeclareRobustCommand{\numimg}{\ensuremath{N}}
\DeclareRobustCommand{\pixind}{\ensuremath{j}}
\DeclareRobustCommand{\nummeas}{\ensuremath{M}}
\DeclareRobustCommand{\datind}{\ensuremath{i}}
\DeclareRobustCommand{\itrind}{\ensuremath{k}}

\DeclareRobustCommand{\actvty}{\ensuremath{\lambda}}
\DeclareRobustCommand{\actvtyvec}{\ensuremath{\bm \actvty}}
\DeclareRobustCommand{\attn}{\ensuremath{\mu}}
\DeclareRobustCommand{\attnvec}{\ensuremath{\bm \attn}}
\DeclareRobustCommand{\imbss}{\ensuremath{v}}

\DeclareRobustCommand{\prompts}{\ensuremath{y}}
\DeclareRobustCommand{\promptsvec}{\ensuremath{\bm \prompts}}
\DeclareRobustCommand{\trues}{\ensuremath{u}}
\DeclareRobustCommand{\truesvec}{\ensuremath{\bm \trues}}
\DeclareRobustCommand{\scatter}{\ensuremath{s}}
\DeclareRobustCommand{\scattervec}{\ensuremath{\bm \scatter}}
\DeclareRobustCommand{\randoms}{\ensuremath{r}}
\DeclareRobustCommand{\randomsvec}{\ensuremath{\bm \randoms}}
\DeclareRobustCommand{\backgr}{\ensuremath{b}}
\DeclareRobustCommand{\backgrvec}{\ensuremath{\bm \backgr}}
\DeclareRobustCommand{\srm}{\ensuremath{\mathsf{A}}}
\DeclareRobustCommand{\srmel}{\ensuremath{a}}
\DeclareRobustCommand{\cdsrm}{\ensuremath{\mathcal{A}}}
\DeclareRobustCommand{\chrf}{\ensuremath{\chi}}
\DeclareRobustCommand{\cdkern}{\ensuremath{\bm \chrf}}
\DeclareRobustCommand{\sensvecel}{\ensuremath{ c}}
\DeclareRobustCommand{\sensvec}{\ensuremath{\bm \sensvecel}}

\DeclareRobustCommand{\pos}{\ensuremath{\bm x}}
\DeclareRobustCommand{\angdir}{\ensuremath{\hat{\bm \theta}}}
\DeclareRobustCommand{\Energy}{\ensuremath{E}}
\DeclareRobustCommand{\raydist}{\ensuremath{\zeta}}

\DeclareRobustCommand{\src}{\ensuremath{q}}

\DeclareRobustCommand{\meas}{\ensuremath{m}}

\DeclareRobustCommand{\radiance}{\ensuremath{\phi}}
\DeclareRobustCommand{\transop}{\ensuremath{\mathcal{T}}}
\DeclareRobustCommand{\propop}{\ensuremath{\mathcal{G}}}
\DeclareRobustCommand{\propLS}{\ensuremath{\mathcal{G}^{\mathrm{LS}}}}
\DeclareRobustCommand{\scatop}{\ensuremath{\mathcal{S}}}
\DeclareRobustCommand{\measop}{\ensuremath{\mathcal{M}}}
\DeclareRobustCommand{\srcop}{\ensuremath{\mathcal{Q}}}
\DeclareRobustCommand{\radon}{\ensuremath{\mathcal{R}}}

\DeclareRobustCommand{\loglik}{\ensuremath{\mathcal{L}}}
\DeclareRobustCommand{\adjsrc}{\ensuremath{h^{\ast}}}
\DeclareRobustCommand{\adjsrcvec}{\ensuremath{\bm h^{\ast}}}
\DeclareRobustCommand{\onevec}{\ensuremath{\bm 1}}
\DeclareRobustCommand{\transp}{^{\!\top}}
\DeclareRobustCommand{\projection}{\ensuremath{\mathcal{P}}}
\DeclareMathOperator*{\argmax}{arg\,max}

\DeclareRobustCommand{\mcop}{\ensuremath{\mathrm{MC}}}
\DeclareRobustCommand{\mcsmpl}{\ensuremath{\Upsilon}}
\DeclareRobustCommand{\prob}{\ensuremath{P}}
\DeclareRobustCommand{\proposal}{\ensuremath{Q}}

\DeclareRobustCommand{\epmax}{\ensuremath{E_\mathrm{max}^\mathrm{P}}}

\theoremstyle{plain}
\newtheorem{theorem}{Theorem}[section]

\theoremstyle{definition}

\theoremstyle{remark}
\newtheorem{remark}[theorem]{Remark}

\providecommand{\rmd}{\mathrm{d}} 
\newcommand{\dbyd}[2]{\frac{\partial {#1}}{\partial {#2}}}

\newcommand{\dontshow}[1]{ } 

\ifreview
  \usepackage[draft,notref,notcite]{showkeys}
  \usepackage{lineno}
  \linenumbers
\fi

\iffinetypo
  \makeatletter
  \microtypesetup{tracking=smallcaps}
  \ifcrimson\def\typo@pagenum{\small\crimsonoldstyle\thepage}\else\def\typo@pagenum{\small\thepage}\fi
  \fancypagestyle{plain}{\fancyhf{}\fancyfoot[C]{\typo@pagenum}}
  \ifcrimson\renewcommand{\plotnums}{\crimsonlining}\fi
  \ifcrimson\AddToHook{env/tabular/begin}{\crimsonlining}\AddToHook{env/tabularx/begin}{\crimsonlining}\fi
  \renewcommand\@makefntext[1]{\setlength{\parindent}{1em}\noindent\llap{\@thefnmark.\enspace}#1}
  \makeatother
\fi

\newcommand*{\versionfiles}{main.tex,preamble.tex,acros.tex,notation.tex,references.bib,%
  sections/01_intro.tex,sections/02_reconbasics.tex,sections/03_MC.tex,sections/04_sm.tex,sections/05_scatter.tex,%
  sections/06_positron.tex,sections/07_atte.tex,sections/08_inv.tex,sections/09_compute_hw.tex,sections/10_outlook.tex,%
  images/fig71_jeaa_tof_scatter.tex,images/fig81_adjoint_born_series.tex,images/fig72_jeaa_scale.tex,%
  images/fig91_hw_trends.tex,images/fig91_hw_trends_a.tex,images/data/pixels_dat.csv,%
  images/fig21_born_series.tex,images/fig22_where_mc_fits.tex,images/fig92_roofline.tex}
\makeatletter
\def\versionid@sums{}
\@for\versionid@file:=\versionfiles\do{%
  \edef\versionid@sums{\versionid@sums\pdf@filemdfivesum{\versionid@file}}}
\def\versionid@seven#1#2#3#4#5#6#7#8\@nil{#1#2#3#4#5#6#7}
\edef\versionid@full{\pdf@mdfivesum{\versionid@sums}}
\edef\versionid{\expandafter\versionid@seven\versionid@full\@nil}
\ifpdftex
  \edef\bt@s{\pdfcreationdate}
  \def\bt@a#1#2#3#4#5#6#7#8#9{
    \def\bt@Y{#3#4#5#6}\def\bt@M{#7#8}\bt@b#9}
  \def\bt@b#1#2#3#4#5#6#7#8#9{
    \def\bt@D{#1#2}\def\bt@min{#3#4*60+#5#6}%
    \bt@o=0 \if#9Z\expandafter\bt@c\else\if#9-\bt@neg=-1 \else\bt@neg=1 \fi\expandafter\bt@d\fi}
  \def\bt@d#1#2'#3#4'#5\relax{\bt@o=\numexpr\bt@neg*(#1#2*60+#3#4)\relax}
  \def\bt@c#1\relax{}
  \newcount\bt@o \newcount\bt@neg
  \expandafter\bt@a\bt@s\relax
  \newcount\bt@t \bt@t=\numexpr\bt@min-\bt@o\relax 
  \newcount\bt@y \bt@y=\bt@Y \newcount\bt@m \bt@m=\bt@M \newcount\bt@dd \bt@dd=\bt@D
  \def\bt@dim{
    \ifcase\bt@m\or31\or\ifnum\numexpr\bt@y-4*((\bt@y-2)/4)=0 29\else28\fi\or31\or30\or31\or30\or31\or31\or30\or31\or30\or31\fi}
  \ifnum\bt@t<0 
    \advance\bt@t 1440 \advance\bt@dd -1
    \ifnum\bt@dd=0 \advance\bt@m -1 \ifnum\bt@m=0 \bt@m=12 \advance\bt@y -1 \fi \bt@dd=\bt@dim\relax\fi
  \else\ifnum\bt@t>1439 
    \advance\bt@t -1440 \advance\bt@dd 1
    \ifnum\bt@dd>\bt@dim\relax \bt@dd=1 \advance\bt@m 1 \ifnum\bt@m=13 \bt@m=1 \advance\bt@y 1 \fi\fi
  \fi\fi
  \edef\buildtime{\the\bt@y-\two@digits{\the\bt@m}-\two@digits{\the\bt@dd}~%
    \two@digits{\the\numexpr(\bt@t-30)/60\relax}:\two@digits{\the\numexpr\bt@t-60*((\bt@t-30)/60)\relax}~UTC}
\else
  \edef\buildtime{\the\year-\two@digits{\the\month}-\two@digits{\the\day}~%
    \two@digits{\the\numexpr(\time-30)/60\relax}:\two@digits{\the\numexpr\time-60*((\time-30)/60)\relax}~local time}
\fi
\makeatother

\ExplSyntaxOn
\int_new:N \g__pmb_affil_int
\prop_new:N \g__pmb_affil_text_prop 
\prop_new:N \g__pmb_affil_num_prop  
\seq_new:N \g__pmb_affil_seq        
\seq_new:N \g__pmb_author_seq       
\seq_new:N \l__pmb_num_seq
\tl_new:N \l__pmb_num_tl
\msg_new:nnn { pmb } { unknown-affil }
  { Affiliation~key~'#1'~is~not~defined~(use~\iow_char:N\\defaffil~in~main.tex). }
\NewDocumentCommand \defaffil { m m } { \prop_gput:Nnn \g__pmb_affil_text_prop {#1} {#2} }
\NewDocumentCommand \addauthor { m m m }
  {
    \seq_clear:N \l__pmb_num_seq
    \clist_map_inline:nn {#3}
      {
        \prop_if_in:NnF \g__pmb_affil_text_prop {##1} { \msg_error:nnn { pmb } { unknown-affil } {##1} }
        \prop_if_in:NnF \g__pmb_affil_num_prop {##1}
          {
            \int_gincr:N \g__pmb_affil_int
            \exp_args:NNnx \prop_gput:Nnn \g__pmb_affil_num_prop {##1} { \int_use:N \g__pmb_affil_int }
            \seq_gput_right:Nn \g__pmb_affil_seq {##1}
          }
        \seq_put_right:Nx \l__pmb_num_seq { \prop_item:Nn \g__pmb_affil_num_prop {##1} }
      }
    \tl_set:Nx \l__pmb_num_tl { \seq_use:Nn \l__pmb_num_seq { , } }
    \seq_gput_right:Nx \g__pmb_author_seq
      {
        \exp_not:n {#1} $^{ \l__pmb_num_tl }$
        \tl_if_blank:nF {#2} { \exp_not:N \, \exp_not:N \orcidlink { \exp_not:n {#2} } }
      }
  }
\NewDocumentCommand \paperauthors { } { \seq_use:Nn \g__pmb_author_seq { ,~ } }
\NewDocumentCommand \paperaffiliations { }
  {
    \seq_map_indexed_inline:Nn \g__pmb_affil_seq
      {
        \int_compare:nNnF {##1} = { 1 } { \\ }
        $^{##1}$ ~ \prop_item:Nn \g__pmb_affil_text_prop {##2}
      }
  }
\ExplSyntaxOff

\newcommand{\maketitlepage}{%
  \begingroup
  \ifreview\nolinenumbers\fi
  \thispagestyle{empty}%
  \vspace*{0.09\textheight}%
  \begin{center}
    {\sffamily\bfseries\huge\linespread{1.1}\selectfont \papertitle\par}
    \vspace{0.065\textheight}%
    {\large\linespread{1.25}\selectfont \paperauthors\par}
    \vspace{0.05\textheight}%
    {\color{black!40}\rule{0.18\textwidth}{0.4pt}\par}
    \vspace{0.03\textheight}%
    {\footnotesize\linespread{1.15}\selectfont \paperaffiliations\par}
    \vspace{4ex}%
    {\footnotesize \papercontact\par}
  \end{center}
  \vfill
  {\footnotesize\centering
    {\color{black!40}\rule{\textwidth}{0.4pt}\par}\vspace{0.8ex}%
    \if\relax\detokenize\expandafter{\paperstatus}\relax\else\paperstatus\\[0.4ex]\fi
    \paperdate
    \ifversionid\ \textperiodcentered\ version \texttt{\MakeLowercase{\versionid}} (build \buildtime)\fi
    \par}
  \newpage
  \endgroup
  \ifreview\linenumbers\fi}

\begin{document}

\defaffil{kul}{Department of Imaging and Pathology, Division of Nuclear Medicine, KU Leuven, Leuven, Belgium}
\defaffil{mgb}{Division of Nuclear Medicine and Molecular Imaging, Department of Radiology, Mass General Brigham and
  Harvard Medical School, Boston, MA, USA}
\defaffil{creatis}{INSA-Lyon, Universit\'e Lyon 1, CNRS, Inserm, CREATIS UMR 5220, U1294, F-69373 Lyon, France}
\defaffil{itefi}{Institute for Physical and Information Technologies “Leonardo Torres Quevedo”, ITEFI, Spanish National
  Research Council (CSIC), Madrid, Spain}
\defaffil{jsc}{J\"ulich Supercomputing Centre (JSC), Forschungszentrum J\"ulich GmbH, J\"ulich, Germany}
\defaffil{lund-mrp}{Department of Medical Radiation Physics, Lund University, Lund, Sweden}
\defaffil{lund-tm}{Department of Translational Medicine, Medical Radiation Physics, Lund University, Lund, Sweden}
\defaffil{ucl-cs}{Department of Computer Science, University College London, London, UK}
\defaffil{inm4}{Institute of Neuroscience and Medicine (INM-4), Forschungszentrum J\"ulich GmbH, J\"ulich, Germany}
\defaffil{ucm}{Nuclear Physics Group, EMFTEL and IPARCOS, Complutense University of Madrid, Madrid, Spain}
\defaffil{idissc}{Health Research Institute of the Hospital Clinico San Carlos (IdISSC), Madrid, Spain}
\defaffil{clb}{Centre L\'eon B\'erard, Lyon, France}
\defaffil{luebeck}{Institute of Medical Engineering, University of Luebeck, L\"ubeck, Germany}
\defaffil{ucl-inm}{Institute of Nuclear Medicine, University College London, London, UK}
\defaffil{ucl-hawk}{UCL Hawkes Institute, University College London, London, UK}

\addauthor{Georg~Schramm}{0000-0002-2251-3195}{kul}
\addauthor{Benjamin~Auer}{0000-0003-0853-5651}{mgb}
\addauthor{Ane~Etxebeste}{0000-0001-6280-9268}{creatis}
\addauthor{Pablo~Galve}{0000-0003-4507-898X}{itefi}
\addauthor{Jens~Henrik~G\"obbert}{0000-0002-3807-6137}{jsc}
\addauthor{Johan~Gustafsson}{0000-0002-1022-8370}{lund-mrp,lund-tm}
\addauthor{\v{Z}eljko~Kereta}{0000-0003-2805-0037}{ucl-cs}
\addauthor{Nils~Krah}{0000-0002-1376-6633}{creatis}
\addauthor{Christoph~Lerche}{0000-0003-2749-2108}{inm4}
\addauthor{Chong~Li}{0000-0002-1849-1528}{inm4}
\addauthor{Joaquin~L.~Herraiz}{0000-0001-7208-8863}{ucm,idissc}
\addauthor{David~Sarrut}{0000-0002-4854-4141}{creatis,clb}
\addauthor{J\"urgen~J.~Scheins}{0000-0003-3474-430X}{inm4}
\addauthor{Jos\'e~Manuel~Ud\'ias}{0000-0003-3714-764X}{ucm,idissc}
\addauthor{Simon~Arridge}{0000-0003-1292-0210}{ucl-cs}
\addauthor{Magdalena~Rafecas}{0000-0001-5691-7756}{luebeck}
\addauthor{Kris~Thielemans}{0000-0002-5514-199X}{ucl-inm,ucl-hawk}
\newcommand{\papercontact}{E-mail: georg.schramm@kuleuven.be}
\newcommand{\paperdate}{Preprint, October 2026} 
\ifjournalversion\renewcommand{\paperdate}{October 2026}\fi
\newcommand{\paperstatus}{Invited roadmap, submitted to \textit{Physics in Medicine \& Biology}; this preprint has not been
  peer reviewed.} 
\ifjournalversion\renewcommand{\paperstatus}{}\fi 
\title{\papertitle} 
\maketitlepage
\markright{\runningtitle} 

\newpage

\begin{abstract}
Quantitative image reconstruction in positron emission tomography and single photon emission computed tomography
depends on how accurately the forward model describes photon emission, transport and detection.
Monte Carlo simulation is the most complete model available, yet it is mostly used as an offline reference
rather than inside the reconstruction.
This roadmap asks what it would take to change that.
The main obstacle is not only simulation time.
Iterative reconstruction requires the expectation of the measured data and the gradient of the data fidelity,
whereas Monte Carlo simulation produces noisy realisations of a complete acquisition.
Graphics-processing-unit implementations, variance reduction and learned surrogate models have made simulations
much faster.
At the same time, modern scanners acquire ever larger data sets with correspondingly larger system models,
and further gains in quantitative accuracy require more detailed physics.
We discuss how fast Monte Carlo simulation can improve the modelling of the system response, scatter, positron range
and the joint estimation of activity and attenuation, how noisy simulations can enter reconstruction algorithms,
and which computing hardware makes this feasible.
Monte Carlo simulation is most valuable where analytic models fail, 
and most practical solutions will combine both.
The most immediate benefit of fast Monte Carlo simulators is the validation and calibration of faster analytic models,
together with more accurate, absolutely quantitative scatter estimates.
Key open problems are the use of noisy simulations as expectations, in particular for sparse data, the
computation of gradients and adjoints, the allocation of a limited simulation budget, 
and the use of scattered events as a source of information rather than a nuisance.
Faster simulators are necessary but not sufficient, and progress requires simulators and reconstruction methods
to be developed together.
\end{abstract}

\noindent\textbf{Keywords:} Monte Carlo simulation, positron emission tomography, single photon emission computed
tomography, image reconstruction, forward model, scatter, positron range

\setcounter{tocdepth}{1}
\tableofcontents
\acresetall 
\newpage


\newsectionpage
\section{Introduction\label{sec:intro}}
\sectionauthor{Georg Schramm, Magdalena Rafecas}

\Ac{pet} and \ac{spect} are quantitative emission tomography modalities.
The reconstructed images display the spatial distribution of the administered radiotracer and can be used for
qualitative image interpretation as well as for quantitative assessment of tracer uptake.
Increasingly, quantitative imaging is also used to derive absorbed doses in radionuclide
therapy~\cite{Ljungberg2016_MIRD26}.
How well this works depends on how faithfully the reconstruction models the physics of photon emission, radiation
transport
and subsequent detection.
\Ac{mc} simulation is the most complete tool available for this purpose, since it can follow radioactive decay,
photon transport through the patient and the detector, and signal formation with few
approximations~\cite{Zaidi1999,sarrut2021c}.
Its use inside image reconstruction was demonstrated decades
ago~\cite{Floyd1985,Rafecas2004a,Lazaro2005PMB,Beekman2002},
and \ac{mc}-based \ac{spect} reconstruction is now commercially available.
Yet in most of \ac{pet} and \ac{spect} practice, \ac{mc} is still used as an offline reference tool rather than as
part of the reconstruction.
This roadmap asks what it would take to change that.

The obstacle is more fundamental than long simulation times.
Each \ac{mc} simulation produces a noisy realisation of the measurement, whereas statistical reconstruction methods
such as \ac{mlem} and its \ac{map} variants require its expectation.
Approximating this expectation by \ac{mc} simulation requires many emission histories, in the limit infinitely
many, for every patient and in some cases for every update of the image estimate.
Current image reconstruction therefore relies on simplified analytic models, pre-computed corrections and fixed
additive terms for effects such as scatter.
For many applications these approaches work well.
They become inaccurate, or inefficient, when the physics is complex, for example for higher-order Compton scatter,
for septal penetration in \ac{spect} collimators, for the continuous bremsstrahlung spectrum in \isotope[90]{Y} imaging,
or for the down-scatter caused by the high-energy emissions of the daughters in the \isotope[225]{Ac} decay chain
in \ac{spect}~\cite{Benabdallah2021}.
Where \ac{mc}-based and analytic models have been compared, the differences are clinically relevant.
In \ac{spect}, \ac{mc}-based reconstruction increased striatal uptake ratios in \isotope[123]{I} brain imaging by up to
25\%~\cite{srm-23}, raised lesion detection in \isotope[111]{In} liver imaging from 31\% to 50\%~\cite{Wikberg2023}, and
increased contrast recovery in \isotope[90]{Y} imaging by up to 42\%~\cite{srm-59} (\cref{sec:FastSysResp}).

Earlier attempts were limited mainly by computing power.
Two developments make it timely to revisit the question now.
First, \ac{mc} simulation has become much faster.
\Ac{gpu} implementations routinely achieve speed-ups of two to three orders of magnitude over \ac{cpu} codes, as
documented in the recent review of \ac{gpu}-based \ac{mc} platforms by \textcite{chi2025}, and variance
reduction and \ac{ai}-based surrogate models add further gains.
Additional speed-ups come from simulators designed for a specific reconstruction task, which simulate only the
interactions that matter for that task; such codes can be competitive even on multi-core
\acp{cpu}~\cite{scheins2021}.
Second, the demand for accurate physics models is growing, in diagnostic as well as in therapeutic imaging.
State-of-the-art \ac{pet} systems combine long axial fields of view with improved \ac{tof}, spatial and energy
resolution, and the accuracy of the forward model has to keep pace with detector technology developments if the
information in the acquired data is to be fully exploited.
Theranostics adds further requirements, since quantitative \ac{spect} and \ac{pet} are used for image-based
dosimetry and therapy verification, often with isotopes whose emission spectra are poorly handled by analytic
models.
At the same time, the reconstruction problems themselves keep growing.
The number of detector elements and image voxels, and with it the size of the \ac{srm}, increases faster than the
memory bandwidth of computers (\cref{sec:comparch}), and \ac{mc} simulation is typically memory-bound.
Together, these developments raise the question of whether, and how, \ac{mc} can replace or at least complement
analytic models inside the reconstruction.

To focus on this question, we deliberately take an optimistic view of the simulators themselves.
Suppose that \ac{mc} codes were fast enough to simulate an acquisition with many detected counts, and hence with
high precision, repeatedly within clinical reconstruction times.
How should they then be used, and for which tasks would they make a difference?
Faster simulators alone do not answer this question, and the problems they raise are easily underestimated.
Part of the difficulty is that ``\ac{mc}-based reconstruction'' means different things to different researchers.
At one end of the spectrum, \ac{mc} only provides a more accurate estimate of the scatter contribution, which is
then added as a fixed term to an otherwise analytic forward model.
At the other end, the complete acquisition is simulated, from the activity, density and material distribution in
the patient to the measured data.
Between these extremes lie many hybrid options (\cref{sec:WhereDoesMCfit} and
\cref{fig:where_mc_fits}).
The far end also raises the question of whether the simulated and measured data should be histogrammed or
list-mode counts, as in current practice, or the digitised detector signals from which these counts are
derived~\cite{Berg2018}.
This leads to questions such as the following.
\begin{itemize}
  \item Where on this spectrum should a given application sit?
        The answer depends on the imaging task, the modality and the size of the problem, and evaluating the
        forward model by simulation in every iteration immediately raises the question of whether the gradient of
        the data fidelity, which requires the adjoint of the (linearised) forward model, has to be simulated as
        well~\cite{yao2018,NimierDavid2020}.
  \item How precise does the simulation have to be?
        The statistical noise of an \ac{mc} estimate of the expectation decreases only with the square root of
        the number of simulated emissions, while the cost grows linearly.
        How this noise propagates into the reconstructed image was 
        first studied for \ac{mc}-computed system
        matrices in \ac{pet}~\cite{Rafecas2004,Rehfeld2006} and 
        \ac{spect}~\cite{srm-78,Lazaro2005PMB},
        and the effect of system-matrix errors in general was 
        analysed in~\cite{Qi2005}.
        If the forward model is instead re-simulated in every iteration, reconstruction becomes a stochastic
        optimisation problem.
        Is a noisy gradient direction estimated from a few simulated photons sufficient far from the solution,
        and how much precision is needed close to it?
  \item Is in-patient scatter a nuisance to be removed, or a signal 
        that carries information, in particular about
        the attenuation~\cite{defrise_time--flight_2012,berker_scattered_2014,Brusaferri2020}?
        And if so, how can this information be exploited in practice?
  \item How much of the physics has to be simulated?
        Photon transport and energy deposition only, or also scintillation light and the photodetector
        response~\cite{Roncali2013},
        or charge transport in semiconductor detectors such as cadmium zinc telluride \cite{srm-38}?
\end{itemize}
We raise these questions here to make clear that combining fast \ac{mc} with reconstruction is not only an
engineering task.
The individual sections return to them for specific applications, and \cref{sec:outlook} discusses them in
more detail.

This paper is a roadmap rather than a comprehensive review.
For each topic, we summarise the state of the art, identify the main bottlenecks and outline the research needed
to overcome them.
It complements the recent review of \ac{gpu}-based \ac{mc} platforms by \textcite{chi2025}, which calls for
a tighter integration of simulators into imaging workflows but addresses reconstruction only in passing.
Our focus is on what reconstruction requires from a simulator, which differs from what most codes are designed
for, namely forward realisations of complete acquisitions.
Reconstruction instead needs the gradient of the data fidelity, i.e., information about how the mismatch between the
estimated and measured data changes when the image is changed. 
Computing this gradient requires expectations, often only for the
detector bins that recorded counts, together with the adjoint of the (linearised) forward model, which maps information
from the measurement space back to the image space.
Both forward projection and back projection (application of the adjoint) are required repeatedly within an iterative
loop to evaluate the gradient of the data fidelity.
We restrict the discussion to \ac{pet} and \ac{spect}, which are clinically established; emerging techniques such
as Compton cameras, as well as treatment verification in particle therapy, are not covered, although much of the
discussion carries over.

The roadmap is written for researchers in image reconstruction and inverse problems, developers of \ac{mc} codes,
detector and computing-hardware experts, and medical physicists working in clinical and preclinical imaging.
Few readers will need every section.
The sections are largely self-contained, so readers can go directly to the topics they are interested in.
\Cref{sec:reconbasics} provides the common background and notation, and its last part
(\cref{sec:WhereDoesMCfit}) serves as a map that places the later sections in context; we recommend reading
it first.

The remaining sections are organised as follows.
\Cref{sec:MCoverview} gives an overview of \ac{mc} simulation in emission tomography and of the acceleration
and variance-reduction techniques that make it fast enough for use in reconstruction.
\Cref{sec:FastSysResp,sec:ScatterEstimation} discuss \ac{mc}-based modelling of the system
response and the estimation of the scatter contribution, for both \ac{pet} and \ac{spect}.
\Cref{sec:PositronRangeModelling,sec:JAAE} are specific to \ac{pet} and address positron range
and the joint estimation of activity and attenuation.
\Cref{sec:InvProb} takes a more mathematical view of solving the inverse problem with \ac{mc} models,
including the role of the adjoint and stochastic reconstruction.
\Cref{sec:comparch} discusses the trends in computing hardware that determine what will be feasible.
Finally, \cref{sec:outlook} returns to the questions raised above and concludes the paper.

\newsectionpage
\section{Forward models and image reconstruction in emission tomography\label{sec:reconbasics}}
\sectionauthor{Simon Arridge,  Kris Thielemans, Georg Schramm, \v{Z}eljko Kereta}

\subsection{Introduction \label{sec:reconChapterIntro}}

Tomography in general, and emission tomography in particular, is an \emph{inverse problem,} i.e., we aim to recover
one or more spatially varying  parameters, given a finite set of data which in emission tomography are counts of photons
(\ac{spect}), or photon coincidences (\ac{pet}), received at detectors, and associated data such as detector position,
deposited energy and timing information.
As with all inverse problems, it is essential to  describe the \emph{forward problem}, which is naturally considered as
the composition of i) a \emph{source} model predicting the generation of photons leading to the observed data, ii) a
physical \emph{propagation} model predicting the trajectories of  photons and/or particles for a given distribution of
parameters and sources, and iii) the \emph{measurement} process dependent on the type and geometrical arrangement of
detectors.
In this section we consider  these components in both continuous and discrete settings; in
\crefrange{sec:MCoverview}{sec:ScatterEstimation} we will see that the propagation and measurement models are amenable
to acceleration and/or improved
accuracy with \ac{mc} approaches.
In \cref{sec:PositronRangeModelling} we consider one aspect of source modelling that is particularly relevant to
\ac{pet}.
\Cref{tab:notation} summarises the notation shared across sections.

\begin{table}[t]
\centering
\footnotesize
\caption{Notation shared across sections.
Bold symbols denote discretised vectors, a bar denotes an expectation, $\transp$ a matrix transpose and $^\ast$ the
adjoint of a continuous operator.
Symbols used in a single section only are defined there.}
\label{tab:notation}
\begin{tabularx}{\linewidth}{@{}l >{\raggedright\arraybackslash}X l@{}}
\toprule
Symbol & Meaning & Introduced \\
\midrule
\multicolumn{3}{@{}l}{\emph{Dimensions and indices}} \\
$\numimg$, $\pixind$ & number of image elements (voxels or basis functions), and their index
  & \S\ref{sec:discretisation} \\
$\nummeas$, $\datind$ & number of data bins (e.g.\ sinogram bins or \aclp{lor}), and their index
  & \S\ref{sec:NatureOfData} \\
$\itrind$ & iteration index; superscript on iterates ($\actvtyvec^{\itrind+1}$), subscript otherwise
  (step size $\gamma_{\itrind}$) & \S\ref{sec:InvProb} \\
\midrule
\multicolumn{3}{@{}l}{\emph{Object and phase space}} \\
$\actvty$, $\actvtyvec$ & activity concentration (continuous, discretised) & \S\ref{sec:TheoreticalForwardModel} \\
$\attn$, $\attnvec$ & linear attenuation coefficient
  & \S\ref{sec:rte} \\
$\imbss_{\pixind}$ & image basis function (e.g.\ voxel, blob) & \S\ref{sec:discretisation} \\
$\pos$, $\angdir$, $\Energy$ & position, propagation direction (unit vector), photon energy & \S\ref{sec:rte} \\
$\src$ & photon source & \S\ref{sec:photongeneration}\\
\midrule
\multicolumn{3}{@{}l}{\emph{Data and forward model}} \\
$\promptsvec$, $\bar{\promptsvec}$ & measured data (counts) and its expectation & \S\ref{sec:NatureOfData} \\
$\truesvec$, $\scattervec$, $\randomsvec$ & unscattered, scattered and random (or other background) events
  & \S\ref{sec:NatureOfData} \\
$\backgrvec$ & additive background in $\bar{\promptsvec} = \srm\actvtyvec + \backgrvec$ (e.g.\ scatter, randoms,
depending on what is included in $\srm$)
  & \S\ref{sec:discretisation} \\
$\cdsrm$, $\chrf_{\datind}$ & continuous-to-discrete forward operator, and its kernel for data bin $\datind$
  & \S\ref{sec:TheoreticalForwardModel} \\
$\srcop$ & source model& \S\ref{sec:photongeneration} \\
$\measop$ & measurement (detection) operator & \S\ref{sec:NonlinearForward} \\
$\srm$, $\srmel_{\datind\pixind}$ & system response matrix (\acs{srm}) and its elements & \S\ref{sec:discretisation} \\
$\sensvec = \srm\transp\onevec$ & sensitivity image & \S\ref{sec:DeterministicRecon} \\
\midrule
\multicolumn{3}{@{}l}{\emph{Photon transport}} \\
$\Omega$, $\partial\Omega$ & domain for photon transport, and its boundary & \S\ref{sec:rte} \\
$\radiance$, $\radiance^\ast$ & radiance, adjoint radiance & \S\ref{sec:rte} \\
$\transop_{\attn}$, $\propop$ & transport operator, propagator $\propop = \transop_{\attn}^{-1}$
  & \S\ref{sec:rte} \\
$\scatop$ & scattering operator & \S\ref{sec:ScatterModel} \\
$\propLS$ & full propagator including scatter (Lippmann--Schwinger) & \S\ref{sec:NonlinearForward} \\
\midrule
\multicolumn{3}{@{}l}{\emph{Reconstruction and \acs{mc}}} \\
$\loglik$ & Poisson log-likelihood & \S\ref{sec:DeterministicRecon} \\
$\adjsrcvec$ & adjoint source, $\adjsrcvec = \promptsvec/\bar{\promptsvec}- \onevec$ & \S\ref{sec:InvProb} \\
$\mcsmpl$ & set of simulated histories & \S\ref{sec:InvProb} \\
$\prob$, $\proposal$ & target and proposal distribution (importance sampling) & \S\ref{sec:MCoverview} \\
$\epmax$ & maximum positron energy & \S\ref{sec:PositronRangeModelling} \\
\bottomrule
\end{tabularx}
\end{table}

\subsection{The nature of data\label{sec:NatureOfData}} 

The data could be stored as a list of all events and their respective properties,  such as detected energy and time of
arrival (``list-mode'' data), or as histograms (often called ``sinogram'' or
``projection'' data).
The latter results in a measurement vector $\promptsvec$. 
In practice, even in list-mode data, the measured data are often discretised, e.g., in detector elements, time-bins,
energy-bins, etc. 
Most of this section is therefore written in terms of histogrammed data. 
Here,
each data index $\datind$ represents a bin in the histogram, with $\prompts_\datind$
the number of counts in the bin. 

\paragraph{\acs{pet}}

In \ac{pet}, positrons are emitted isotropically, propagate a short distance before annihilating with an electron,
predominantly generating a \emph{pair} of 511\,keV photons emitted in approximately opposite directions. 
Each data index $\datind$ is associated with a pair of (approximately antipodal) detectors.
A coincidence event, contributing to $\prompts_{\datind}$, is recorded
when both members of this pair detect an event within a sufficiently small time coincidence window.
It is then inferred that the annihilation occurred along the line connecting the two detectors.
Many \ac{pet} scanners record the difference in arrival time of the two photons, known as \acl{tof}, giving extra
information on the location
of the annihilation along the line.
All current \ac{pet} scanners use an energy window around 511\,keV, with some capable of having several energy windows.

\paragraph{\acs{spect}}

In \ac{spect} single photons are emitted isotropically from the radionuclide.
In some cases, \ac{spect} scanners are also used to measure bremsstrahlung photons originating from high-energy charged
particles emitted by certain radionuclides.
Collimators are used to infer directional information of the $\gamma$ photons.
In most \ac{spect} systems, the physical detectors are rotated to cover several angles.
The detection itself often relies on continuous crystals or semiconductors.
In practice, detector position, detection location and detected photon energy are then discretised.  \Ac{spect} scanners
often record data in several energy windows.
This is useful for estimation of scatter contributions,  for radionuclides with several $\gamma$ emissions and for
studies with multiple radionuclides.

\paragraph{}
It is usually assumed that the observations are the sum of three 
components: true (unscattered) events $\truesvec$, scattered 
events $\scattervec$, and other background counts $\randomsvec$ (including accidental coincidences or ``randoms'' in
\ac{pet} and/or background radiation) as
\begin{equation}
  \promptsvec    = \truesvec    + \scattervec    + \randomsvec.
\label{eq:fwd_model1}
\end{equation}

\begin{remark}
    For both \ac{pet} and \ac{spect} the acquired data $\promptsvec 
    \in \mathbb{R}^\nummeas$
     are inherently random.
The observed counts $\prompts_{\datind}$ are (nearly always) modelled as independent Poisson random variables with mean
$\bar{\prompts}_{\datind}$.
\end{remark}

Due to the finite spatial, energy and timing resolution of the detectors, the measured detection position, photon energy
and (difference in) arrival time are subject to uncertainties.
The energy uncertainty limits the ability to identify scatter events, and the spatial and timing uncertainties limit the
accuracy with which the annihilation/emission location can be determined.
The forward model for $\bar{\promptsvec}$ can take these effects into account.

\subsection{Continuous-to-discrete forward model in emission tomography\label{sec:TheoreticalForwardModel}}

The radiotracer activity concentration $\actvty(\pos)$ is a continuous function of $\pos\in \mathbb{R}^3$, while the
acquired data $\promptsvec$ is (usually)
discrete.
The continuous-to-discrete forward operator $\cdsrm$ is thus a vector-valued linear integral operator with kernel
$\cdkern(\pos)$, mapping continuous activity to discrete measurements. 
The background $\randomsvec$ usually carries little information on the activity distribution compared with the unscattered
events.
It is therefore commonly not modelled as part of $\cdsrm$, but treated as a nuisance: an additional, independent Poisson term
whose non-zero mean $\bar{\randomsvec}$ is estimated beforehand.
Since the expectation is linear, the expected data are the sum of the counts modelled by $\cdsrm$ and
$\bar{\randomsvec}$, which results in an \emph{affine} forward model
\begin{equation}
\bar{\promptsvec} = {\mathbb{E} [\promptsvec]} = \cdsrm\,\actvty \, + \bar{\randomsvec}:= \int_{\Omega} \cdkern(\pos)
\actvty(\pos)\, \rmd \pos\, + \bar{\randomsvec}.
\label{eq:Cts2Discrete1}
\end{equation}
Rows $\chrf_{\datind}(\pos)$ of the kernel define the sensitivity of detection bin $\datind$ to activity at position
$\pos$.
In the simplest approximation, each $\chrf_{\datind}$ is an indicator function, supported on a sub-volume of $\Omega$
that contributes to the measurement value $\prompts_{\datind}$.
Their nature depends on the modality and scanner-specific geometry. 

\paragraph{\acs{pet}} The line segment connecting the two detectors for bin $\datind$  constitutes the \acf{lor}, which
in practice has a width determined by the detectors. 
Thus, the sensitivity region is a tube-like volume in $\Omega$, often termed \ac{tor} or \acl*{vor}.
Furthermore, some scanners combine several detector pairs into a single bin $\datind$ for space and computational
reasons.

\paragraph{\acs{spect}}
The acceptance (or sensitivity) region determined by the collimator geometry defines a cone-like volume in $\Omega$, and
constitutes the $\datind$-th row of the corresponding integrating kernel.

\medskip
\noindent
Realistically,
$\cdkern(\pos)$ should model various other effects such as attenuation, scatter, collimator acceptance, detector blur,
etc. 
As mentioned above, these can be computed in steps: positron tracking in \ac{pet} (or charged particle tracking in
bremsstrahlung \ac{spect}), photon generation, photon transport and detector modelling.
In the next few subsections, we will elaborate on these steps.

\subsection{Photon generation model\label{sec:photongeneration}}
The source term includes effects such as  positron range in \ac{pet} and bremsstrahlung in \ac{spect} as well as
multiple gamma-ray emissions for some radionuclides (including prompt gammas in \ac{pet}).
We will sometimes use the notation
$\srcop(\actvty)$ for the generation of (directional) photons in \ac{spect} and (almost) antiparallel pairs in \ac{pet},
where
    \begin{equation}\label{eq:srcmodel1}
    \src = \srcop^{\rm SPECT} (\actvty) \,, \qquad
        \begin{pmatrix}
            \src_{+} \\\src_{-}
        \end{pmatrix}= \srcop^{\rm PET}(\actvty),
    \end{equation}
with $\src(\pos,\angdir)$ a source of photons at position $\pos$ travelling in direction $\angdir$,
a vector on the unit sphere $\mathbb{S}^2$, where we have
dropped energy or multiple photons from the notation for simplicity.

\subsection{Photon transport and the \acl{rte}}\label{sec:rte}

To be able to include absorption and scatter during photon transport,
we now introduce a fundamental model of photon (or particle) propagation, described in terms of \emph{radiance}, also
called \emph{angular particle density}, 
governed by the \acf{rte}, also known as the \emph{Boltzmann Transport Equation}~\cite{chandrasekhar50,case67,thomas99}.
This model also appears in Computer Graphics where it is known as the \emph{Rendering
Equation}~\cite{Jensen2001,Pharr2023}.

The \emph{radiance} $\radiance(\pos,\angdir)$, within a domain $\Omega$ with boundary $\partial \Omega$,  is defined as the density of photons (particles) at position $\pos$
travelling in direction $\angdir$, with units $L^{-3} SR^{-1}$ (i.e. per unit volume per steradian). 
In the presence of attenuation, 
unscattered (also known as ``uncollided'' or ``ballistic'') photons obey the steady-state transport equation
\begin{equation}
\underbrace{\left(\angdir \cdot \nabla_{\pos} + \attn(\pos)\right)}_{\transop_{\attn}} \radiance(\pos, \angdir) =
\src(\pos, \angdir),
\label{eq:rte_noscatter}
\end{equation}
where
$\src$ is the photon source term,  $\attn(\pos)$ is a linear attenuation coefficient of the medium, $\angdir \cdot
\nabla_{\pos}$ is the directional derivative in the photon propagation direction (rate of change per unit path length),
and ${\transop_{\attn}}$ is the transport operator in presence of attenuation.

\Cref{eq:rte_noscatter} states that along a ray in direction $\angdir$, photons are generated by the source $\src$
and attenuated at rate $\attn$.

\paragraph{Propagation operator}
The inverse of the transport operator defines a propagation operator $\propop = \transop_{\attn}^{-1}$ (where we do not
explicitly include the dependence on $\attn$ in $\propop$), so that a
 general solution to \cref{eq:rte_noscatter}
finds the radiance at point $\pos_1$  due to a source at $\pos_0$ in the direction $\angdir_{01} = \widehat{\pos_1 -
\pos_0 }$, i.e. a unit vector from $\pos_0$ to $\pos_1$, as
\begin{equation}\label{eq:RTEpropagator}%
    \radiance(\pos_1,\angdir_{01}) = \propop_{01} \src(\pos_0,\angdir_{01}) := \src(\pos_0,\angdir_{01}) \exp
    \left[-\int_{\raydist_0}^{\raydist_1} \attn(\raydist){\rm d}\raydist\right] .
\end{equation}
This is the \emph{Beer--Lambert} law.

\paragraph{Radon and attenuated Radon transforms}
In (non-bremsstrahlung) \ac{spect}, the source $\src(\pos,\angdir)$ has no directionality and is proportional
to $\actvty(\pos) $.
Then the  solution to \cref{eq:rte_noscatter} is given by the \emph{Attenuated Radon Transform}
\begin{equation}
     \radiance(\pos,\angdir) =  \radon(\attn) \actvty := \int_{\raydist_0}^{\raydist_1} \actvty(\raydist)\exp\left[-
     \int_{\raydist}^{\raydist_1} \attn(\raydist'){\rm d}\raydist' \right] {\rm d} \raydist,
     \label{eq:AttenuatedRadonTransform}
\end{equation}
where $\raydist$ parametrises the arc length along the ray, through $\pos$ in direction $\angdir$, with $\raydist_0$ and
$\raydist_1$ denoting the entry and exit points of the ray in $\Omega$.
This is the relevant forward model for \ac{spect},  when scatter is ignored.

\paragraph{Adjoint transport equation}
The adjoint of the forward transport operator, which satisfies the \emph{Adjoint transport equation}, is given by
\begin{equation}
\underbrace{\left(-\angdir \cdot \nabla_{\pos} + \attn(\pos)\right)}_{\transop^{\ast}_{\attn}}
\radiance^{\ast}(\pos,{\angdir}) =  \src^{\ast}(\pos, \angdir),
\end{equation}
with solution
\begin{equation}\label{eq:adjointRTEpropagator}
    \radiance^{\ast}(\pos_0,\angdir_{10}) = \propop^{\ast}_{10} \src^{\ast}(\pos_1,\angdir_{10}) :=
    \src^{\ast}(\pos_1,\angdir_{10}) \exp \left[-\int_{\raydist_1}^{\raydist_0} \attn(\raydist){\rm d}\raydist\right] .
\end{equation}

\noindent
The adjoint propagates the information in the backwards direction along a given ray (i.e.\ $\raydist$ runs backwards in
\cref{eq:adjointRTEpropagator}), and 
is a key component 
in the backprojection step of standard iterative reconstruction schemes (see below).

\subsection{Modelling scatter\label{sec:ScatterModel}}

In emission tomography (inelastic) scattering  corresponds to a change in both energy and direction, leading to an
augmented \ac{rte} of the form:
\begin{align}
\left(\angdir \cdot \nabla_{\pos} + \attn_{\rm a}(\pos,\Energy) + \attn_{\rm s}(\pos,\Energy)\right)
\radiance(\pos,\angdir,\Energy) \hspace{6cm}  \nonumber \\
- \int \attn_{\rm s}(\pos,\Energy' ) 
\underbrace{\int_{\mathbb{S}^{n-1}} \Theta(\angdir,\angdir',\Energy,\Energy') \radiance(\pos,\angdir',\Energy') {\rm
d}^{n-1}\angdir'}_{{\scatop}\radiance} \rmd \Energy'  = \src(\pos,\angdir,\Energy),
\label{eq:RTE2_MultiEnegy}
\end{align}
where $\attn_{\rm a}$ and $\attn_{\rm s}$ are the absorption and scattering coefficients respectively,
$\Theta(\angdir,\angdir')$ is the phase function, i.e. an angular probability density function of a photon scattering
from direction $\angdir'$ to direction $\angdir$, and $n\in\{2,3\}$.
The phase function is related to the \emph{differential scattering cross section} by normalisation: $\Theta =
\frac{1}{\sigma} \frac{\rmd \sigma}{\rmd \angdir}$.

Note that all attenuation-related quantities depend only on the location $\pos$ and photon energy, while the phase
function $\Theta$ depends only on photon directions and energies.
It is generally assumed that for X- and gamma-ray photons $\attn_{\rm a}$ is negligible compared to $\attn_{\rm s}$;
thus in the following, unless explicitly stated, we use $\attn$ to mean the scattering term only.
Although extensively studied, \cref{eq:RTE2_MultiEnegy} has an analytic solution only in simple geometries and with
homogeneous parameters. 
Furthermore, classical numerical methods, such as finite differences or finite elements, become computationally
intractable due to a relatively high dimensionality ($\mathbb{R}^3$ in space and $\mathbb{S}^2$ in direction).  

\paragraph{Klein--Nishina formula.}
Of particular interest in emission tomography is \emph{Compton scattering} which arises from the interaction of a photon
with electrons in an atom and gives rise to a change in angle and change in energy. 
For photon scattering through an angle $\cos \theta = \angdir \cdot \angdir'$, the differential cross section is given
by the Klein--Nishina~formula  for unpolarised incident photons:
\begin{equation} 
\qquad \frac{d\sigma}{d\angdir} = \frac{r_e^2}{2} \left(\frac{E}{E'}\right)^2 \left[ \frac{\Energy}{\Energy'} +
\frac{\Energy'}{\Energy} - \sin^2\theta \right], 
\label{eq:KleinNishina}
\end{equation}
where  $r_e$ is the classical electron radius,  $\Energy'$ is the incident photon energy, 
and $\Energy$ is the scattered photon energy. 
After imposing conservation of energy and (relativistic) momentum on the photon–electron collision, and treating the
target electron as free and initially at rest, $\Energy$  can be written as
\[ 
    \Energy = \frac{\Energy'} {1+\dfrac{\Energy'}{m_e c^2}(1-\cos\theta)}. 
\] 
\paragraph{Series solution for \acs{rte}}
A solution to the \ac{rte} can be expressed through the \emph{Method of Successive Approximations}~\cite{sobolev63},
which is formally developed from a Neumann series expansion of \cref{eq:RTE2_MultiEnegy}, and is written as
\begin{equation}
    \radiance = \propop_{01}\src_{01} + 
    \sum_k \propop_{k1} \attn_k\scatop_{0k1}\propop_{0k}\src_{0k} + 
    \sum_{k,k'} \propop_{k'1} \attn_{k'} \scatop_{kk'1} \propop_{kk'} \attn_k \scatop_{0kk'} \propop_{0k}\src_{0k} + 
    \ldots
    \label{eq:SuccessiveAprroximations}
\end{equation}
where $\src_{ik}$ is shorthand for the photon source at $\pos_i$ (in the direction of $\pos_k$)
and $\scatop_{ikj}$ represents the
probability that a photon propagating from $\pos_i$ to $\pos_k$ scatters into the
direction towards $\pos_j$, see \cref{fig:BornSeries1}.
\begin{figure}[t]
    \centering
    \resizebox{0.9\linewidth}{!}{
\definecolor{mcblue}{RGB}{31,119,180}
\begin{tikzpicture}[
  font=\small,
  path/.style={-{Latex[length=2mm]}, thick, mcblue},
  vtx/.style={circle, fill=black, inner sep=1.1pt},
  src/.style={circle, fill=mcblue!25, draw=mcblue, inner sep=1.6pt},
  det/.style={rectangle, fill=black!15, draw=black!60, minimum width=2mm, minimum height=4mm, inner sep=0pt},
  lab/.style={font=\footnotesize},
]
\begin{scope}[shift={(0,0)}]
  \filldraw[black!6, draw=black!35] (1.1,0.8) ellipse[x radius=1.75cm, y radius=1.3cm];
  \node[font=\scriptsize\itshape, black!55] at (1.0,1.72) {patient, $\attn(\pos)$};
  \node[src] (a0) at (0,0) {};
  \node[det, rotate=-35] (a1) at (3.2,1.85) {};
  \draw[path] (a0) -- (a1);
  \node[lab, above left] at (1.6,0.92) {$\propop_{01}$};
  \node[lab, below] at (a0.south) {$\pos_0$};
  \node[lab, right=1pt] at (a1.east) {$\pos_1$};
  \node[font=\footnotesize, align=center] at (1.6,-1.0) {(a) unscattered};
\end{scope}
\begin{scope}[shift={(4.6,0)}]
  \filldraw[black!6, draw=black!35] (1.1,0.8) ellipse[x radius=1.75cm, y radius=1.3cm];
  \node[src] (b0) at (0,0) {};
  \node[vtx] (bk) at (2.0,0.45) {};
  \node[det, rotate=-35] (b1) at (3.3,1.85) {};
  \draw[path] (b0) -- (bk);
  \draw[path] (bk) -- (b1);
  \node[lab, above] at (1.0,0.25) {$\propop_{0k}$};
  \node[lab, above left] at (2.65,1.15) {$\propop_{k1}$};
  \node[lab, below] at (b0.south) {$\pos_0$};
  \node[lab, below right] at (bk) {$\pos_k$};
  \node[lab, below=7pt] at (bk) {$\scatop_{0k1}$};
  \node[lab, right=1pt] at (b1.east) {$\pos_1$};
  \node[font=\footnotesize, align=center] at (1.6,-1.0) {(b) single scatter};
\end{scope}
\begin{scope}[shift={(9.2,0)}]
  \filldraw[black!6, draw=black!35] (1.1,0.8) ellipse[x radius=1.75cm, y radius=1.3cm];
  \node[src] (c0) at (0,0) {};
  \node[vtx] (ck) at (2.35,0.5) {};
  \node[vtx] (ckk) at (1.25,1.35) {};
  \node[det, rotate=-35] (c1) at (3.05,2.3) {};
  \draw[path] (c0) -- (ck);
  \draw[path] (ck) -- (ckk);
  \draw[path] (ckk) -- (c1);
  \node[lab, above] at (1.1,0.28) {$\propop_{0k}$};
  \node[lab, right] at (1.85,1.02) {$\propop_{kk'}$};
  \node[lab, above left] at (2.2,1.85) {$\propop_{k'1}$};
  \node[lab, below] at (c0.south) {$\pos_0$};
  \node[lab, below right] at (ck) {$\pos_k$};
  \node[lab, below=7pt] at (ck) {$\scatop_{0kk'}$};
  \node[lab, left] at (ckk) {$\pos_{k'}$};
  \node[lab, anchor=north east] at (1.3,1.25) {$\scatop_{kk'1}$};
  \node[lab, right=1pt] at (c1.east) {$\pos_1$};
  \node[font=\footnotesize, align=center] at (1.75,-1.0) {(c) double scatter};
\end{scope}
\end{tikzpicture}}
    \caption{First three terms of the successive-approximation series in \cref{eq:SuccessiveAprroximations} for
    (a) unscattered, (b) single-scattered and (c) double-scattered photons.
    $\propop_{ij}$ propagates from $\pos_i$ to $\pos_j$, and $\scatop_{ikj}$ describes 
    scattering at $\pos_k$.
    The factors $\attn_k$ of the equation are omitted for clarity.}
    \label{fig:BornSeries1}
\end{figure}

\subsection{The continuous-to-discrete forward model as a composition\label{sec:NonlinearForward}}

Let us reconsider the continuous-to-discrete operator 
as a composition of a \emph{source}, \emph{propagator} and a \emph{measurement} operator, i.e.
\begin{align}
\bar{\promptsvec} &= \cdsrm(\attn)\actvty  + \bar{\randomsvec} = \measop \circ \propLS(\attn)\circ \srcop(\actvty)   +
\bar{\randomsvec} \label{eq:FwdOpLippmannSchwinger} \\
    &=  \underbrace{\measop \circ \propop \circ\srcop(\actvty) }_{\bar{\truesvec} = \cdsrm^{(0)}\actvty } +
    \underbrace{\measop\circ \left( \propop\circ \attn \scatop  + \ldots + \left( \propop\circ \attn \scatop \right)^k
     + \ldots \right)\circ \propop\circ \srcop(\actvty)}_{\bar{\scattervec} = \left(\cdsrm^{(1)} + \ldots +\cdsrm^{(k)}
    + \ldots \right)\actvty} + \bar{\randomsvec},
\end{align}
where $\propLS(\attn) =  \left[\transop_{\attn} - \attn \scatop \right]^{-1}$ is the \emph{Lippmann--Schwinger} integral
operator. 
Under this notation we see that     \vspace{-0.5em}
    \begin{equation}\label{eq:scatterterms}
\bar{\truesvec} = \cdsrm^{(0)}(\attn) \actvty\, ,    \qquad    \bar{\scattervec} = \sum_{k=1}^{\infty}
\cdsrm^{(k)}(\attn) \actvty \,,      
    \end{equation}
where the superscript $k$ explicitly indicates the number of scattering events in each term of the series.

\subsection{Discretisation\label{sec:discretisation}}

In practical image reconstruction implementations, both the kernel $\cdkern$ and activity $\actvty$ are discretised by
projection onto a set of basis functions $\{\imbss_{\pixind}\}_{\pixind=1}^{\numimg}$.
For (quasi-)continuous measured data, a similar approach can be used for discretisation.
This yields the discrete \ac{srm} $\srm$ and a vectorised activity image $\actvtyvec$.
The $(\datind,\pixind)$ entry of $\srm$ quantifies the
contribution of basis function $\imbss_{\pixind}$ to data point $\prompts_{\datind}$.
The entries of $\srm$ and of the vectorised activity image are
\begin{align*}
    \srmel_{\datind,\pixind} := \langle \chrf_{\datind},\, \imbss_{\pixind} \rangle\,, \qquad \actvty_{\pixind} :=  \langle
    \actvty,\, \imbss_{\pixind} \rangle,
\end{align*}
where $\langle\cdot,\cdot\rangle$ denotes the inner product in $L^2(\Omega)$. 
Vectors of measurements and activity functions are then denoted as $\promptsvec\in\mathbb{R}^{\nummeas},\,
\actvtyvec\in\mathbb{R}^{\numimg}$.
The usual choice of basis functions is the set of indicator functions on (cuboid) voxels, though alternatives such as blobs or
splines are also used~\cite{Lougovski2015}.
Note that most \ac{mc} simulators accept a definition of the activity image in terms of geometric shapes and their
intensity, as well as a vectorised activity image.

Accordingly, the resultant discretised forward model is typically expressed in terms of the expected measurement vector
$\bar{\promptsvec}$ as the affine mapping
\begin{align}
     \bar{\promptsvec} = \srm {\actvtyvec}+ \backgrvec\,.
     \label{eq:expected_prompts}
\end{align}
\noindent
Here, $\bar{\prompts}_{\datind}$ is the expected measurement by detector element $i$, and $\backgrvec$ is a term that is
pre-estimated before the reconstruction.
\begin{remark}
    In current practice, $\srm$ usually models only unscattered events (i.e. $\srm \equiv \srm^{(0)}$), and $\backgrvec$
    corresponds to the sum of expected scatter and randoms/background events: $\backgrvec = \bar{\scattervec} +
    \bar{\randomsvec}$, see \cref{eq:fwd_model1}.
    The topic of ``scatter estimation'' (\cref{sec:ScatterEstimation}) should therefore be interpreted as a
    data-preprocessing step that estimates a ``fixed'' $\bar{\scattervec}$ for any given dataset.
    See also \cref{sec:JAAE}. 
\end{remark}

\begin{remark}
  In the presence of attenuation the forward operator $\cdsrm$, and thus the \ac{srm}, depend on $\attn$.
  Where appropriate, such as in \cref{sec:JAAE}, we sometimes add 
  a dependence to the \ac{srm} as $\srm (\attn)$ to make this explicit.
  Note that in \ac{pet}, positron range also depends on the density and tissue composition.
\end{remark}

\subsection{Practical implementation of the forward model\label{sec:ModelPipeLine}}

It is often convenient to factorise the forward model into several parts.
Although such a ``factorised'' system matrix is not entirely accurate due to discretisation or other approximations, it
allows for optimisation of computation and memory requirements, as well as easier adjustments based on the patient
and/or the detector performance.
A commonly used factorised model is
\begin{align}
\srm = \srm_{\rm norm} \, \srm_{\rm sin\_blur} \, \srm_{\rm att\_proj} \, \srm_{\rm img\_blur} \, \srm_{\rm img\_warp},
\label{eq:factorised_srm}
\end{align}
where the components are
\begin{itemize}
    \item \emph{\Acf{rm}} $\srm_{\rm sin\_blur}$ and $\srm_{\rm img\_blur}$ in detector space and reconstruction space,
    respectively. 
    \item \emph{Propagator} $\srm_{\rm att\_proj} $ for unscattered photons (including attenuation).
    This is the discretisation of the propagation operator. 
    \item \emph{Normalisation} $\srm_{\rm norm}$, a diagonal matrix encoding the detection efficiencies, including
    dead-time and optionally measurement duration and radioactive decay. 
    \item \emph{Motion} $\srm_{\rm img\_warp}$, a warping/resampling matrix to take motion into account.
    Note that the attenuation image should be adjusted as well. 
\end{itemize}
This factorisation is reminiscent of \cref{eq:FwdOpLippmannSchwinger} for the continuous-to-discrete case.
For \ac{pet}, the attenuation of both annihilation photons can be combined (assuming collinearity), leading to a
factorised
model for the unscattered coincidences:
\begin{align}
\srm_{\rm att\_proj} = \srm_{\rm att} \, \srm_{\rm proj}, 
\label{eq:factorised_srm_pet}
\end{align}
with $\srm_{\rm proj} $ a propagator for the unscattered photons and $\srm_{\rm att}$  a diagonal matrix encoding the
total attenuation on each \ac{lor}.
The detector-space blur $\srm_{\rm sin\_blur}$ is an approximation that is
adequate for \ac{pet}, where the \ac{lor} approximately fixes the direction of incidence, but not for the
depth-dependent collimator--detector response in \ac{spect}, which can only be separated from the patient part
through the phase space of the photons reaching the collimator (position, direction and energy).

Factorisation \eqref{eq:factorised_srm} can be useful to separate object-dependent terms (such as attenuation and density) from
system-dependent terms.
The latter are usually  precomputed, for instance using \ac{mc}.
\Cref{sec:FastSysResp} will discuss this factorisation as well as alternative approaches in more detail.

\begin{remark}\label{remark:fwd_dependencies}
    The forward model depends on the detector (and collimator).
    However, in \ac{pet} and \ac{spect}, it also depends on the physical properties of the object being imaged.
    The scattering and absorption probabilities, and
hence $\mu$, depend on the material and the photon
energy.
    In addition, the cross sections related
    to positron range and bremsstrahlung generation also depend on material properties, including atomic number $Z$ and density,
    as well as any external magnetic fields
    in \ac{pet}/\ac{mr}.
    Note that this dependency on the object needs to be taken into account when motion is present.
    Finally, the photon source also depends on the radionuclide and its decay chain.
    For accurate \ac{mcs}, all this information needs to be made available to the simulator.\\
    In current practice, $\mu(E)$ and other material properties are derived from \ac{ct} or \ac{mr} images.
    However, this mapping is not unique, e.g. when contrast fluids are used, unless multi-spectral \ac{ct} is available
    \cite{wang_integration_2026}.
\end{remark}

\subsection{Reconstruction in emission tomography using a deterministic forward model\label{sec:DeterministicRecon}}

Current emission reconstruction methods in \ac{pet} and \ac{spect} mainly use a deterministic forward model, i.e., both
the system matrix $\srm$ and background term $\backgrvec$ are fixed.
These methods are based on either \acl*{ml} or \ac{map} estimators.
In particular, given the Poisson statistics of photon counting, the log-likelihood $\loglik$ of the observed data
$\promptsvec$, given the activity vector $\actvtyvec$, is given by
\begin{align}
    \label{eq:loglikelihood}
    \loglik(\promptsvec, \srm, \backgrvec, \actvtyvec) 
    =\log p(\promptsvec\mid \actvtyvec) = \sum_{\datind=1}^{\nummeas} \left[\prompts_{\datind} \log
    \bar{\prompts}_{\datind}(\actvtyvec) - \bar{\prompts}_{\datind}(\actvtyvec) \right]+ {\rm{const}},
\end{align}
where ``const'' indicates terms that do not depend on $\actvtyvec$.
It is worth noting (and adopting the convention $0 \log 0=0$) that the first term in the sum only contains elements
where $\prompts_{\datind}>0$ 
and its calculation therefore only needs the corresponding rows of $\srm$ and $\backgrvec$.
The second term in the sum can be expanded as
\begin{align}
     \sum_{\datind=1}^{\nummeas} \bar{\prompts}_{\datind}(\actvtyvec)=
     \sum_{\pixind=1}^{\numimg} \underbrace{\left[\sum_{\datind=1}^{\nummeas}
     \srmel_{\datind\pixind}\right]}_{\sensvecel_{\pixind}} \actvty_\pixind +
     \sum_{\datind=1}^{\nummeas}\backgr_{\datind}.
     \label{eq:SensVec}
\end{align}
The vector whose elements $\sensvecel_{\pixind}$ are specified within the square brackets is sometimes called the
``sensitivity image'', as it is proportional to the probability of detection of activity at voxel $\pixind$, although
this
terminology can be confusing if not all physical effects are included in $\srm$.
We will denote it as $\sensvec$.

Maximising \cref{eq:loglikelihood} with respect to $\actvtyvec$ yields the maximum likelihood solution.
Often, a non-negativity  constraint $\actvtyvec\geq0$ is used.
The corresponding solution can be obtained using the classical \ac{mlem} algorithm:
\begin{equation}
    \actvtyvec^\mathrm{new}= \frac{\actvtyvec^\mathrm{old}} {\sensvec} \srm\transp\left( \frac{\promptsvec}{\srm
    \actvtyvec^\mathrm{old}+\backgrvec}\right)\,,
    \label{eq:mlem}    
\end{equation}
where element-wise division and multiplication are implied.
Restricting the update to use only a ``subset'' of the data, corresponding for instance to a subset of projections or
detector elements, yields \ac{osem}, a standard reconstruction method.  
However, the inverse problem is ill posed in general.
In particular when including resolution modelling this leads to ringing artefacts \cite{Nuyts2014UnconstrainedSolution}.
Various regularisation strategies are used to stabilise the solution, see for instance
\cite{Qi2006,ehrhardt2021MultimodalityImagingStructurePromoting, arridge2021OverviewSynergisticReconstruction}.

\begin{remark}\label{remark:mismatched_projectors}
    While the above equations are written in terms of $\srm$, some authors have used mismatched forward and
    backprojection operators.
    A common example is where the forward operator includes full resolution modelling, but the backprojector does not.
    This has been shown to increase convergence rate, but potentially at the expense of increased noise near
    convergence.
    \textcite{zeng2000unmatched} analysed this set-up for \ac{wls} reconstruction using Landweber iterations, showing
    good
    results at early iterations under certain conditions for the projector pair.
    However, their analysis
    also shows that the algorithm no longer converges to the \ac{wls} minimum.
    Similar results are reported there for \ac{mlem} with mismatched projectors.
    It is less clear how mismatched projectors perform in penalised reconstruction algorithms.\\    
    Another common case of mismatched projectors is where the forward operator is implemented using \ac{mcs}, while the
    backprojector is analytic (and deterministic).
    This is further discussed in \cref{sec:InvProb}.
\end{remark}

Many algorithms for (constrained) optimisation can be used, often based on the gradient of the log-likelihood
\begin{align}
    \label{eq:gradientloglikelihood}
    \nabla_{\actvtyvec} \loglik(\promptsvec, \srm, \backgrvec, \actvtyvec) 
    =
    \srm\transp\left( \frac{\promptsvec}{\srm \actvtyvec+\backgrvec} - \onevec \right)
    =
    \srm\transp\left( \frac{\promptsvec}{\srm \actvtyvec+\backgrvec}\right) - \sensvec
    \,.
\end{align}
\Ac{mlem} can thus be seen to be a type of preconditioned gradient ascent.

The objective function can be written as a sum of subsets or mini-batches, enabling the use of stochastic gradient
descent, or its many variants, e.g. variance-reduced modifications, such as \acl*{svrg} \cite{johnson2013accelerating}
or \acl*{saga} \cite{defazio2014saga}, see \cite{Twyman2023}.
For a review, see~\cite{Ehrhardt_2025}.

It is worth noting that if the \ac{srm} $\srm$ is factorised and the first factor is a diagonal matrix, \textit{i.e.}
$\srm = D \tilde{\srm}$, that factor can be moved to the background term,
\begin{align*}
    \prompts_\datind \log \left( \srm_\datind \actvtyvec + \backgr_\datind \right) = 
    \prompts_\datind \log \left( \tilde{\srm}_\datind \actvtyvec + d_\datind^{-1} \backgr_\datind \right) +
     {\rm{const}},
\end{align*}
which allows writing the reconstruction in terms of $\tilde{\srm}=D^{-1}\srm$, $\tilde{\backgrvec}=D^{-1}\backgrvec$ and
$\sensvec$.
This is particularly useful when pre-computing $\tilde{\backgrvec}$ and $\sensvec$.

The above equations can be extended to the case of list-mode data.
The theoretical derivation assumes an impulse-valued Poisson random process and it can be written in terms of continuous
measurement variables \cite{Barrett1977}.
Gradients are computed as a sum over events $e$
\begin{align}
    \label{eq:gradientlist-modeloglikelihood}
    \dbyd{\loglik}{\actvty_{\pixind}} = \sum_e \srmel_{e\pixind}\left( 
          \frac{1}{\sum_{\pixind'} \srmel_{e\pixind'} \actvty_{\pixind'}+\backgr_e}\right) - \sensvecel_\pixind
    \,.
\end{align}

\subsection{Where does \acl{mc} fit?}\label{sec:WhereDoesMCfit}

In current practice the forward model developed above is evaluated deterministically: both
the \ac{srm} $\srm$ and the background term $\backgrvec$ in \cref{eq:expected_prompts} are often
based on simplified analytic models
and kept fixed during reconstruction. 
\Ac{mc} enters this picture not as a separate model, but as a
stochastic realisation of the same physics: each simulated photon history samples the
propagator terms $\propop_{kj}\scatop_{ikj}\propop_{ik}$ of the
successive-approximations series in \cref{eq:SuccessiveAprroximations}, intrinsically modelling Compton scattering
(\cref{sec:ScatterEstimation}),  and can in addition
capture effects that the analytic model omits, such as positron range in heterogeneous materials
(\cref{sec:PositronRangeModelling})
and advanced detector effects such as septal penetration, detector scatter and local energy deposition. 

\begin{figure}[tb]
  \centering
  \resizebox{\linewidth}{!}{
\definecolor{mcblue}{RGB}{31,119,180}
\definecolor{mcorange}{RGB}{214,110,20}
\begin{tikzpicture}[
  font=\small,
  >={Latex[length=2.2mm]},
  stage/.style={rectangle, rounded corners=2pt, draw=black!55, fill=black!6,
                text width=46mm, align=center, minimum height=11mm, inner sep=4pt},
  phys/.style={rectangle, rounded corners=2pt, draw=mcblue, fill=mcblue!10,
               align=center, font=\footnotesize, inner sep=4pt},
  numer/.style={rectangle, rounded corners=2pt, draw=mcorange, fill=mcorange!18,
                align=center, font=\footnotesize, inner sep=4pt},
  note/.style={rectangle, rounded corners=2pt, draw=black!45, fill=black!3,
               align=left, inner sep=6pt, font=\footnotesize},
  flow/.style={->, thick, black!70},
  link/.style={->, mcblue, dashed, thick},
  linko/.style={->, mcorange, dashed, thick},
]
\def\xl{2.65}\def\xr{10.55}

\node[phys, text width=68mm] (legP) at (3.6,0)
  {\textbf{MC as physics engine}\\ complex integrand, huge domain (accuracy)};
\node[numer, text width=68mm] (legN) at (11.25,0)
  {\textbf{MC as numerical estimator}\\ simple integrand, huge domain (feasibility)};

\node[stage] (img) at (\xl,-1.55) {\textbf{Activity image}\\[2pt]\mbox{$\actvtyvec$}};
\node[stage] (fwd) at (\xl,-4.15)
  {\textbf{Forward}\\ {\footnotesize(expectation of measured data)}\\[2pt]
   \mbox{$\bar{\promptsvec}(\actvtyvec,\attnvec)=\srm(\attnvec)\actvtyvec+\backgrvec$}};
\node[stage] (cmp) at (\xl,-6.95)
  {\textbf{Compare}\\[2pt]\mbox{$\adjsrcvec=\promptsvec/\bar{\promptsvec}(\actvtyvec,\attnvec)-\onevec$}};
\node[stage] (adj) at (\xl,-8.85) {\textbf{Adjoint step}\\[2pt]\mbox{$\srm\transp\adjsrcvec$}};
\node[stage] (upd) at (\xl,-11.1)
  {\textbf{Update}\\ {\footnotesize preconditioned gradient ascent}\\[3pt]
   \mbox{$\actvtyvec^{+}=\actvtyvec+\mathrm{diag}\!\left(\actvtyvec/\sensvec\right)\nabla_{\actvtyvec}\loglik$}\\[2pt]
   {\footnotesize with}\ \ \mbox{$\nabla_{\actvtyvec}\loglik=\srm\transp\adjsrcvec$}};

\draw[flow] (img) -- (fwd);
\draw[flow] (fwd) -- (cmp);
\draw[flow] (cmp) -- (adj);
\draw[flow] (adj) -- (upd);
\draw[flow] (upd.west) -- ++(-0.35,0) |- (img.west);

\node[phys, text width=76mm, align=left] (bF) at (\xr,-4.15)
  {\textbf{MC realisations of the forward model}\\[3pt]
   (1)\ \ \mbox{$\bar{\promptsvec}(\actvtyvec,\attnvec)=\srm(\attnvec)\actvtyvec+\backgrvec^{\mcop}(\actvtyvec,\attnvec)$}\\
   \hspace*{1.6em}scatter as a nuisance, from MC\\[2pt]
   (2)\ \ \mbox{$\bar{\promptsvec}(\actvtyvec,\attnvec)=\mcop(\actvtyvec,\attnvec)$}\\
   \hspace*{1.6em}on-the-fly MC step, no stored SRM \mbox{$\srm$}\\[2pt]
   (3a)\ \mbox{$\bar{\promptsvec}(\actvtyvec,\attnvec)=\srm^{\mcop}\actvtyvec$}\\
   \hspace*{1.6em}precompute and store the full SRM \mbox{$\srm$}\\[2pt]
   (3b)\ \mbox{$\bar{\promptsvec}(\actvtyvec,\attnvec)=\srm_{\mathrm{rest}}(\attnvec)\,\srm_{\mathrm{PR}}^{\mcop}(\attnvec)\,\actvtyvec$}\\
   \hspace*{1.6em}precompute one SRM factor (e.g.\ positron range)};
\draw[link] (bF.west) -- (fwd.east);

\node[align=left, font=\footnotesize, black!65, text width=76mm] (data) at (\xr,-6.95)
  {measured data \mbox{$\promptsvec$}\\ (scattered and unscattered events in the energy window(s))};
\draw[flow, black!55] (data.west) -- (cmp.east);

\node[phys, text width=76mm] (bAT) at (\xr,-8.85)
  {\textbf{Adjoint \mbox{$\srm\transp$}}\\ MC or analytical backprojection};
\draw[link] (bAT.west) -- (adj.east);

\node[numer, text width=76mm, anchor=south] (bS) at (\xr,-11.05)
  {\textbf{Sensitivity \mbox{$\sensvec=\srm\transp\onevec$}}\\ sample the data space and back-project};
\node[phys, text width=76mm, anchor=north] (bS2) at (\xr,-11.25)
  {or: simulate emissions per voxel and count the detected events};
\draw[linko] (bS.west) -- ++(-0.45,0) |- ([yshift=1.5mm]upd.east);
\draw[link] (bS2.west) -- ++(-0.45,0) |- ([yshift=-1.5mm]upd.east);

\node[note, text width=147mm, anchor=north] (dec) at (7.4,-12.75)
  {\textbf{Things to consider}\\[3pt]
   \mbox{$\bullet$}\ \ (In-patient) scatter can be regarded as \emph{nuisance} or \emph{signal}; the choice probably depends on
   what is being estimated (activity or attenuation).\\[2pt]
   \mbox{$\bullet$}\ \ Current MC simulators favour the forward path; a cheap analytical adjoint mismatches forward and
   back-projection.\\[2pt]
   \mbox{$\bullet$}\ \ MC is stochastic: the variance perturbs the likelihood and its gradients \mbox{$\Rightarrow$} variance reduction
   is needed.\\[2pt]
   \mbox{$\bullet$}\ \ Storing the full SRM (3a) is bounded by the problem size (voxels \mbox{$\times$} data bins): feasible for some
   SPECT systems, prohibitive for most TOF-PET systems.\\[2pt]
   \mbox{$\bullet$}\ \ For sparse data, list-mode MLEM needs the forward model only in bins with counts (far fewer than the
   total); ideally an MC simulator would evaluate only those LORs, not all of them.};
\end{tikzpicture}}
  \caption{
  Possible use cases of Monte Carlo techniques in emission tomography reconstruction, shown as one \ac{mlem}
  iteration in preconditioned-gradient-ascent form.
  Blue: \ac{mc} as a physics engine; orange: \ac{mc} as a numerical estimator.
  See text for details.}
  \label{fig:where_mc_fits}
\end{figure}
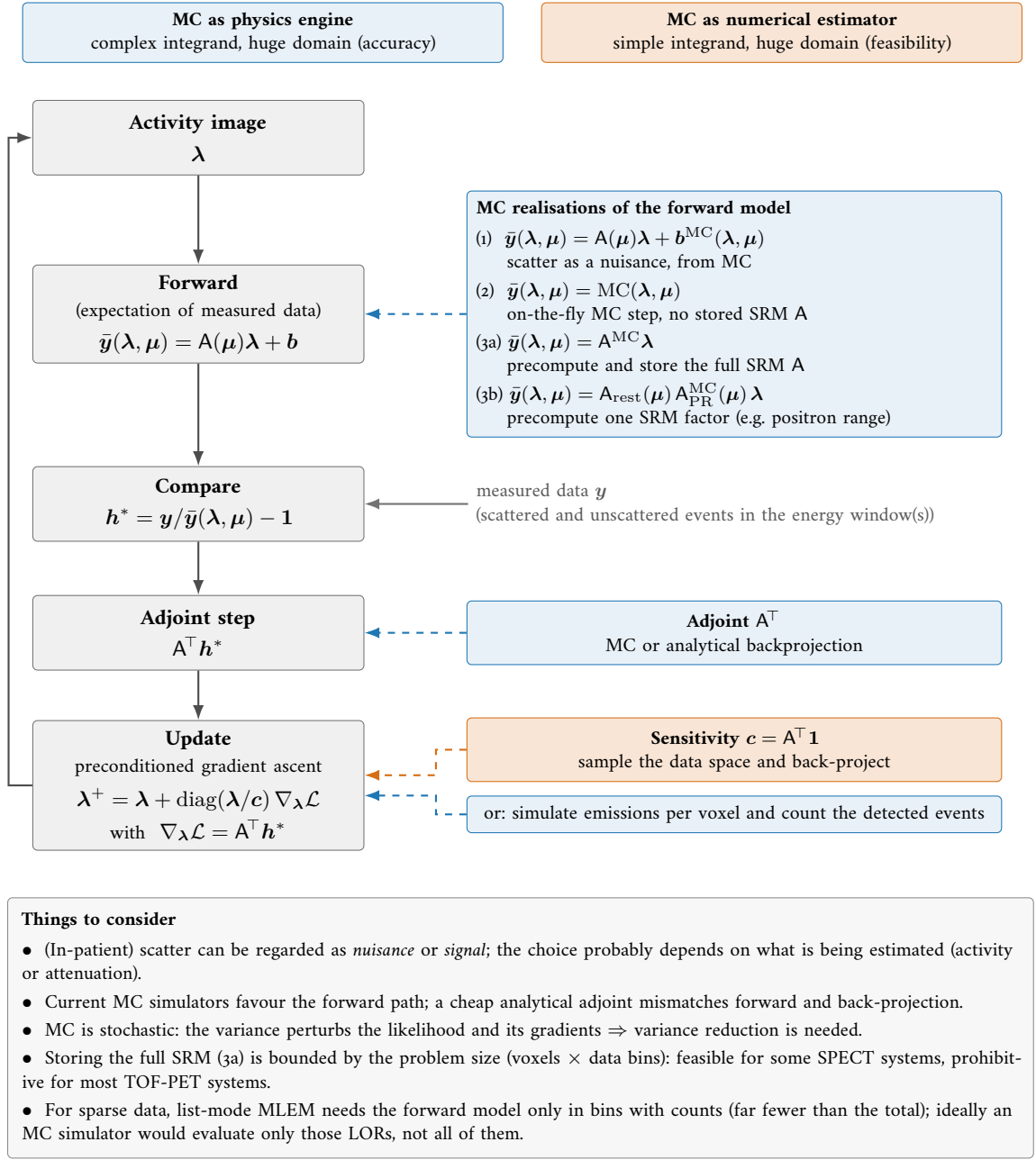

The relevant question is therefore not whether \ac{mc} describes the forward model, but which of its factors or
terms are best replaced by an \ac{mc} estimate, and what this implies for the
reconstruction. 
In doing so, it is useful to distinguish two roles that \ac{mc} can play.
In the first, it acts as a \emph{physics engine}; here the appeal is the fidelity with which it
samples photon transport, better capturing (higher-order) scatter and other effects that
analytic models struggle to approximate accurately. 
In the second, it acts as a \emph{numerical estimator}
of a high-dimensional sum or integral whose integrand may itself be known analytically, the
appeal here being feasibility rather than accuracy.
Mathematically, both roles are \ac{mc} integration, since the expected data are themselves a high-dimensional
integral over emission points, directions and interaction paths.
Both integrals run over very large domains.
They differ in the integrand, which in the first case encodes the full
transport physics, whereas in the second it is simple and often known analytically.
The first three uses below are of the former kind; the sensitivity image at the end can be estimated in either
role.
We consider them from the outside of the affine model in \cref{eq:expected_prompts} inwards, an
ordering that also reflects increasing difficulty.
\Cref{fig:where_mc_fits} summarises
these entry points and the two roles.

The \emph{physics-engine} role can be stated compactly.
A photon \ac{mc} simulation samples individual photon paths through the series in \cref{eq:SuccessiveAprroximations},
drawing a free-flight distance from the attenuation law ($\propop$) and a new direction and energy from the phase
function ($\scatop$) at each scatter, so that averaging histories estimates $\radiance$ without evaluating the nested,
high-dimensional integrals explicitly.
Its appeal is therefore greatest where the analytic series becomes intractable, namely for higher-order scatter and
for the additional physics that the analytic model omits.

\paragraph{Estimating the background (scatter) term.}
The background term $\backgrvec$ collects contributions treated as nuisance, namely scatter
in both \ac{spect} and \ac{pet}, and additionally randoms in \ac{pet}. 
Estimating the scatter component of $\backgrvec$ with
\ac{mc} promises a more accurate $\backgrvec$ than analytical methods, which in particular
struggle with higher-order scatter.
The main challenge is that a low-variance estimate is
required, which demands either long simulations or effective variance reduction
(\cref{sec:varred}). 
This use of \ac{mc}, in or near the reconstruction loop, is treated in detail in
\cref{sec:ScatterEstimation}.

Whether scatter should be treated as a nuisance at all, however, depends on what is being
reconstructed.
When ``only'' the activity is estimated, the scattered events carry little
information about $\actvtyvec$ (especially in \ac{pet} due to its limited energy resolution), and absorbing them into a
fixed
pre-estimated $\backgrvec$ discards little, at least unless multiple energy windows are used. 
In the joint estimation of activity and attenuation from emission data, in contrast, the
scattered photons are  informative about the attenuation $\attn$, so treating them as
nuisance forgoes useful information; it then becomes attractive to move scatter into the
non-nuisance part of the forward model. 
The reconstruction must then evaluate gradients of the Poisson log-likelihood
in \cref{eq:loglikelihood} with respect to $\attn$ through the scatter term, which is where a differentiable, 
scatter-aware \ac{mc} forward model becomes valuable.
This is developed in \cref{sec:JAAE}.

\paragraph{Precomputing the system matrix.}
A second option is to use \ac{mc} to precompute parts of the \ac{srm} $\srm$. 
The factorisations in \cref{eq:factorised_srm,eq:factorised_srm_pet} 
are useful here, as they separate system-dependent factors, which can be precomputed once, 
from object-dependent factors, which depend on the attenuation $\attn$ and density and 
must be re-estimated for every acquisition. 
The remaining challenges are practical: $\srm$ can be too large to store,%
\footnote{For a state-of-the-art
whole-body \ac{tof} \ac{pet} scanner with a timing resolution of about 200\,ps and an axial field of view of
about 25\,cm (some $10^9$ \acp{lor}, a few tens of \ac{tof} bins and $10^6$--$10^7$ voxels), the full
uncompressed system matrix has of the order of $10^{17}$ elements.
For \ac{lafov} scanners without a restricted acceptance angle, this number grows approximately with the cube
of the axial length, since the number of \acp{lor} grows quadratically and the number of voxels linearly.}
it becomes non-sparse once scatter is included, and its object-dependent parts must be
recomputed per acquisition. 
Fast and accurate construction of the system response is the
subject of \cref{sec:FastSysResp}, while the associated storage and memory
constraints are discussed in \cref{sec:comparch}.

\paragraph{Sampling within the reconstruction loop.}
Most ambitiously, \ac{mc} can be used inside the iterative reconstruction itself, either
only for the complete forward step $\srm\actvtyvec + \backgrvec$ while retaining an analytical
adjoint (backprojection) $\srm\transp$, or for both the forward and the adjoint. 
This raises several open questions, which we frame here and address in \cref{sec:InvProb}. 
First, a stochastic forward model is incompatible with the deterministic Poisson log-likelihood
in \cref{eq:loglikelihood}, so that a tractable surrogate that allows the computation of the gradient
(or a descent direction) is required.
Second, most current \ac{mc} simulators are optimised for the forward
direction and not for the adjoint. 
Third, the variance and bias of \ac{mc} gradient estimates interact with stochastic optimisation schemes
which are commonly used to accelerate reconstruction.
Stochastic reconstruction and the consequent requirements for an adjoint \ac{mc} are developed in
\cref{sec:StochasticReconstruction,sec:AdjointMCrole}.

\paragraph{Estimating the sensitivity image.}
The sensitivity image $\sensvec = \srm\transp \onevec$ (\cref{eq:SensVec}) is a sum over the whole data space, which
can be very large, for instance for detectors with high spatial, energy and/or timing resolution.
\Ac{mc} can estimate it in either of the two roles.
As a numerical estimator, it samples elements of the data space, instead of enumerating them, and back-projects
them with the transpose of the \ac{srm}, $\srm\transp$; non-uniform sampling can reduce the variance~\cite{qi2006sens}.
As a physics engine, it simulates emissions from each voxel and counts the detected
events~\cite{Rafecas2004a,Lazaro2005PMB}, which requires no adjoint \ac{mc} (\cref{sec:AdjointMCrole}) and can be
preferable when the data space is too complex to be sampled efficiently.
Both estimates are noisy, and the emission-based estimate does not in general match the back-projector used in the
reconstruction.
Both errors propagate into the reconstructed image~\cite{qi2004sens,Qi2005}.

\medskip

Beyond these roles inside the reconstruction, a further --- and perhaps trivial, but
important --- use of \ac{mc} is to validate and improve the analytical models on which
emission tomography routinely relies.
Because a sufficiently detailed \ac{mc} simulation can
serve as a controlled ground truth, it is a natural reference against which approximate
analytical models (for scatter, positron range, detector response, and the like) can be
checked, calibrated, and refined.

Common to all of these uses is a single prerequisite: the \ac{mc} simulation must be both
accurate enough to justify replacing the analytic model, and fast enough to be run, often
repeatedly, within a reconstruction.
This turns the question of where \ac{mc} fits from a
conceptual one into a practical one, and the practical hurdle is steep, as a direct,
brute-force \ac{mc} simulation is typically far too slow for this setting. 
The following
section therefore reviews \ac{mc} simulation in emission tomography, together with the
variance-reduction and acceleration techniques that aim to make it fast enough to be
useful in the roles outlined above (\cref{sec:MCoverview}).

\newsectionpage
\section{Monte Carlo simulation and its acceleration\label{sec:MCoverview}}
\sectionauthor{David Sarrut, Ane Etxebeste, Nils Krah}

\subsection{State of the art\label{sec:mcsota}}

Pioneering papers employing \ac{mc} simulations for modelling and analysis in emission tomography emerged in the early
1980s for \ac{spect}~\cite{beck1982} and the early 1990s for \ac{pet}~\cite{haynor1990}.
Since then, \ac{mc} simulation has become an indispensable tool, offering a robust and versatile method for accurately
modelling the intricate physical processes governing photon transport within a patient or phantom and subsequent
detection by the imaging system.
As described in \cref{sec:reconbasics}, \ac{mc} methods provide a stochastic realisation of the transport processes and
scattering series (\cref{sec:ScatterModel,sec:WhereDoesMCfit}).
Its strength lies in its capacity to handle sophisticated models and complex scenarios where traditional closed-form
mathematical solutions are inadequate.
By simulating the stochastic interactions of radiation, \ac{mc} methods allow for a detailed understanding of phenomena,
such as scattering and absorption, as photons propagate through both the human body and the detector elements (e.g.,
collimators in \ac{spect}, crystal rings in \ac{pet}).
For \ac{pet} specifically, \ac{mc} simulations also capture the timing information required for coincidence modelling.

Several \ac{mc} simulation software tools have been developed and utilised over the years.
\Ac{simind} was first introduced in 1989~\cite{ljungberg1989} and has been continuously developed since then.
It is specifically designed for \ac{spect} cameras and includes various \acp{vrt} to accelerate simulations.
\Ac{simind} also offers a wide range of pre-modelled industrial \ac{spect} systems ready for use.
\Ac{simset} was developed around 1993~\cite{harrison1993} and supports both \ac{spect} and \ac{pet} simulations, also
incorporating \acp{vrt} and efficient handling of voxelised volumes.
\Ac{gate} was first released in 2004~\cite{jan2004} and provides a broad array of imaging systems, including both
clinical and preclinical \ac{spect} and \ac{pet} systems~\cite{sarrut2014a}.
\ac{gate} continues to be developed~\cite{sarrut2026,krah2026} and includes specific modules for Compton
cameras~\cite{etxebeste2020}.
It is built on the \ac{geant4} general-purpose \ac{mc} toolkit.
However, \ac{gate}’s reliance on accurate, general-purpose \ac{cpu}-based particle tracking results in high
computational demands;
achieving sufficient statistics for complex acquisitions can often require days of computation on standard clusters. 

In parallel, numerous other specialised \ac{mc} tools, mostly for \ac{pet}, were developed from the 1990s onwards to
investigate specific aspects of system design, detector physics or data generation, such as PETSIM~\cite{thompson1992},
Eidolon~\cite{zaidi1999a}, SORTEO~\cite{reilhac2004}, PeneloPET~\cite{espana2006} and GAMOS~\cite{arce2008}.
More recently, the high computational demand of general-purpose codes has motivated accelerated simulators such as
\ac{pps}~\cite{scheins2021} and the \ac{gpu}-based codes discussed in \cref{sec:varred}.

In practice, \ac{mc} simulation in emission tomography models the complete physical chain from nuclear decay to
electronic signal acquisition.
At the source level, this includes single- and multi-energy gamma emissions in \ac{spect}, positron range and
non-collinear annihilation in \ac{pet}, photon cascades  associated with non-pure positron emitters (e.g., three-photon
imaging), and complex decay spectra from $\alpha$- and $\beta^-$-emitting theranostic radionuclides.
In terms of transport, it accounts for photon and electron interactions in heterogeneous patient tissues, including
photoelectric absorption, multiple-order Compton and Rayleigh scattering, and bremsstrahlung, as well as interactions
within passive extracorporeal structures (patient bed, headrests, and gantry shielding).
Within the detection system, \ac{mc} methods account for collimator septal penetration, scatter, and X-ray fluorescence
in \ac{spect}, along with scintillation physics, inter-crystal scatter, depth-of-interaction, and optical photon
tracking in \ac{pet}.
Finally, the electronic digitisation chain models energy and timing blurring, dark counts, dead-time, and pulse pile-up,
before sorting events into energy-windowed projections or coincidence streams (classifying prompt, true, scatter, and
random pairs with time-of-flight tagging). 

\subsection{Variance reduction and acceleration approaches\label{sec:varred}}

The primary objective of a Monte Carlo simulation in emission tomography is to estimate an expectation value
(observable) $\mathbb{E}_{\prob}[f] = \int f(\omega) \prob(\omega)\,\mathrm{d}\omega$,
where $\prob$ denotes the physical probability
density function governing particle transport and $f$ is the scoring function (e.g., deposited energy or bin detection).
In analogue simulations with $H$ independent simulated particle histories (sample paths) $\{\omega_h\}_{h=1}^H$ drawn
directly
from $\prob$, the sample-mean estimator is given by:

\begin{equation}
\hat{f}_H = \frac{1}{H} \sum_{h=1}^H f(\omega_h), \quad \omega_h \overset{\mathrm{i.i.d.}}{\sim}  \prob.
\label{eq:analog_estimator}
\end{equation}

Because only a fraction of emitted photons reach the detectors in emission tomography, analogue simulation is
computationally inefficient and suffers from high variance.
\Acp{vrt}, such as importance sampling~\cite{haynor1991}, reduce estimator variance by sampling from a modified proposal
distribution $\proposal$ that artificially increases the likelihood of scoring events.
To guarantee an unbiased estimate, each sample is assigned a statistical weight
$w(\omega) = \prob(\omega)/\proposal(\omega)$:

\begin{equation}
\hat{f}_H = \frac{1}{H} \sum_{h=1}^H w(\omega_h) f(\omega_h), \quad \omega_h \overset{\mathrm{i.i.d.}}{\sim} \proposal.
\label{eq:vrt_estimator}
\end{equation}

Whenever $\proposal$ is chosen such that $\mathrm{Var}_{\proposal}[w f] < \mathrm{Var}_{\prob}[f]$, the required number
of simulated histories for a given target precision is reduced.

\paragraph{\Acl{fd}}

\Ac{fd} and \ac{cfd} overcome low geometric sensitivity by deterministically projecting photons towards the detector
surface at emission or scatter sites~\cite{beck1982, haynor1990, haynor1991}.
To preserve unbiasedness, the photon’s statistical weight is multiplied by the solid angle fraction and the survival
probability along the forced trajectory.
While \ac{fd} provides significant acceleration over analogue tracking, it presents a practical trade-off: projecting to
too few target bins introduces ray artefacts and aliasing, whereas using too many targets increases computational
overhead.

\paragraph{Coincidence multiplexing in PET}
In \ac{pet}, multiplexing of back-to-back photon pairs from annihilations has been shown to speed up simulations
by a factor of approx.~4--15 \cite{scheins2021}.
In general, this \ac{vrt} can be easily integrated into all \ac{pet} \ac{mcs} implementations and is suitable on
\ac{gpu} and \ac{cpu}
hardware.

\paragraph{\Acl{arf}} 

The \ac{arf} technique replaces the computationally demanding simulation of photon transport through collimators and
detector crystals with pre-calculated probability models~\cite{song2005, descourt2010}.
By querying \aclp*{lut} or trained \aclp*{ann}~\cite{sarrut2018}, the simulator determines spatial detection
probabilities based on the incident photon's direction and energy, effectively bypassing detailed collimator
interactions.
This approach yields speed-up factors of 20--180\,\texttimes, though it requires generating dedicated models for each
collimator-detector geometry and energy window configuration~\cite{cajgfinger2018}.

\paragraph{Deep generative models}

To circumvent particle tracking in voxelised media, deep generative architectures, such as conditional
\aclp*{gan}~\cite{goodfellow2014}, have been adapted to learn phase-space distributions of exiting radiation directly
from activity and attenuation maps.
When combined with \acp{arf}, this approach facilitates the rapid synthesis of projection datasets, achieving
acceleration
factors exceeding 100$\times$~\cite{sarrut2021d, saporta2022a}.
Current limitations include the need to train specific generative models for distinct patient anatomies and acquisition
settings.

\paragraph{Hardware acceleration}

Complementing algorithmic \acp{vrt}, hardware acceleration exploits massive thread parallelism on \acp{gpu} as well as
vectorised multi-core \ac{cpu} architectures~\cite{smith1993, scheins2021}.
Notable \ac{gpu}-based implementations include \ac{ggems}~\cite{garcia2016}, 
\ac{mcgpupet}~\cite{herraiz2023}, 
\ac{gpet}~\cite{lai2019}, and \ac{umcpet}~\cite{galve2024}, reviewed in~\cite{chi2025}.
These specialised codes have demonstrated speed-ups ranging from 70 to 100$\times$ compared to standard \ac{cpu}-based
general-purpose engines, particularly when deployed in tandem with algorithmic \acp{vrt}.
In parallel, optimised multi-threaded \ac{cpu} codes with \acl*{simd} vectorisation (such as
\ac{pps}~\cite{scheins2021}) demonstrate high computational throughput without \ac{gpu} memory transfer overheads.

In summary, fast specialised simulators and \acp{vrt} are essential for practical forward modelling and iterative
reconstruction, while general-purpose platforms remain the gold standard for validation.
Emerging cross-disciplinary concepts, such as photon replay~\cite{yao2018} (storing random seeds to reproduce detected
paths for sensitivity and Jacobian estimation), represent additional avenues to further accelerate specialised emission
tomography workflows.

\subsection{Challenges and needs\label{sec:mcchall}} 

\Ac{mc} simulations are characterised by high computational demands, requiring the tracking of a massive number of
particle histories to achieve low statistical noise.
In analogue simulations (without \acp{vrt}), simulating a standard therapeutic administration of 7.4~GBq of
\isotope[177]{Lu} for a single 15-second \ac{spect} projection at 24 hours post-injection requires tracking
approximately $10^{10}$ gammas~\cite{sarrut2021d}.
Similarly, a phantom acquisition with 30~MBq \isotope[99m]{Tc} sphere activity plus background necessitates simulating around
$2\times10^{9}$ gammas for a single 15-second projection~\cite{vergnaud2023}.
In \ac{pet}, the burden is equally high: simulating a 3-minute bed position for a patient injected with 250~MBq of
\isotope[18]{F} involves tracking roughly $4.5\times10^{10}$ decays.
Achieving sufficient statistics for high-resolution imaging can require days of computation on standard hardware,
creating a critical conflict between physical accuracy and the time constraints essential for clinical routine.

Integrating Monte Carlo simulations directly into iterative reconstruction frameworks (introduced in
\cref{sec:reconbasics,sec:InvProb}) remains a non-trivial mathematical and software engineering challenge.
While \ac{mc} methods provide an accurate forward model, standard particle tracking is stochastic and non-differentiable
in its classical imperative implementation.
Consequently, an active area of research is the development of \emph{differentiable Monte Carlo simulators}
(\ac{diffmc})~\cite{Zhang23_DiffMC, Catumba25DiffMC}, alongside differentiable formulations in transmission and
particle imaging~\cite{Lee2015, aehle2023exploration}.
From a software architecture perspective, incorporating automatic differentiation into established \ac{mc} libraries
(e.g., \ac{geant4}, \ac{egsnrc}) is fundamentally hindered by discrete branching and boundary checks, which create
discontinuities in derivative evaluation.
Furthermore, reverse-mode automatic differentiation across long stochastic tracks entails substantial memory
requirements.
Overcoming these barriers requires mathematical shifts towards path-space formulations and variance-reduced gradient
estimators, or the use of uncorrelated stochastic adjoint models (discussed in \cref{sec:InvProb}).

Beyond direct reconstruction, Monte Carlo simulations provide an essential advantage for deep learning: the generation
of idealised, ground-truth training data that cannot be acquired experimentally, such as noise-free projections or
isolated scatter components~\cite{leube2024, kaprelian2025}.
A key challenge remains the domain gap between simulated physics and physical scanner acquisitions, where discrepancies
in detector response or anatomical modelling can degrade the clinical generalisability of networks trained on synthetic
datasets.

Finally, \ac{mc} tools must maintain physical fidelity across complex and non-standard emission scenarios.
This includes theranostic and multi-isotope regimens involving complex decay schemes (e.g., cascade gammas, high-energy
emissions from $\alpha$-emitters, and bremsstrahlung backgrounds from $\beta$-emitters), where simplified energy-window
scatter corrections and standard analytical approximations are inadequate.

\subsection{Opportunities and goals\label{sec:mcopp}}

Overcoming current computational and architectural limitations presents major opportunities to advance both the
algorithmic foundations and clinical applications of \ac{mc} simulation in emission tomography.

\paragraph{Hardware heterogeneity and scalability}

While \acp{gpu} remain central to acceleration, future performance gains will increasingly rely on scaling across
heterogeneous, multi-node \acl*{hpc} architectures combining multi-core \acp{cpu}, \acp{gpu}, and specialised \ac{ai}
hardware
(e.g., \acp{tpu})~\cite{chi2025}.
A primary goal is developing modular simulation kernels that scale seamlessly from local workstation accelerators to
distributed clusters.
This computational throughput will enable previously intractable large-scale studies, including dynamic 4D
pharmacokinetic modelling across total-body (\acs{lafov}) \ac{pet} systems and population-scale virtual clinical trials.

\paragraph{Real-time range verification in particle therapy}

An emerging clinical opportunity for ultra-fast \ac{mc} lies in real-time range verification for proton and heavy-ion
therapy.
Non-equilibrium nuclear reactions produce secondary emission signatures—such as prompt gammas and short-lived
$\beta^+$-emitters detected via Compton cameras~\cite{etxebeste2020} or in-beam \ac{pet}~\cite{parodi2002}.
Fast \ac{mc} engines capable of sub-second execution will allow on-the-fly generation of expected emission maps from the
treatment delivery plan, enabling instantaneous comparison with measured counts to verify range and identify anatomical
shifts during beam delivery.

\paragraph{Patient-specific learned priors and digital twins}

Deep learning regularisation in emission tomography typically relies on offline networks pre-trained on generic
population cohorts, creating susceptibility to domain shifts or hallucination artefacts in rare pathologies.
Ultra-fast simulation offers the opportunity to generate physics-accurate, patient-specific synthetic realisations
on-the-fly during the scanning and reconstruction workflow.
By tailoring synthetic data directly to the patient's individual attenuation anatomy, count rate, and scanner geometry,
scan-specific neural regularisers or digital twins can be optimised dynamically, improving reconstruction robustness.

\paragraph{Direct stochastic forward solvers and uncertainty quantification}

Within a 5--10 year horizon, ultra-fast simulation throughput may allow bypassing pre-calculated system matrices
entirely (as explored in \cref{sec:InvProb}), performing forward projections on-the-fly using the stochastic simulation
kernel itself.
This enables exact modelling of non-linear physical phenomena (e.g., detector dead-time and pulse pile-up) and
high-order scattering without memory bottlenecks.
Furthermore, embedding fast stochastic forward models within Markov Chain Monte Carlo frameworks
($\text{MC}^3$)~\cite{bal2013} opens practical pathways for full Bayesian posterior sampling and voxel-wise uncertainty
quantification, moving beyond conventional point estimates.

\paragraph{Open science and standardised benchmarks}
The sustainable adoption of fast simulation tools requires an open, community-driven ecosystem.
Key priorities include establishing standardised digital phantoms, public multi-modal benchmark datasets from modern
scanners, and common phase-space interfaces.
Establishing shared benchmarks is crucial to ensure transparent, reproducible comparisons between general-purpose codes,
dedicated \ac{gpu} engines, and hybrid \ac{mc}-\ac{ai} surrogates.

\subsection{Final remarks\label{sec:mcFinalRemarks}}

The trajectory of \ac{mc} simulation in emission tomography reflects a continuous drive to balance physical fidelity
against computational tractability.
While early developments relied primarily on algorithmic variance reduction, contemporary methods combine massively
parallel hardware with machine learning surrogates to expand the practical scope of stochastic transport modelling.

Looking forward, key research directions include the integration of machine 
learning directly within the simulation kernel, for instance learned 
importance sampling or adaptive selection of \ac{vrt} parameters during the 
run, and fully differentiable simulation architectures~\cite{kato2020, Zhang23_DiffMC}. 
Combined with extreme hardware scaling, ultra-fast Monte Carlo promises to transform from an offline benchmarking tool
into an active, on-the-fly engine for adaptive image reconstruction, uncertainty quantification, and real-time
theranostic guidance.

\newsectionpage
\section{System response modelling with Monte Carlo simulation\label{sec:FastSysResp}}
\sectionauthor[SPECT]{Benjamin Auer}
\sectionauthor[PET]{Pablo Galve, Jos\'e Manuel Ud\'ias, Magdalena Rafecas}

This section is about the \ac{srm} $\srm$ of the forward model described in \cref{eq:expected_prompts} and how \ac{mc}
can be used to compute it for both \ac{spect} and \ac{pet}.
As introduced in \cref{sec:reconbasics}, $\srm$ commonly describes the detection probability of unscattered events,
while patient scatter is modelled as part of the background $\backgrvec$ (\cref{sec:ScatterEstimation}).
Stored element by element, $\srm$ is intractable for most systems, even if only non-zero elements are considered, so in
practice it is factorised into scanner-specific and patient-specific factors (\cref{sec:ModelPipeLine}).
The scanner-specific factors, i.e.\ the collimator and detector response, are the natural target for \ac{mc}.
They can be computed once per system and capture effects that analytical models describe least well.
In terms of \cref{fig:where_mc_fits}, this corresponds to cases~(3a) and~(3b), in which the matrix or one of its factors
is precomputed and stored.
Evaluating the full forward model by \ac{mc} in every iteration, case~(2), is treated in \cref{sec:InvProb}.
\Ac{mc} computes $\srm$ by stochastic numerical integration: each element (\cref{sec:discretisation}) is estimated as
the fraction of emissions sampled from an image basis function $\imbss_{\pixind}$ that are detected in a given data
element.
In both modalities, the \ac{mc}-estimated \ac{srm} is therefore itself noisy, and its size, in addition to the speed of
the simulation, is the main limitation.

In the remainder of this section, the \ac{spect} part (\cref{sec:srspect}) focuses on the collimator--detector response,
and the \ac{pet} part (\cref{sec:srpet}) solely on the detector response.

\subsection{\acs{spect}\label{sec:srspect}}

\subsubsection{Background}

As detailed in \cref{sec:discretisation,sec:DeterministicRecon}, the \ac{srm} $\srm$ encodes the probability of
detecting photons emitted from each image voxel in each detector pixel, and its accuracy directly governs the fidelity
with which the true activity distribution is estimated via reconstruction.
A critical component of the \ac{srm} $\srm$ is the system response, characterised by the \acp{psf}, representing the
collimator-detector's response to point sources of activity positioned at different locations within the \ac{fov}.
It includes geometric acceptance through the collimator, septal penetration, attenuation and scatter within the
collimator and shielding, intrinsic blurring in the detector crystal and electronics readout, and backscatter from
components behind the detector, such as photomultiplier tubes. 

With the system response separated through the phase space of photons reaching the collimator, the expected
unscattered counts $\bar{\truesvec} = \measop\circ\propop\circ\srcop\,\actvty$ (\cref{sec:NonlinearForward}),
the photon source $\srcop$ is normally
proportional to activity, primary photon attenuation within the object belongs to $\propop$ that is often calculated
analytically, whereas the scanner-specific collimator-detector response belongs to $\measop$ and can be estimated by
\ac{mc}.
Ideally, primary photon attenuation would be included directly in a patient-specific \ac{srm} precomputed via \ac{mc}
and applied during reconstruction as forward- and back-projector \cite{Floyd1985,Lazaro2005PMB,peterson2017monte}.
However, this approach remains too slow, while also adding the necessity for the \ac{srm} to be re-estimated for each
patient.
In practice, attenuation is typically estimated between each object voxel-detector pixel pair by using a single central
ray connecting the voxel centre to the detector pixel~\cite{srm-71, srm-72, srm-73}.
This approach was found to be of reasonable accuracy~\cite{srm-71}.
These attenuation factors are then applied as multiplicative factors to a precomputed collimator-detector response table
\cite{dejong2001, Beekman2002}.
Photon scatter within the patient is discussed in detail in \cref{sec:ScatterEstimation}. 

Three main approaches have been proposed to model the system response, each with its own strengths and limitations.
The first is analytical modelling, where the system response is obtained from geometric ray-tracing combined with a
depth-dependent 2D Gaussian function whose parameters are derived from collimator and detector
specifications~\cite{srm-2,srm-3}.
This is the current reference standard in commercial and open-source reconstruction software, offering fast and
computationally efficient reconstruction well-suited for \isotope[99m]{Tc} with parallel-hole collimators.
However, it fails to capture septal penetration, collimator scatter, lead X-ray fluorescence, backscatter, and
energy-dependent non-Gaussian \ac{psf} tails that arise with higher-energy
radionuclides~\cite{srm-14,srm-24,Polson2025}, where a more complete system response model is desired.

The second approach measures the system response experimentally using point sources at multiple \ac{fov}
locations~\cite{srm-5,srm-6,srm-7}, providing the highest precision by directly incorporating the characteristics of the
actual scanner hardware.
Since measurement at every voxel location is impractical, \acp{psf} are acquired at sparse positions, fitted to an
analytical function, and interpolated~\cite{srm-8,srm-9,srm-57}.
This is time-intensive, often requiring multiple days of acquisition.
It has enabled sub-millimetre tomographic spatial resolution in small-animal multi-pinhole
systems~\cite{srm-5,srm-6,srm-7,srm-8,srm-11}, but is limited in clinical practice by larger \ac{fov} and the relatively
short isotope half-lives.
Long-lived surrogate sources such as \isotope[57]{Co} and \isotope[75]{Se} have been proposed to extend measurement
feasibility~\cite{srm-61}, and more recently \isotope[99]{Mo}-microsphere sources to extend the effective acquisition
time for \isotope[99m]{Tc} to approximately 11~days~\cite{srm-60}, though these remain approximations for isotopes with
distinct energy-dependent penetration and scatter characteristics.
The large number of measurements required, combined with the challenges of fitting and interpolation, makes the approach
difficult to automate and to transfer across radionuclides, scanners, and collimators.

The third approach uses \ac{mcs} to model the system response by transporting photons through a computerised scanner
model incorporating all relevant physical phenomena.
\Ac{mcs} is now widely recognised as the gold standard for system response modelling in \ac{spect}, particularly with
the growing need for quantitative imaging in theranostics, provided the simulation incorporates sufficient realism in
geometry, materials, collimator, and detector components.
Established simulation platforms such as \ac{gate}/\ac{geant4} and \ac{simind} have been extensively validated for this
purpose~\cite{sarrut2021c,ljungberg1989}.
\Ac{mcs} can precisely capture all \ac{psf} components, including tails critical for higher-energy imaging.
For \isotope[123]{I} imaging with a low-energy high-resolution parallel-hole collimator, up to 50\% of detected counts
in the photopeak originate from higher-energy emissions, via septal penetration and Compton scatter in the back
compartment behind the crystal, shielding and collimator, which are difficult to capture
analytically~\cite{srm-14,srm-15}.
Similarly, \ac{mcs}-based approaches demonstrated better performance compared to analytic models for
\isotope[90]{Y}~\cite{srm-59}, \isotope[111]{In}~\cite{Wikberg2023}, \isotope[177]{Lu}~\cite{Polson2025,srm-28},
\isotope[131]{I}~\cite{srm-24,srm-28}, \isotope[223]{Ra}~\cite{srm-16}, and emerging alpha-emitter imaging such as
\isotope[225]{Ac}~\cite{srm-28}.
For multi-pinhole systems, the \ac{psf} varies in all three spatial dimensions and is more difficult to approximate by a
Gaussian function, so \ac{mcs}-based modelling represents one of the most accurate approaches to characterise the full
system response~\cite{srm-20,srm-63}.

\subsubsection{Status}
\Ac{mcs}-based system response modelling has been shown to outperform analytic approaches in quantitative accuracy and
image fidelity.
In a multicentre validation across eight scanners at five institutions, \textcite{srm-22} demonstrated a coefficient of
variation below 5\% for larger spheres in \isotope[99m]{Tc} phantom studies, indicating enhanced
reproducibility of \ac{mcs}-based reconstruction.
For \isotope[123]{I} brain imaging, \ac{mcs}-based reconstruction increased striatal uptake ratios by up to 25\% over
analytic modelling in clinical studies~\cite{srm-23} and improved recovery coefficients for both \isotope[123]{I} and
\isotope[131]{I} in phantom studies~\cite{srm-24,srm-62}.
For [\isotope[111]{In}]In-octreotide liver imaging, \ac{mcs}-based reconstruction increased lesion detection from 31\%
to 50\% compared with analytic reconstruction in a clinical observer study~\cite{Wikberg2023}.
For \isotope[177]{Lu}, \ac{mcs}-based reconstruction reduced bias in activity recovery and variability across regions of
interest~\cite{Polson2025,srm-28}.
For \isotope[90]{Y} bremsstrahlung liver imaging, full \ac{mcs} system response modelling increased contrast recovery by
up to 42\% and was consistently preferred in blinded clinical evaluation~\cite{srm-59}.
For \isotope[166]{Ho} liver radioembolisation, \ac{mcs}-based system response improved image quality and
quantification~\cite{srm-30}.
For \isotope[223]{Ra}, quantitative \ac{spect} with \ac{mcs}-based system response modelling has been demonstrated with
medium- and high-energy collimators~\cite{srm-16}.

Two main strategies have been used to incorporate \ac{mcs}-derived system response into iterative reconstruction.
The first precomputes the full system matrix by simulating the system response to a point source in each voxel, or to
each voxel uniformly filled with activity, and storing the complete sparse matrix for use in reconstruction (see
approach 3a of \cref{fig:where_mc_fits}).
In small-animal pinhole \ac{spect}, this achieves approximately 1~mm spatial resolution with activity recovery close to
the reference value of 1, though image quality depends critically on the number of photons simulated to form the
matrix~\cite{srm-17,srm-18}.
For a clinical multi-pinhole brain \ac{spect} system, sparse storage reduced the theoretical 6.4~TB \ac{mcs}-based
\ac{srm} down to 36~GB while maintaining image quality, demonstrating superior performance over analytic
approaches~\cite{srm-19,srm-63}.
This approach is well-suited to complex pinhole geometries where the \ac{psf} varies spatially in all three dimensions
but remains memory and storage intensive and challenging for extended-\ac{fov} systems.
Preclinical scanners are not exempt from these limitations.
Even though their \ac{fov} is smaller, their sharper spatial resolution requires much finer voxel and detector sampling
to characterise the \ac{psf} with sufficient precision, resulting in practice in large \ac{srm} as well.

The second and more widely available strategy uses precomputed \ac{mcs}-derived kernels or response tables incorporating
some degree of analytical approximation to reduce the computational cost, applied during forward and back-projection at
each iteration (see approach 3b of \cref{fig:where_mc_fits}).
Implementations range from distance- and energy-dependent \ac{mcs}-based kernels applied as depth-dependent convolution
in projection space~\cite{srm-23,srm-24,srm-28,srm-37,srm-50,srm-62} to \ac{mcs}-derived tables encoding the full
angular- and energy-dependent collimator-detector response sampled stochastically during
reconstruction~\cite{ryden2018,song2005,descourt2010,saporta2022a,srm-34}.
Common approximations include modelling collimator scatter and backscatter via Gaussian functions rather than full
\ac{mcs} transport, or representing the detector response analytically rather than through \ac{mcs}-based photon
tracking.
This strategy is substantially more memory- and storage-efficient and has been implemented in commercial platforms,
including Hermes Medical Solutions (Hermes Medical Solutions AB, Stockholm,
Sweden)~\cite{srm-50,srm-21,srm-22,srm-23,srm-24} and Voximetry (Voximetry, Inc., San Francisco, CA,
USA)~\cite{ryden2018}, as well as in the open-source PyTomography framework~\cite{srm-37}.

Similarly to including object scatter within reconstruction (see \cref{sec:ScatterEstimation}), the dual-matrix approach
has been proposed for system response modelling (\cref{sec:WhereDoesMCfit}).
For high-energy radionuclides, the inclusion of the \ac{psf}'s long tails due to penetration and scatter through the
collimator, shielding, and backscatter compartments, in both the forward- and back-projector can considerably slow down
the reconstruction.
For example, \textcite{Gustafsson2018} utilised a full system response modelled by \ac{mcs} in the forward projector,
including the geometric response, septal penetration, collimator scatter, and backscatter, whereas the backprojector was
based only on an analytical model accounting for the geometric and intrinsic response.
This approach was found accurate within $\sim$6\% of the reference total activity in phantoms for \isotope[99m]{Tc},
\isotope[177]{Lu}, and \isotope[131]{I}~\cite{Gustafsson2018}.
\textcite{srm-46} developed a dual-matrix approach in which the full \ac{psf}, including the long tails, is
modelled only in the forward projector, while the backprojector uses a truncated version without these tails.
For \isotope[18]{F} imaging on a small-animal multi-pinhole scanner, images reconstructed with this technique were found
qualitatively and quantitatively equivalent to those obtained with identical forward and back-projectors while speeding
up the reconstruction by 5 to 11-fold \cite{srm-46}.

\subsubsection{Current and future challenges}
Despite compelling evidence of its advantages, \ac{mcs}-based system response modelling has not been adopted as standard
practice.
The challenges are largely practical and fall into five areas: (1) computational burden of \ac{mc} simulation and
reconstruction; (2) growing complexity of modern scanner geometries and their simulation models; (3) limited
availability and lack of standardisation of data formats for newer generation systems; (4) insufficient documentation of
collimator, detector, and scanner specifications from manufacturers; and (5) limited availability of commercial and
open-source reconstruction software incorporating \ac{mcs}-based system response.

Computational burden remains the primary barrier at two stages: generating the \ac{srm} or kernels via \ac{mcs}, and
incorporating these into iterative reconstruction.
\Ac{spect} sensitivity is inherently low ($10^{-5}$--$10^{-4}$, compared with $\sim$1--10\% for \ac{pet}), as the vast
majority of photons are absorbed within the collimator septa, requiring simulation of billions of photons to achieve
sufficient statistics for system response modelling.
Full system matrix calculation requires computing clusters, with the equivalent of more than 5000 single-\ac{cpu} days
reported for small-animal multi-pinhole systems~\cite{srm-17}.
\Ac{mcs}-derived kernels must be precomputed at multiple source-to-collimator distances per radionuclide and collimator
configuration, requiring approximately 12~minutes per energy-distance response on a single \ac{cpu}~\cite{srm-50}.
At the reconstruction stage, full \ac{mcs}-based \isotope[123]{I}/\isotope[131]{I} reconstruction requires approximately
4.5~minutes versus 38--49~seconds for analytic alternatives~\cite{srm-24}.
Additionally, \ac{mcs}-derived system matrices and kernels always carry residual statistical noise due to the inherently
Poisson nature of simulation, which propagates into reconstructed images and creates a direct trade-off between system
response accuracy and computation time~\cite{srm-18}.
The latter was illustrated in a [\isotope[111]{In}]In-octreotide observer study, where \ac{mcs}-based system response
improved lesion detectability but also increased image noise, reducing observer confidence and necessitating
post-filtering that partially offset the spatial resolution gains~\cite{Wikberg2023}.
Statistical robustness of the system response can be improved by increasing the number of simulated events.
The required number of events depends on the radionuclide and system configuration, with higher-energy radionuclides
with long \ac{psf} tails requiring more events and thus longer computation times.
\Ac{vrt} (\cref{sec:varred}), bootstrapping, and \ac{pca} have all been proposed to improve the \ac{srm}'s statistical
robustness \cite{srm-78, srm-79,srm-63}.
Note that symmetries can be exploited to reduce simulation time, for example by using rotators to replicate the response
for one location to its symmetric positions.
However, the same photon histories are then reused, so that the statistical noise becomes correlated across matrix
elements and thus translates into structured image artefacts.
Finally, full system matrix storage requires tens to hundreds of gigabytes depending on the radionuclide and voxel-pixel
resolution, even with sparse compression~\cite{srm-19}, requiring high-memory high-storage computing infrastructure for
use in iterative reconstruction.

The complexity of modern scanners introduces additional modelling challenges.
Primitive-based models (e.g., cylinders, boxes, and spheres) available in simulation platforms cannot accurately
represent sophisticated scanner components and geometries.
Mesh-based modelling addresses this limitation by representing scanner components as triangulated surfaces, but
introduces slower photon simulation and requires manual mesh optimisation to achieve practical simulation
times~\cite{srm-20}.
Semiconductor detectors require charge transport modelling including hole trapping, charge sharing, and depth-dependent
charge collection, which produce low-energy spectral tailing that is not yet fully supported in current \ac{mcs}
platforms~\cite{srm-38}.
Validated simulation models for conventional clinical and preclinical systems, including semiconductor scanners and
novel collimator designs, are not widely available, and no standardised online repositories exist.

Software availability and data standardisation remain significant obstacles.
To the best of our knowledge, on the clinical side only Hermes Medical Solutions, Voximetry, and Mediso Medical Imaging
Systems (Mediso Ltd., Budapest, Hungary)~\cite{mediso_anyscan_spect} currently offer commercial \ac{mcs}-based
reconstruction, limited to conventional parallel-hole scanners and selected isotopes with specific energy windows.
On the preclinical side, Mediso Medical Imaging Systems also provides a \ac{mcs}-derived solution for one of their
small-animal multi-pinhole scanners~\cite{mediso_nanoscan_spect}.
PyTomography~\cite{srm-37} supports \ac{mcs} kernel-based reconstruction for conventional scanners but not full
\ac{mcs}-based system matrix approaches.
\Acl*{stir}~\cite{srm-68} incorporates pinhole reconstruction with analytical system response only.
For newer semiconductor scanners, \ac{mcs}-based reconstruction is not available from manufacturers or third-party
platforms, as data formats are proprietary, acquisition geometries are non-standard, and scanner and collimator
specifications are insufficiently documented.

\subsubsection{Development needed to meet the challenges}
Addressing the challenges described above requires advances across five areas: (1) computational acceleration of
\ac{mcs} and reconstruction; (2) system matrix compression and efficient memory management; (3) expansion and
dissemination of validated scanner models through standardised repositories and simulation platforms; (4) broader
implementation in commercial and open-source reconstruction software; and (5) clinical validation studies demonstrating
diagnostic utility in practice.

The slow speed of \ac{mcs} is the primary challenge to wider adoption, limiting both the calculation of system matrices
and kernels and their incorporation into iterative reconstruction.
For system response modelling, as described in \cref{sec:varred}, combining \ac{gpu}-acceleration with \ac{vrt}
represents the most promising path to reduce simulation cost.
\Ac{gpu}-accelerated \ac{spect} simulation has demonstrated $\sim$70-fold speedup over \ac{cpu}~\cite{garcia2016}.
\Ac{vrt} including forced detection~\cite{srm-64}, analytical detector modelling~\cite{cajgfinger2018}, precalculated
angular response table approaches~\cite{song2005,descourt2010}, and free-flight simulation~\cite{srm-42} have each
demonstrated orders-of-magnitude speedup.
Neural networks have been trained to generate angular response tables from low-statistic simulations, reducing
computational time by a factor of up to 400 compared with analogue simulation~\cite{sarrut2018}.
While several of these \ac{vrt} are available in \ac{mcs} platforms, their utilisation remains limited, as simulation
times may still be too long for system response modelling in routine practice.
Recent developments in simulation platforms, including modern Python-based interfaces, are expected to facilitate
multithreading and \ac{gpu} integration~\cite{srm-69}, which are not yet widely available, and thus further reduce
computation times.

For reconstruction, approaches incorporating hybrid analytic-\ac{mcs} system response modelling have been shown to
achieve clinically feasible runtimes.
For example, \textcite{ryden2018} achieved approximately 3~minutes for a \isotope[177]{Lu} clinical reconstruction
using four \acp{gpu}, where the full angular- and energy-dependent collimator-detector response including septal
penetration was modelled via precomputed \ac{mcs} tables, while collimator scatter and backscatter were approximated by
a Gaussian kernel.
\textcite{srm-50} demonstrated a complementary approach using distance-dependent precomputed \ac{mcs}-based
kernels applied via convolution during reconstruction, achieving greater than 10-fold speedup over direct simulation and
generating near noise-free projections in approximately 1.5~minutes per projection on a single \ac{cpu} for
\isotope[99m]{Tc} and \isotope[123]{I}.
\textcite{srm-28} proposed a reduced-order one-dimensional rotationally symmetric system response operator derived
from \ac{mcs}, reducing reconstruction time from 12.5 to 5~minutes for 200 \ac{mlem} iterations using \ac{gpu}
acceleration, while maintaining quantitative accuracy for \isotope[177]{Lu}, \isotope[131]{I}, and
\isotope[225]{Ac}.
However, all these methods rely on approximations in the system response modelling to achieve faster reconstruction
times.

Full \ac{mcs}-based reconstruction is the most complete approach for accurately modelling the system response, as it
does not rely on analytical approximations, but is also the most demanding in terms of computational and memory
resources, since the use of large matrices slows down reconstruction.
Full system matrix compression strategies will be essential, including dual-matrix dual-voxel
decomposition~\cite{srm-46}, \ac{gpu}-accelerated list-mode matrix generation~\cite{srm-48}, geometric symmetry
exploitation~\cite{srm-32}, and dimensionality reduction techniques such as \ac{pca}, which was initially applied by
\textcite{srm-78} to improve statistical robustness of the \ac{srm}.
Deep learning offers an additional avenue, with \textcite{srm-67} proposing and \textcite{srm-54}
validating synthesis of full projection sets from sparse \ac{mcs}-based \isotope[177]{Lu} acquisitions while maintaining
dosimetry accuracy.

Expansion of validated \ac{mcs} models to cover modern \ac{spect} systems is critical, alongside making these and
conventional scanner models openly available in standardised repositories within simulation platforms.
Comprehensive tutorials on simulation implementation and integration into reconstruction software will further
facilitate dissemination within the field.
Mesh-based modelling provides the geometric foundation for extending models to complex systems and enhancing realism,
but automated mesh optimisation pipelines, a prerequisite for reasonable simulation times, are not yet widely available
and currently require manual expert processing~\cite{srm-19,srm-20}.
Semiconductor charge transport modelling requires dedicated development, with initial validation demonstrated only for
handheld and cardiac-dedicated scanners~\cite{srm-38,srm-66}, while comprehensive models for clinical full-ring
semiconductor scanners have not yet been investigated.
Making scanner geometry, collimator and detector specifications, as well as material definitions openly accessible would
substantially lower the barrier to \ac{mcs}-based system response implementation in third-party software.

Broader software implementation of existing methods is needed.
To the best of our knowledge, no platform currently supports full \ac{mcs}-based system matrix reconstruction, which
represents the most flexible approach as it is not collimator-specific.
Commercial availability of hybrid \ac{mcs}-based approaches remains limited to \ac{mcs}-based system response for
selected isotope-energy window-collimator combinations.
Engagement with scanner manufacturers to standardise data formats and provide accessible scanner specifications is
essential to enable \ac{mcs}-based reconstruction for semiconductor and modern scanners.

Finally, wider adoption in routine settings will require stronger evidence from larger clinical and preclinical studies.
The emergence of theranostics, particularly \isotope[177]{Lu}-based and other radionuclide therapies, represents the
most immediate opportunity to demonstrate, beyond phantom validation, the clinical and preclinical benefit of more
accurate system response modelling.
As these methods become more widely available, evidence from routine clinical and preclinical settings, including
diagnostic performance and dosimetry accuracy, will drive adoption across the field.

\subsubsection{Concluding remarks}
The evidence base is now sufficient to justify the transition of reconstruction incorporating \ac{mcs}-based system
response modelling from a research tool to the clinical and preclinical standard for quantitative \ac{spect},
particularly with the emergence of theranostics.
Analytic Gaussian models are not accurate enough for higher-energy radionuclides where non-Gaussian \ac{psf} tails
dominate, and \ac{mcs}-based reconstruction has been shown to reduce systematic bias, improve precision and image
quality, and enhance lesion detectability.
The primary obstacles to wider adoption remain practical: slow computation time, limited availability of simulation
models for conventional and modern scanners, data access restrictions, and limited availability in established
open-source and commercial software. 

The roadmap towards routine adoption requires accelerating \ac{mcs}-based simulation and reconstruction through \ac{gpu}
computing and \ac{vrt}, developing and disseminating validated simulation models for both conventional and modern
scanners, expanding open-source reconstruction platforms to support multiple data formats and \ac{mcs}-derived
algorithms, distributing online accessible benchmarking datasets for the most common scanners and collimators, and
engaging scanner manufacturers to open access to their system specifications and data formats. 

\subsection{\texorpdfstring{\acs{pet}}{PET}\label{sec:srpet}}

\subsubsection{Status}
In \ac{pet}, $\srm$ takes the factorised form of \cref{eq:factorised_srm} with the propagator of
\cref{eq:factorised_srm_pet}.
Its accuracy is crucial for image quality, optimising resolution and reducing artefacts~\cite{Iriarte2016}.
Where \ac{mcs} enters depends on which factor of the factorised model of \cref{eq:factorised_srm} carries the \ac{rm},
and on how much of the \ac{srm} is made explicit.
Four families have been proposed, from the most approximate to the most explicit:
\begin{itemize}
    \item \emph{Image-space blurring}, $\srm_{\rm img\_blur}$ alone: a space-variant kernel applied in the image
    domain~\cite{Reader2003, Kotasidis2010, Arias-Valcayo2023, Zhang2017}.
    Fast and undemanding of the scanner description, but it averages over the direction of incidence and therefore
    cannot reproduce directionally dependent effects such as the block structure of the detector.
    \item \emph{Detector-space blurring}, $\srm_{\rm sin\_blur}$ alone: the same kernel applied in the projection
    domain~\cite{Qi1998, Alessio2006, Panin2006}, which recovers part of that dependence at the cost of a
    scanner-specific description.
    \item \emph{Both}, splitting the blurring according to where the physics acts: positron range in $\srm_{\rm
    img\_blur}$, non-collinearity and detector response in $\srm_{\rm sin\_blur}$~\cite{Terstegge1996, Zhou2011}.
    \item \emph{Included in the propagator} $\srm_{\rm proj}$ of \cref{eq:factorised_srm_pet}, either as explicit
    detection probabilities for every voxel--\ac{lor} pair~\cite{Rafecas2004a, Southekal2011, Zhang2010, Cabello2012,
    Meng2019, Deng2021}, feasible only for systems with few elements, or as \acp{tor}, the tube-shaped sensitivity
    volumes of \cref{sec:TheoreticalForwardModel}, described by wide analytical functions~\cite{Herraiz2006, Pratx2011,
    Lougovski2015, Shopa2023}.
    Both sacrifice the sparseness of $\srm_{\rm proj}$ and rely on symmetries to remain storable~\cite{Mumcuoglu1996,
    Rafecas2004a, Herraiz2006, Zhang2010, Scheins2011}. 
    Symmetries that the voxel grid does not share can only be applied through interpolation, which may affect image
    quality.
    Alternative image elements, such as polar pixels or blobs, have been used to exploit them instead~\cite{Cabello2012,
    saha2014iterative}.
    Moreover, symmetric elements are replicated from the same photon histories, so their noise is correlated
    and may translate into structured image artefacts rather than incoherent noise.
    A stochastic sampling strategy might be used to sample several line integrals per event from the distribution of the
    \ac{lor} endpoints~\cite{Moehrs2008, Gillam2013, Vo2025}, suppressing aliasing from inhomogeneous sampling at
    intermediate cost.
\end{itemize}

Whichever family is chosen, its factors must be populated, and three sources are available. 
Experimental measurement --- point-source acquisitions across the \ac{fov}~\cite{Alessio2005, Tohme2009, Gong2017,
Zhao2023} --- reproduces the response of the actual hardware without a detailed scanner description, but the
acquisitions are laborious, they entangle effects such as positron range that one may wish to model separately, and in
practice they have been limited to kernels describing $\srm_{\rm img\_blur}$ or $\srm_{\rm sin\_blur}$ macroscopically;
$\srm_{\rm norm}$ is always based on measurements rather than modelled since it contains internal efficiency factors.
Analytical models are fast enough to be computed on the fly, are straightforward to factorise, and describe the blurring
factors well through parametrised kernels.
Pushed into the propagator itself, however, they must formulate every effect explicitly, and have mostly covered
solid-angle acceptance and attenuation within the crystal~\cite{Huesman2000, Pratx2011, Shopa2023, Matulic2024}.
The resulting integrals rarely have a closed form and are evaluated by deterministic or random
sampling~\cite{Moehrs2008, Gillam2013, Vo2025}.
In inter-crystal scatter, where the detected position depends on a sequence of interactions rather than on a single
path, each additional interaction adds dimensions to the integrals to be evaluated~\cite{Roser2026}.
\Ac{mc} simulation is subject to neither restriction: 
it can reproduce non-collinearity and the full detector response together, limited only by the fidelity of the scanner
model, and it can populate any of the four families above.
\Cref{fig:comparison_of_SRM_models} illustrates the difference this makes.

\ifjournalversion
\begin{figure}[!t]
    \centering
    \includegraphics[width=1.\linewidth]{images/fig41_srm_mc_comparison_journal.pdf}
    \caption{(a) Brain activity reconstruction from a patient injected with [\isotope[18]{F}]FDG (221~MBq for 89~kg)
    with the Siemens hybrid MR-BrainPET showing \ac{mlem} without 
    \ac{psf} (top) and with a \ac{mc}-based sinogram \ac{psf} (bottom). 
    Adapted from \cite{Xu2019}, licensed under CC BY 3.0.
    (b) [\isotope[18]{F}]FDG cardiac mouse study with the IRIS PET/CT preclinical scanner (cardiac region magnified
    below).
    The images compare a geometric model without \ac{srm} (left), an analytical model of the detector penetration with
    multi-ray tracing (middle), and a \ac{mc}-based sinogram \ac{psf} (right).  
    Reprinted from Physica Medica, Vol.~57, A.~Pilleri et al., A Monte Carlo detector response model for the IRIS PET
    preclinical scanner, pages 107--114, Copyright (2019), with permission from Elsevier~\cite{Pilleri2019}.
    (c) Simulated point source resolution (radial, tangential and axial \ac{fwhm}) measured in the 6R-SuperArgus scanner
    for a fixed image \ac{psf}, spatially variant image \ac{psf} and pure-\ac{mc} based projection/backprojection. 
    Adapted from \cite{Arias-Valcayo2023}, \textcopyright IOP Publishing Ltd.
    Reproduced with permission. All rights reserved.}
    \label{fig:comparison_of_SRM_models}
\end{figure}
\else
\begin{figure}[!t]
    \centering
    \includegraphics[width=1.\linewidth]{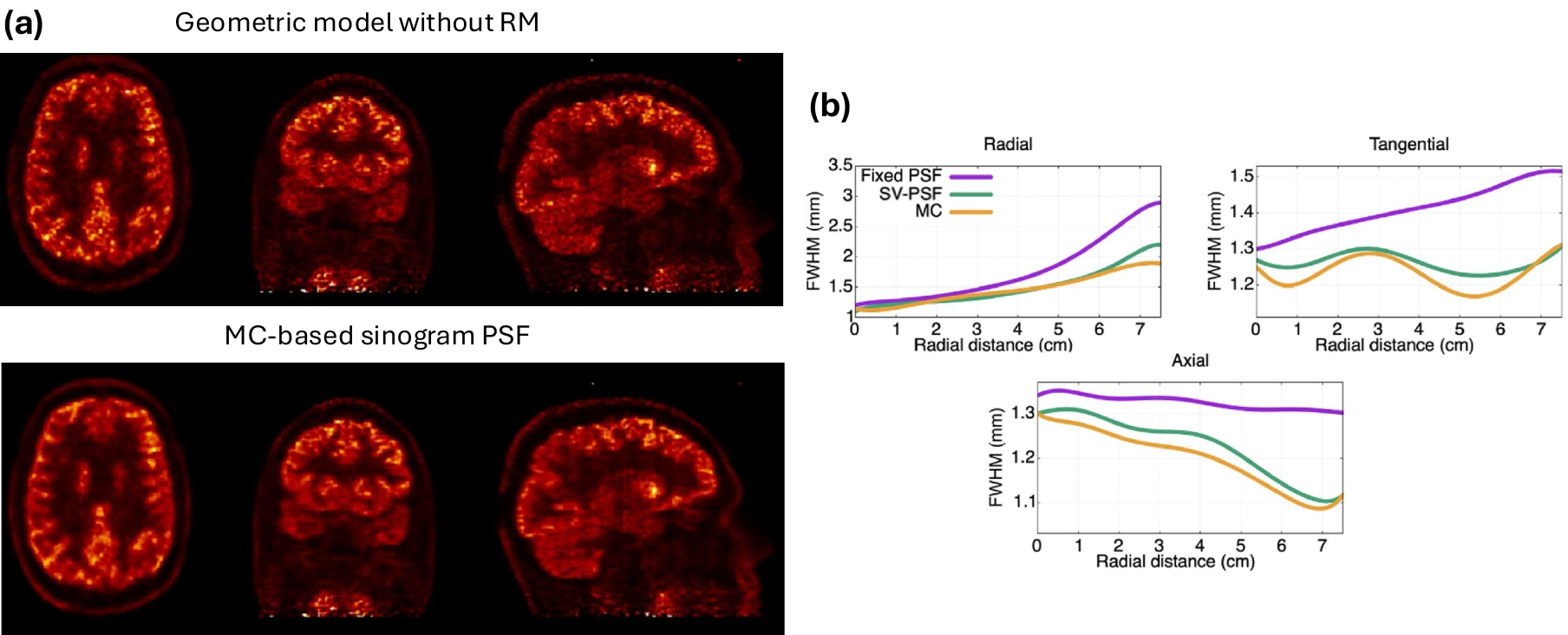}
    \caption{(a) Brain activity reconstruction from a patient injected with [\isotope[18]{F}]FDG (221~MBq for 89~kg)
    with the Siemens hybrid MR-BrainPET showing \ac{mlem} without 
    \ac{psf} (top) and with a \ac{mc}-based sinogram \ac{psf} (bottom). 
    Adapted from \cite{Xu2019}, licensed under CC BY 3.0.
    (b) Simulated point source resolution (radial, tangential and axial \ac{fwhm}) measured in the 6R-SuperArgus scanner
    for a fixed image \ac{psf}, spatially variant image \ac{psf} and pure-\ac{mc} based projection/backprojection. 
    Adapted from \cite{Arias-Valcayo2023}, \textcopyright IOP Publishing Ltd.
    Reproduced with permission. All rights reserved.}
    \label{fig:comparison_of_SRM_models}
\end{figure}
\fi

\subsubsection{Current and future challenges}

As mentioned above, the first challenge facing \ac{srm} estimation is the computational cost of extensive \ac{mc}
simulations.
Avoiding \ac{psf} simplifications, estimating the \ac{srm} based on either \acsp{tor} or explicit matrix elements
requires launching thousands of events to be histogrammed at every non-null voxel-LOR element, which may take days to
acquire enough counts in current \ac{cpu}-simulators such as \ac{gate}.
If the \ac{srm} is factorised to account only for the factors that are intrinsic to the scanner, such as
non-collinearity, depth of interaction or inter-crystal scatter, one may argue that this estimation just needs to be
done once, but its cost still limits the number of tests that can be run while the \ac{srm} itself is being developed.
\Ac{gpu}-based simulators and \ac{cpu}/\ac{gpu} clusters reduce this cost by orders of magnitude
(\cref{sec:MCoverview}), but this
alone does not make \ac{mc}-based \ac{srm} modelling practical, for reasons that have less to do with the speed of the
simulation than with the \ac{srm} itself.

A second cost of estimating the \ac{srm} by \ac{mc} is that the result is itself a noisy estimate: the noise of the
simulation propagates into the reconstructed image. 
How much statistics is enough to achieve a highly detailed \ac{srm} remains an empirical question which has been
targeted by different authors using multiple approaches. 
\textcite{Rafecas2004} found that between 7\,000 and 14\,000 detected coincidences per pixel were needed to assure good
\ac{snr} and contrast in a 2D preclinical system modelled with GEANT.
Longer simulations, however, mostly add non-zero elements from highly improbable detections, whose statistical quality
is very difficult to improve, and discarding elements below 1\% of the maximum weight was harmless while reducing the
matrix size by up to 40\%~\cite{Rafecas2004}.
\textcite{Xu2019} simulated some $2\times10^{4}$ $\gamma$-rays per entry of a \ac{gate}-derived look-up table, at
roughly
2\,000~h per cluster node, or five days distributed over sixteen nodes.
In \textcite{Pilleri2019}, the number of photons simulated per \ac{tor} was
adaptively selected by inverse binomial sampling, so that every estimated
probability reached a prescribed relative precision.
They required about 4.5~days per detector module on a single workstation with GEANT4~\cite{Pilleri2019}, and only
retained elements down to 60\% of the total detection probability of each row purely to meet a memory budget.

Other authors went further and included the attenuation map inside the \ac{srm}, which extends the physics the model
must carry and forfeits the scanner symmetries.
Rehfeld and Alber~\cite{Rehfeld2006} did so for a single-ring scanner with idealised detectors, as the matrix of a 3D
scanner could not be stored.
Including patient scatter raised the fraction of non-zero elements from 2\% to 40\%, and the noise of the matrix became
comparable to that of a 30-min acquisition at $10^{4}$ simulated emissions per voxel.
In this low-scatter case, a dual-matrix variant with scatter only in the forward projector converged faster and
performed slightly better than using the full matrix in both directions.
\textcite{Southekal2011} introduced a detailed model of this kind for a rat-brain-dedicated system,
estimating the \ac{srm} elements explicitly with detector response, positron physics, attenuation and scatter; the
matrix took close to 10~days on eight \ac{cpu} cores to simulate $10^{7}$ decays per voxel with \ac{simset}, so a common
attenuation map had to be reused across animals. 
The same idea was later implemented on \ac{gpu} with \ac{umcpet}~\cite{Galve2022, Arias-Valcayo2023}.
In~\cite[ch.~6]{Galve2022} a patient-specific \ac{srm}, with attenuation and scatter inside the simulation, was
generated in 12~hours at $1.2\times10^{3}$ events per \ac{lor}, and reconstruction over a restricted region of the
\ac{fov} of a preclinical scanner took under 5~hours; it improved on a spatially variant \ac{psf} at matched noise,
although the statistical noise of the simulation propagated into the image and had to be contained with a strongly
regularised initial estimate and post-filtering.
Backprojection was kept analytical there precisely for that reason: evaluating the adjoint by \ac{mc} would have doubled
the reconstruction time and injected the simulation noise a second time.

Several of these works report that the benefit saturates well before the simulation budget is exhausted.
Beyond 20\,000 detected coincidences per voxel the accuracy of the matrix barely improved~\cite{Rafecas2004}; raising
the statistics fivefold gained only about 5\% in bias and nothing in precision, the latter being dominated by the
uncertainty in the data rather than by the statistics of the matrix~\cite{Southekal2011}; and essentially the same mean
squared error was reached with a look-up table built from ten times fewer events, at about 200~h per node instead of
2\,000~\cite{Xu2019}.
Each of these thresholds was, however, established for one scanner and one object.
Qi and Huesman proposed a general criterion, requiring the variance that errors in the \ac{srm} induce in the forward
projection to stay below a small fraction of the Poisson variance of the data~\cite{Qi2005}.

Regarding \ac{mc} simulations, there are still physical parameters to be investigated.
Recently, \textcite{Toussaint2024} discussed the accuracy of the non-collinearity in most of the \ac{mc} models, usually
introduced as a 0.5\textdegree{} \ac{fwhm} Gaussian distribution which may underestimate the real values.
\Ac{prc} is also under scrutiny, facing the heterogeneity of the patient body and isotope energy dependence
(\cref{sec:PositronRangeModelling}).
Accounting for positronium formation is also an open topic in \ac{pr}.
A further question under consideration is annihilation photon entanglement \cite{Moskal2025, Kumar2025, Eslami2024},
which is not considered in most simulation packages, and may be relevant in the \ac{srm}.

The need for higher sensitivity, resolution and extended \acp{fov} has led to a dramatic rise in the number of \ac{srm}
elements, as
is the case in \ac{lafov} \cite{Prenosil2022, Spencer2021} or some brain-dedicated \cite{Li2024, Akamatsu2022} systems,
adding computational challenges for complex models.
Atypical non-cylindrical geometries proposed for brain-dedicated scanners \cite{Akamatsu2022, Allen2024} also pose
challenges for \ac{srm} storage given the reduced number of symmetries.
\Ac{tof} and \ac{doi} add further dimensions to the problem: both improve the quality of the data, and both multiply the
number of elements the \ac{srm} has to carry and the events needed to populate them. 
Some vendors reduce the size of the acquired data by mashing or by increasing the span, but this does not reduce the
cost of modelling the \ac{srm} whenever the compression itself has to be modelled.
The growth of the data space and its consequences for memory are discussed in \cref{sec:comparch}.

The time consumption in projection and backprojection for complex \ac{srm} models might become a limitation, given that
this is the main time-consuming step in reconstruction.
From this arises a trade-off between computational speed and image quality, especially considering the high performance
of standard image-based \ac{psf}.
At this point, many authors introduced methods to accelerate \ac{srm} projection \cite{Schramm2024, Cui2011,
Herraiz2011, Pratx2009}.

Finally, the improvement in resolution brought by an accurate \ac{srm} is typically modest compared with \ac{psf}
implementations (see \cref{fig:comparison_of_SRM_models}\ifjournalversion\else{} as well as figure~6
in~\cite{Pilleri2019}\fi).
Moreover, the observed resolution over a homogeneous background is
correlated with image contrast~\cite{Gong2016}, which can be hidden by a poor \ac{snr}. 
Patient motion also imposes a comparable bound, since a resolution limit set by movement cannot be recovered by a better
\ac{srm}~\cite{Wang2024, Kyme2021, Lamare2022}.
None of these are \ac{srm} modelling problems, but they may establish an external bottleneck for image quality
improvement derived from the \ac{srm}.
Regularisation helps to contain the image noise arising both from the data and from the \ac{srm}~\cite{Qi2005, Ahn2003,
Wang2015, Gong2019}.

\subsubsection{Requirements needed to meet the challenges}

Setting aside minor differences between them, the accelerated \ac{pet} simulators reviewed in \cref{sec:MCoverview} are
fast enough to bring \ac{srm} estimation using \ac{mcs} within reach.
They differ, however, in how explicitly the detector is tracked --- a hard requirement for \ac{srm} modelling --- and in
the accuracy of the underlying physical models. 
They have been validated against other simulators~\cite{lai2019, galve2024, scheins2021} and on aggregate observables
--- energy and crystal-index distributions, NEMA scatter fraction --- which constrain the model as a whole rather than
each of its factors separately.
For \ac{srm} modelling the relevant question is narrower, namely whether the detector sub-model itself is correct, and
it is also the hardest to answer: modern detectors differ widely in readout, multiplexing, light sharing and \ac{doi}
encoding, and most accelerated simulators stop at the energy deposited in the crystal, leaving scintillation light
transport and the electronics either parametrised or omitted.
Speed translates directly into two of the challenges above: a faster simulator buys the statistics that keep the noise
of the \ac{srm} below the noise of the data, and it makes it affordable to retain physical effects in the model that
would otherwise be approximated away.
The technology is ready to be implemented; what remains out of reach is less the simulation itself than its use inside
the reconstruction loop and the storage of the resulting \ac{srm}.
This rests, however, on the assumption that the physics models and geometry of accelerated simulators are accurate
enough for \ac{srm} modelling, which has not yet been verified systematically.

Another limitation is the complexity and computational cost of highly detailed models of the \ac{srm}.
\Ac{psf} modelling in image space is fast and can be easily generalised to any scanner design.
This has led to general reconstruction frameworks, such as parallelproj~\cite{Schramm2024}, CASToR~\cite{Merlin2018} or
\ac{stir}~\cite{Thielemans2012}, which provide thin projectors combined with convolution kernels to model scanner
resolution.
These frameworks are typically modular, so more complex models such as \acp{tor} or explicit \ac{srm} element estimation
can also be implemented, though the user must supply an exact scanner specification and may need to adapt the projection
routines~\cite{Herraiz2006, He2021}.
Furthermore, \acp{tor} and explicit voxel--\ac{lor} models are inherently slower, since projection and backprojection
require a higher number of computational operations, and may be limited by memory in systems with a high number of
\acp{lor} and additional \ac{tof} and \ac{doi} information.
The implementation of image \ac{psf}-based \ac{srm} might look more attractive in these scenarios, either using actual
measurements or simulations, and memory issues might be reduced by combining these methods with phenomenological
functions fitted to complex \ac{mc} simulations~\cite{He2021, Shopa2023}.

The works discussed above illustrate how much ground has been covered: system responses that once required computing
clusters and days of simulation are now within reach of single \acp{gpu} or multi-core systems for small scanners with a
few
elements.
Large clinical scanners still pose a challenge given the growth of the \ac{srm} and the time it adds to every
reconstruction iteration. 

\subsubsection{Conclusions}

\Ac{mcs}-based \ac{srm} models have demonstrated better performance in image reconstruction, but they are limited by the
time and memory needed to estimate the \ac{srm} explicitly, and by the cost of applying it during projection and
backprojection.
\Ac{gpu}-\ac{mc} simulation packages have already demonstrated the performance required for \ac{srm} estimation, so the
computational burden of the simulation is largely overcome.
The memory needed still forces the use of simplified models for large scanners, but a wide range of scanners can already
benefit from \ac{gpu}-accelerated \ac{mc}-\ac{srm} (e.g.\ small preclinical and organ-dedicated systems). 
Where the \ac{srm} can be stored --- case~(3a) of \cref{fig:where_mc_fits} --- its transpose provides a matched adjoint
at no additional simulation cost, so the adjoint mismatch discussed in \cref{sec:InvProb} need not arise.
Pure \ac{mc}-models can be kept as a gold standard for specific applications, such as \ac{ai} training or limiting
cases in diagnosis.

Looking to the next five to ten years, simulation speed is unlikely to remain the limiting factor.
The obstacle is the size of the matrix being simulated: the \ac{srm} has been growing faster than the memory systems
that must hold it (\cref{sec:comparch}). 
As a consequence, the development of new representation strategies --- compression, symmetry, learned or
phenomenological surrogates, or evaluation on the fly instead of storage --- is likely to remain an active research
topic.

\newsectionpage
\section{Scatter estimation with Monte Carlo simulation\label{sec:ScatterEstimation}}
\sectionauthor[SPECT]{Johan Gustafsson}
\sectionauthor[PET]{J\"urgen J. Scheins}

Scattered events can be handled in reconstruction in two fundamentally different 
ways, irrespective of whether the scatter is modelled analytically or by \ac{mc}.
They differ in where the scatter expectation enters the forward model
$\bar{\promptsvec} = \srm\actvtyvec + \backgrvec$ (\cref{sec:WhereDoesMCfit} and
\cref{fig:where_mc_fits}).

\paragraph{Scatter as nuisance.}
In this case, the scatter expectation $\bar{\scattervec}$ is part of the background $\backgrvec$, and the system model
describes unscattered events only, $\srm = \srm_{\rm \trues}(\attnvec)$.
In the gradient $\nabla_{\actvtyvec}\loglik = \srm\transp(\promptsvec/\bar{\promptsvec}) - \sensvec$ 
(\cref{eq:gradientloglikelihood}), the scatter estimate therefore
enters only through the ratio, and scattered events are never 
considered in the backprojection.
The estimation of the expected scatter $\bar{\scattervec}$, either before or within the iterative process with 
improving activity distribution accuracy, addresses a computational trade-off.
The scatter background is dominated by low spatial frequencies, and \ac{mlem} recovers these within a few iterations, 
so a small number of updates is usually sufficient for a stable scatter estimate.
This is how \ac{mc}-based scatter estimates are currently typically used in both \ac{spect} and \ac{pet}.

\paragraph{Scatter as signal.}

The scatter expectation becomes part of the system model,
$\srm = \srm_{\rm \trues}(\attnvec) + \srm_{\rm \scatter}(\attnvec)$, so that scattered events contribute to the
activity
update\footnote{In the notation of \cref{sec:NonlinearForward} we would have $\srm_{\rm \trues} \equiv \srm^{(0)} $ and
$ \srm_{\rm \scatter} \equiv \sum_{k=1}^{\infty }\srm^{(k)}$.}.
This requires the adjoint $\srm_{\rm \scatter}\transp$, i.e.\ back-projection along broken, patient-specific paths.
That is difficult with \ac{mc}, since simulators favour the forward direction and storing $\srm_{\rm \scatter}$ is
rarely feasible.
The approach has been demonstrated in \ac{spect}~\cite{Lazaro2005} and
\ac{pet}~\cite{rehfeld2007parallelizable,conti2012reconstruction,hemmati2017compton,schaar2019modelling}.
For the activity distribution, however, the gain might be small because of the domination of the  low-frequency
information.
The attenuation, in contrast, shapes the scatter distribution directly.
Scattered events can therefore contribute considerably more to the joint estimation of activity and attenuation
in \ac{pet}, which is discussed in \cref{sec:JAAE}.
The adjoint needed in both cases is treated in \cref{sec:InvProb}.

\bigskip

This section is concerned with the first option, i.e.\ with how \ac{mc} can make the scatter estimate more accurate
and fast enough to be used in or near the reconstruction loop.
The \ac{spect} part (\cref{sec:scatterspect}) contrasts window-based, analytical model-based and \ac{mc}-based
estimates, including reconstruction with \ac{mc} in the forward projector and a scatter-free back-projector.
The \ac{pet} part (\cref{sec:scatterpet}) traces the transition from (tail-fitted) single scatter simulation to full
\ac{mc} scatter simulation.

\subsection{\acs{spect}\label{sec:scatterspect}}

\subsubsection{Background}
The go-to solution for scatter-correction in \ac{spect} has long been various forms of window-based methods, in
particular the \acl*{dew} and \ac{tew} methods \cite{Jaszczak1984,Ogawa1991}.
In these approaches, the scatter in the main energy window is estimated from one or two adjacent scatter windows, and
the resulting scatter estimate is added to the forward-projected source estimate in the iterative  reconstruction as the
term $\backgrvec$ in \cref{eq:mlem}.
Although such methods have indeed demonstrated sufficient accuracy for many cases \cite{Zeintl2010}, they are
 associated with some limitations.
There is an inherent conflict between the need for the signal in the scatter windows to be representative of scatter in
the main window, requiring narrow scatter windows, and the need for high \ac{snr} in the scatter estimate, requiring
broad energy windows \cite{Ljungberg1994}.
Thus, an improved \ac{snr} will come at the cost of increased bias and vice versa.
Window-based methods are typically insufficient for imaging of radionuclides with complicated photon emission spectra,
e.g.\ bremsstrahlung \ac{spect} of \isotope[90]{Y}, where there are no energy windows where the signal is scatter only.
Similar concerns are valid for the application of window-based scatter-estimation methods for gamma-camera systems with
pixelated semiconductor detectors, where charge-sharing causes a low-energy tail of primary events from the full-energy
peak \cite{Pourmoghaddas2015,Mann2015}, thus contaminating scatter windows with primary events. 

One property of window-based methods is that they cannot distinguish between scatter from the patient and photons that
result from scattering processes in the camera head before detection.
In contrast, the classical definition of scatter is that it refers to photons resulting from scattering in the patient,
while scattering in the camera head is part of the collimator-detector response.
For window-based methods, these cannot be separated, thus resulting in a more complex book-keeping of effects in the
case of collimator-detector response modelling, and the exact meaning of scatter is not always agreed upon in the
literature.
In the current description, scatter will refer to scatter from the patient only.

In response to the shortcomings of window-based methods, model-based scatter estimation methods have been developed, for
example, the \ac{esse} method \cite{Frey1996} and the method by \textcite{Wells1998}, with the aim of
calculating the scatter component (in projection space) from a current estimate of the source distribution and a
description of the scattering medium, typically obtained from \ac{ct}.
These methods are typically exact for first-order scatter and approximate for higher scatter orders.
Generalisation of scatter models to become exact for higher scatter orders and for general geometries quickly becomes
computationally intensive, and the only feasible evaluation method is then to use \ac{mc} simulations.

\begin{figure}[!t]
    \centering
    \includegraphics[width=0.75\linewidth]{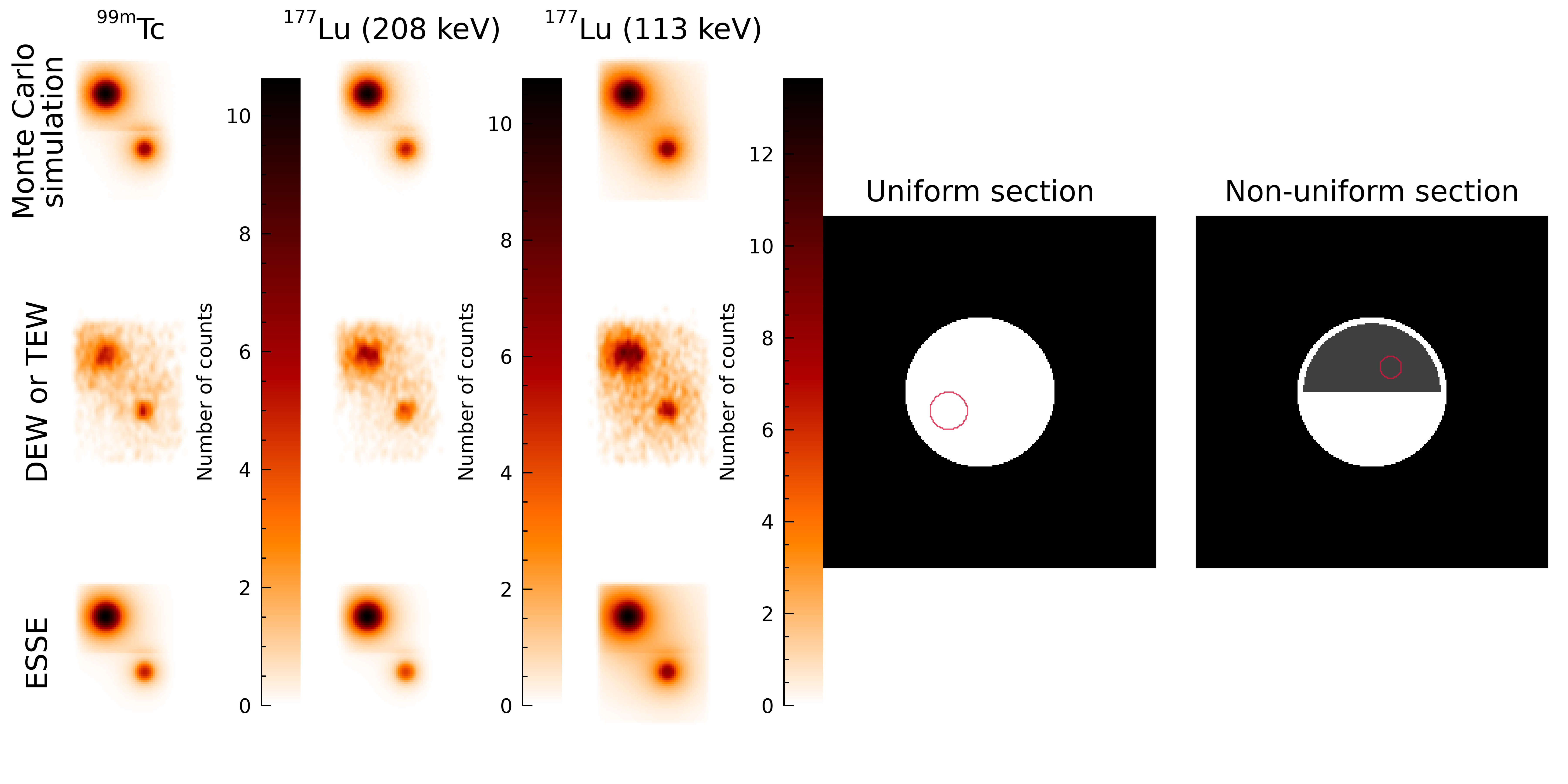}
    \caption{Comparison of scatter-estimation methods for \ac{spect} of \isotope[99m]{Tc} and \isotope[177]{Lu} using
    Monte Carlo simulated projections (left).
    The simulated source distribution and attenuating and scattering geometry, with one uniform and one non-uniform
    section, are illustrated on the right.
    The red circles indicate the activity-filled spheres.}
    \label{fig:scatter_spect}
\end{figure}

\begin{figure}[!t]
    \centering
    \includegraphics[width=0.75\linewidth]{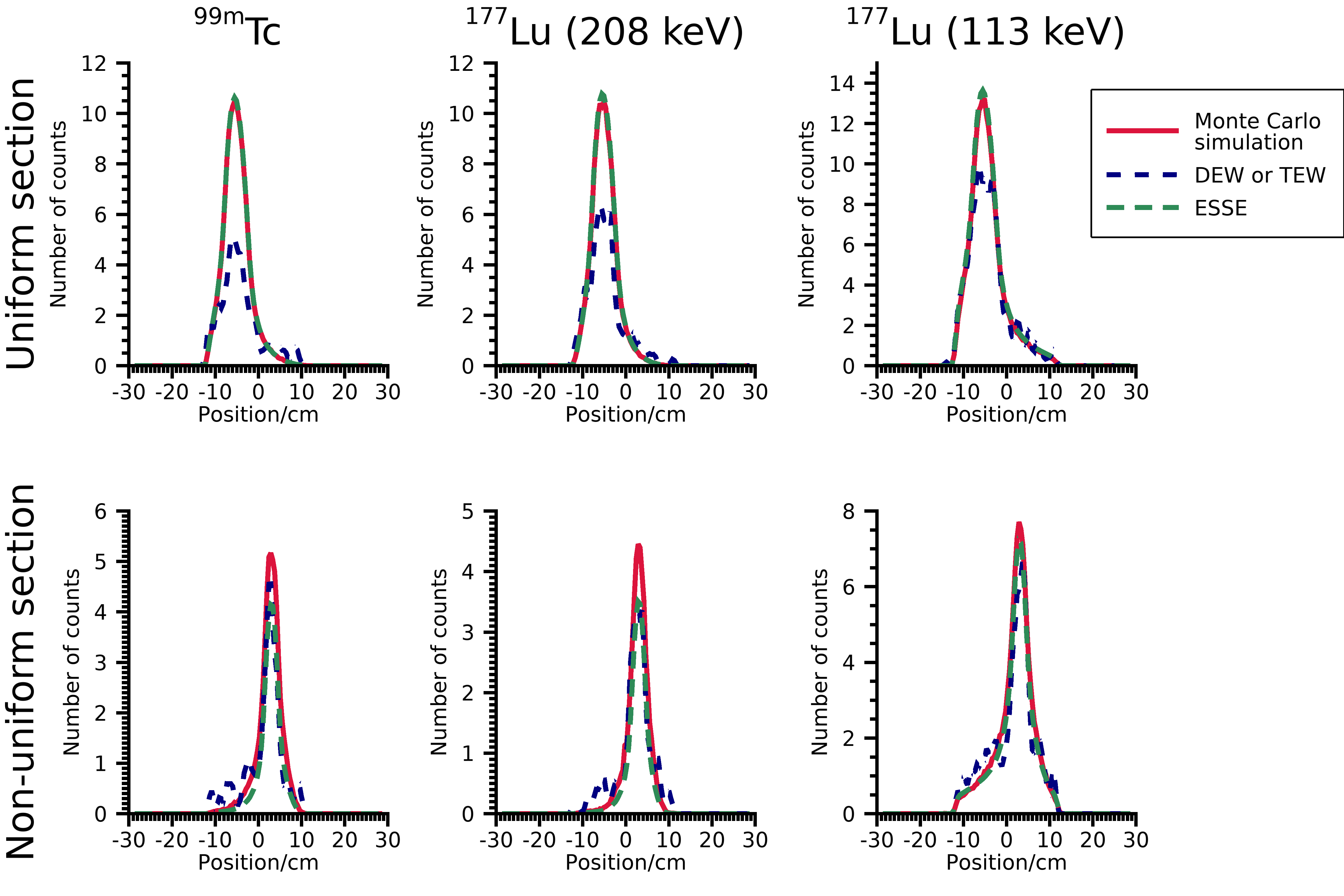}
    \caption{Profiles through the spheres in the projections of \cref{fig:scatter_spect}.}
    \label{fig:scatter_spect_plt}
\end{figure}

A comparison of \ac{mc}-modelled scatter estimation with window-based methods is presented in \cref{fig:scatter_spect}.
 The cylindrical water phantom is divided into two sections: one with a uniform density of
1.0\,g\,mL\textsuperscript{\textminus1} and a non-uniform section where the upper half has a lower density of
0.25\,g\,mL\textsuperscript{\textminus1}.
A large sphere of volume 113\,mL filled with activity is placed in the uniform section, and a smaller sphere of 34\,mL
is placed in the non-uniform section, as indicated by the red circles.
Simulations were performed using the \ac{simind} Monte Carlo program \cite{ljungberg1989}  for 30\,s per projection and
activity concentrations of 0.1\,MBq\,mL\textsuperscript{\textminus1} (\isotope[99m]{Tc}) or
 1.0\,MBq\,mL\textsuperscript{\textminus1} (\isotope[177]{Lu}).
The camera was equipped with a low-energy high-resolution (\isotope[99m]{Tc}) or medium-energy general-purpose
(\isotope[177]{Lu}) collimator and 20\% energy windows centred at 140\,keV (\isotope[99m]{Tc}) or 113\,keV and 208\,keV
(\isotope[177]{Lu}).
The \ac{esse} method follows the implementation in \textcite{Gustafsson2026}.
The worse scatter conditions for \isotope[177]{Lu} at 113\,keV can be appreciated, as can the difference in \ac{snr} for
the window-based methods compared with the model-based ones.
The quantitative difference between methods is illustrated in \cref{fig:scatter_spect_plt}.
The model-based methods perform generally better than the window-based method.
The model-based methods agree well in the uniform section, but the \ac{esse} method underestimates scatter in the
non-uniform section.  

\subsubsection{Status}

The idea of using \ac{mc} simulations for evaluating the projection model in \ac{spect} was initially explored by
\textcite{Floyd1985} for two-dimensional problems, and the dramatic improvement in computational power since the
1980s, in combination with development of models that can be efficiently evaluated, has made its application for
three-dimensional reconstruction fully realistic with clinically acceptable computation times
\cite{Beekman2002,ryden2018,sohlberg2008,Bexelius2018}.

\Ac{mc}-based \ac{spect} reconstruction programs can be broadly categorised into those that employ the Monte
Carlo-evaluated
model in the forward projection only 
\cite{Beekman2002,sohlberg2008,ryden2018,Moore2006,Gustafsson2018}, 
while a simpler model excluding scatter typically evaluated using non-\ac{mc} methods is used in the back-projection 
(dual-matrix reconstruction),
and programs where the system matrix is precomputed and used for both forward- and back-projection
\cite{Floyd1985,Lazaro2005}.
The underlying theory for dual-matrix is relatively well understood \cite{Zeng2000}.
Thus, the omission of scatter in the back-projector is unlikely to pose a problem in practice and is a way to avoid the
storage requirements when pre-computing the system matrix, which for the scatter component needs to be performed
separately for each reconstruction.
Rather than treating scatter as an additive component as in \cref{eq:expected_prompts}, a dual-matrix framework may
still compute the scatter component from the current source estimate, which is thus also refined as the source estimate
improves.
However, the scatter component is still treated as a nuisance in the sense that it is not utilised when updating the
source estimate. 

Either both the primary and scatter components in the projector can be evaluated by \ac{mc} techniques
\cite{Gustafsson2018,ryden2018} or the evaluation of the primary component may be fully separated from the evaluation of
the
scatter component and the results combined \cite{Dewaraja2006,Dewaraja2017}.
Since scatter is a low-frequency component of the projections, a common strategy for model-based scatter compensation is
to model scatter explicitly in some iterations only and to reuse the scatter projections in subsequent iterations
\cite{Kadrmas1998}, which has also been applied to \ac{mc}-based scatter evaluation \cite{Dewaraja2006,Dewaraja2017}.

The low fraction of emitted photons in \ac{spect}  that end up being detected in the camera head (typically of the order
of 1:10 000) leaves substantial room for applying \acp{vrt} (\textit{cf.} \cref{sec:varred}).
A technique of particular relevance to scatter estimation is the use of \ac{fd} \cite{ljungberg1989}, especially when
combined with computationally efficient detector models, \textit{e.g.} in the form of \ac{cfd} \cite{dejong2001}.
Initially, this technique was introduced for geometric collimator models, i.e., ignoring penetration scatter in the
camera head, but has since been expanded to models including these effects by pre-simulation of distance- and
energy-dependent kernels with application to \ac{mc}-based reconstruction of  \isotope[90]{Y} bremsstrahlung imaging
\cite{elschot2013} and imaging of \isotope[166]{Ho} \cite{elschot2013b}.

A major step forward for making Monte Carlo-based reconstruction feasible in clinical routine has been the transition
from \ac{cpu} to \ac{gpu} \cite{ryden2018}.
The combination of \ac{gpu} with computationally efficient \ac{mc} codes has made Monte Carlo-based reconstruction
clinically realistic \cite{sohlberg2008,Bexelius2018}, and such reconstruction is today commercially available.
However, transferring existing \ac{mc} codes to \ac{gpu} has in general turned out to be non-trivial and to require
rather
substantial changes to the model for optimal performance \cite{Bexelius2018}.
An alternative route for acceleration of \ac{mc} derived scatter estimates is to combine \ac{mc} simulations with deep
learning.
Either fully supervised training based on \ac{mc} simulations \cite{Xiang2020} or short \ac{mc} simulations combined
with
a weakly supervised network to improve \ac{snr} \cite{Kim2023} have been used.
However, these techniques, while rapidly developing, are still in their infancy. 

\subsubsection{Current and future challenges \label{sec:spect_merit}}
\Ac{mc}-based scatter compensation is now clinically available but not yet established as the standard method for
\ac{spect}.
Hence, there is a need for studies that clarify when these more advanced methods bring a benefit in clinical practice.
The technical benefits of model-based scatter compensation, and \ac{mc}-based scatter compensation in particular, have
been clearly demonstrated in quantitative imaging \cite{Dewaraja2006} and with respect to \ac{snr} in cardiac imaging
\cite{Xiao2006,Xiao2007,Dietze2020}, but relatively few studies have demonstrated how these technical advantages also
lead to an improved clinical benefit compared with the simpler window-based methods.
\textcite{Wikberg2023} compared the detectability of liver lesions in [\isotope[111]{In}]In-octreotide
\ac{spect} imaging for different reconstruction methods, including their in-house \ac{mc} reconstruction software
\cite{ryden2018}.
However, they did not compare with standard window-based scatter-compensation. 

\Ac{mc}-based \ac{spect} scatter compensation is often described as slow but accurate.
This is an unfortunate description of the problem as it is rather the model for scatter that is accurate, whilst the
\ac{mc} method is a relatively fast evaluation method for models of a certain high complexity.
There is thus a need to understand what level of model refinement is necessary for different purposes.
There is reason to believe that the main advantage of \ac{mc}-based scatter compensation will be for quantitative
imaging, for example, in the context of peri-therapeutic imaging for dosimetry in radionuclide therapy.
The radionuclides used for these purposes typically have less favourable imaging properties than their diagnostic
counterparts.
A number of papers have presented the adaptation of \ac{mc}-based reconstruction to such radionuclides
with unfavourable scatter properties, e.g., \isotope[90]{Y} \cite{Dewaraja2017,elschot2013} or \isotope[166]{Ho}
\cite{elschot2013b}.
The trade-off between the need for model accuracy and computational speed for particularly difficult radionuclides is
still not fully understood.
Ultimately, there will be a limit when improved model accuracy will not transfer to better source estimates in practice.

A problem sometimes raised for model-based scatter compensation, including the use of \ac{mc} methods, is the problem of
scatter from outside the \ac{fov}.
This contribution is intrinsically included in window-based methods, which is a potential advantage for these methods.
For cardiac \ac{spect}, \textcite{Xiao2010} found that scatter from activity in the liver was a small problem
for single-nuclide imaging but a major problem for dual \isotope[99m]{Tc}/\isotope[201]{Tl} imaging with \ac{mc}-based
scatter compensation.
However, for that particular case the problem could be circumvented by extending projections by replicating rows at the
edge of the \ac{fov}.
Similarly, the problem of extra-FOV activity, including the scatter, has been investigated for an experimental
brain-\ac{spect} system, and the problem was shown to be modest \cite{auer2021}.    

\subsubsection{Development needed to meet the challenges}
Despite \ac{mc}-based scatter modelling arguably being technically superior to other methods (i.e., window-based methods
or other model-based methods), further studies about its clinical utility are urgently needed to demonstrate that the
method brings meaningful improvement also for patients.
A problem for this development seems to be the difficulty of employing different scatter-estimation methods within the
same
software frameworks.
Thus, modularisation of \ac{mc} scatter estimation routines so that they can be combined within existing frameworks
would be a welcome development in order to isolate the effect of the scatter-compensation method.
A prerequisite for such research to be conducted is also that \ac{spect} systems will remain relatively open such that
projection data and basic geometry specifications are available to users.
This poses a challenge for the field due to the trend of new gamma-camera types becoming commercially available
\cite{Erlandsson2009,Desmonts2020}.
For the future, \ac{mc}-based scatter compensation will need to be implemented for these new geometries.     

An improved understanding of the trade-off between the need for model accuracy and evaluation speed is of interest for
\ac{mc} scatter modelling to become clinically available also for radionuclides with poor imaging properties, e.g.,
\isotope[90]{Y}.
Ultimately, there will be a limit where further model refinement will not result in meaningful improvement of the
reconstructed image, and systematic studies comparing models of gradually increasing complexity would be welcome in this
respect.
Inspiration for such work may be found in the recent publication by \textcite{Polson2025}, where \ac{mc}-based
reconstruction was compared with conventional reconstruction with \ac{tew} scatter compensation within the PyTomography
framework.
Further acceleration of \ac{mc}-based scatter estimates may also be helped by the combination with machine learning.
How to make best use of machine learning should be an important area for future studies.

Understanding the effect of scatter from extra-FOV activity on \ac{spect} images reconstructed with model-based scatter
compensation versus window-based methods would be useful in order to dismiss or confirm this concern against the use of
\ac{mc}-based scatter modelling.
In principle, such studies should be straightforward, but the problem lies in the dependence on geometry, energy and
acquisition settings.
Thus, general statements may be difficult to achieve.
Inspiration for a detailed study of the problem and its potential solution can be found in the technical report by
\textcite{Sauve2025}.

\subsubsection{Concluding remarks}
\Ac{mc}-modelling for scatter compensation has been firmly proven to be both clinically realistic and
advantageous for quantitative-imaging tasks, with the main benefit being in the realm of peri-therapeutic imaging in
molecular radiotherapy, which often faces the problem of complex photon-emission spectra.
From a theoretical perspective, there should also be a gain for diagnostic tasks due to an improved \ac{snr} compared
with window-based methods and, potentially, better contrast from improved scatter compensation.
However, the clinical relevance of such an improvement remains to be seen.    

\subsection{\acs{pet}\label{sec:scatterpet}}

\subsubsection{Introduction}

In \ac{pet}, Compton scattering causes deflections of emitted 511\,keV gamma photons due to interactions with matter (in
the subject as well as passive infrastructure material) on the pathway to the detector elements, thus invalidating the
spatial information of affected gamma pairs originating from the $e^+e^-$ annihilation process. 
This has two consequences: the scattered photons are missing on the \ac{lor} of their emission, which reduces the
probability of detecting true coincidences in the corresponding detector pair (attenuation), and they may instead be
detected in other detector pairs, where they contribute scattered coincidences as background.
The effect of attenuation can be estimated and corrected by using appropriate attenuation maps (see also
\cref{sec:jeaabg}), whereas the 
estimation of the scatter background for all detector pairs depends on the underlying attenuation map (i.e.\ electron
density and material composition) as well as the emission distribution. 
The scatter background comprises approx.~30--35\,\% for brain-dedicated \ac{pet} scanners and more than 50\,\% in
whole-body scanners, 
thus representing a predominant source of bias if not considered appropriately.
In principle, scattered gamma photons can be identified and discriminated by the measured energy of the registered
photons due to the associated energy loss during Compton scattering.
However, state-of-the-art \ac{pet} scanners achieve an energy resolution between 10\,\% and 20\,\%, thus requiring the
application of a relatively wide energy window to ensure high sensitivity.
Here, the energy window setting determines the trade-off between true count sensitivity and Compton background
contamination. 

Technically, the nature of processes induced by Compton scattering makes accurate analytical calculations of scattering
background hard to achieve 
due to the complexity of occurring multi-scatter cascades (cf.~\cref{sec_PETscatter_status}). 
The theoretical foundation is given in \cref{sec:ScatterModel}. 
While full Monte Carlo scatter simulations offer the highest accuracy and robustness for this purpose, their known huge
computational burden has hindered routine use, making the concept of Single Scatter Simulation in a variety of different
implementations the most widely adopted (fast) alternative to date. 
However, this paradigm is currently changing due to the continuously increasing computing resources of \ac{cpu}/\ac{gpu}
hardware in combination with progress in code development/optimisation.
This is currently shifting the trade-off between effort and accuracy in favour of full Monte Carlo simulations (see also
\cref{sec:WhereDoesMCfit}).

\subsubsection{Status\label{sec_PETscatter_status}}

Quantitative \ac{pet} imaging requires an accurate scatter correction and both strategies, i.e.~(semi-)analytical
approaches as well as full \ac{mcss} to 
estimate Compton scattering background have been considered and investigated since the 
beginning of \ac{pet}.
Here, by design \ac{mcss} is known to provide superior accuracy while requiring 
tremendous computing resources which were economically unreachable for clinical and routine applications in the past.
For this reason, high effort was spent to develop and optimise 
analytical approaches as practical solution.
In this context, \ac{sss} \cite{watson_single_1996},  \cite{ollinger_model-based_1996}, \cite{werling2002} represents
one of the most widespread approaches in \ac{pet} until today.
Single scatter estimates can be adjusted to the measured data by tail-fitting, thus partly compensating for higher order
scatter and  contributions originating from 
outside the scanner's field of view. 
In addition, a modified scatter correction is needed 
for \ac{tof} \ac{pet} which was achieved by extending \ac{sss} to address also \ac{tof} bins~\cite{werner2006},
\cite{watson2007}.

In conclusion, the approximate 
\ac{sss} method still represents a good compromise for the majority of  
clinical applications regarding the trade-off between accuracy and effort.
Further improvement of the accuracy requires the consideration of double scattering or 
even multiple scattering which becomes more relevant for cases with increased scatter fraction~\cite{Markiewicz_2007}.
The \ac{sss}
 can be extended to double scatter correction~\cite[e.g.][]{ollinger_model-based_1996,watson_double_2020}.
In \textcite{ollinger_model-based_1996}, 1D Gaussian convolution kernels are applied on \ac{sss} to model multiple
scatter
while \textcite{Markiewicz_2007} evaluate integrals for double scatter, but report exploding computational effort
for higher order scattering.
Obviously, such extensions bring analytical approaches close
to a break-even point regarding the trade-off between benefit and complexity (\cref{sec:ScatterModel}).

Consequently, the quest for accuracy immediately leads to \ac{mc} simulations for scatter correction.
Now, 
the increasing availability of cost-effective \ac{cpu}/\ac{gpu} hardware with an enormous boost of numerical 
throughput is progressively supporting \ac{mcss} as the ultimate technique.
\Ac{mcss} studies show higher accuracy 
due to the appropriate handling of double and multiple scattering 
\cite{adam1999,zaidi1999a,holdsworth2001investigation,jan2004,reilhac2004}. 
Recently, \acp{gpu} have been utilised to boost full \ac{mcss} and convey it closer to routine usage
\cite{bert2013,Ma_Scatter_TMI2020}.
Meanwhile very fast \ac{mc} implementations for Compton scatter correction running on \ac{gpu} hardware
are available \cite{herraiz2023,galve2024}.
But also multi-core \ac{cpu} hardware has been 
demonstrated to be a competitive alternative~\cite{scheins2021,harrison1993}, because \acp{gpu} have limitations to
reach
peak
performance for task-parallel problems as addressed by \ac{mc} simulations. 

In summary, the currently available \ac{mcss} \ac{gpu}/\ac{cpu} implementations  
can generate sufficient statistics for \ac{pet} scatter corrections of the order of minutes 
thus making their routine utilisation possible and attractive.

\subsubsection{Current limitations and future challenges}

Technically, \ac{mcss} is now on the verge of becoming feasible in routine/clinical applications 
where fast processing with moderate computing resources is of great importance.
However, new application paradigms, e.g. \ac{lafov} \ac{pet} and dual tracer \ac{pet},
pose further challenges for \ac{mcss}.
First, long axial field-of-view scanners address significantly more
lines-of-response \cite{Bayerlein_2024}, thus causing additional load in terms of
memory management and performance. 
Second, dual radionuclide imaging \cite{andreyev2011}  has a higher complexity for 
the \ac{mcss} task to obtain quantitative results due to the additional prompt gamma energy 
and occurrence of triple photon states. 
In this context, accurate \ac{mcss} is capable of resolving ambiguities and of appropriately quantifying 
both interfering tracer distributions.
That means, further improvements in terms of numerical
throughput and handling of dual radionuclide simulations are desirable and still needed.

At the time of writing, the high \ac{mc} throughput is usually achieved at the expense of some 
simplifications and approximations when comparing results to references like \ac{geant4}/\ac{gate}.
For example, \textcite{herraiz2023} report a huge production rate of roughly 7 billion 
coincidences/minute while discarding the detector simulation and having maximum deviations of the order of 10\% with
respect to \ac{gate}/PeneloPET.
Here, further reduction of 
deviations and including a detector simulation while preserving the production rate 
is needed.

Furthermore, highly accurate \ac{mcss} requires accurate attenuation maps, which remains 
a difficulty for \ac{pet} systems which cannot provide a matching \ac{ct} scan or transmission
scan, e.g. \ac{mr}-\ac{pet} hybrid scanners.
So far, \ac{mr}-based attenuation maps 
can introduce significant inaccuracies which propagate into quantitative bias.
Finding robust solutions to this known problem is currently an active field of 
research \cite{lillington2020PETMRIAttenuation,krokos_2023}.
See also \cref{sec:JAAE}.
 
Finally, any \ac{mcss} requires also an emission distribution as input and such 
input is usually provided by a preliminary reconstruction without scatter correction. 
Effectively, this causes some bias which can be minimised by repeating \ac{mcss} with
successively improved emission distribution images, i.e. incorporating scatter
corrected emission images in an extended processing.
However, such an iterative 
approach of scatter estimation with subsequent image reconstruction in a loop
induces additional computational burden.

\subsubsection{Targets to meet challenges \label{sec:PET_tagets}}

Although the performance bottlenecks progressively diminish,
the reported advantages of \ac{mcss} in comparison to state-of-the-art \ac{sss} 
are not immediately triggering the widespread introduction of this approach into the clinical 
workflow.
Existing approval procedures and regulations require some attention and 
cause additional effort until such an innovation can be released. 
In this context, the merit for patients with respect to diagnosis and/or therapy 
needs to be shown more clearly as one of the next steps (see also \cref{sec:spect_merit}).
For future prototypes and upcoming new scanners, the utilisation of 
the \ac{mcss} can be  reasonably expected~\cite{Bayerlein_2024}.
\Ac{pet} neuroscience applications
using dedicated human brain scanners for dynamic imaging and neuroreceptor modelling 
represent an excellent target because of the immediate benefit due to 
improved quantification accuracy.

The relevance of full \ac{pet} detector simulations to further improve the precision of the 
scatter correction is not yet fully clear and needs to be studied, because scatter background in \ac{pet} is
uncollimated,
thus inducing modified detection efficiencies with respect to collimated coincidences.
Meanwhile, fast full detector simulations have become possible~\cite{scheins2021} to investigate the relevance of the
detector
response on \ac{mcss} accuracy.

The development of novel variance-reduction techniques can also boost
the \ac{mcss} throughput to make it even more suitable for standard applications (cf. \cref{sec:varred}).

\subsubsection{Concluding remarks}

The continuous  progress of computing hardware in terms of data throughput as well as the 
availability and improvement of efficient \ac{pet} \ac{mc} simulation tools is currently inducing a transition 
from analytical approaches (e.g.~\ac{tof}-\ac{sss}) towards \ac{mcss}.

\newsectionpage
\section{Positron range modelling with Monte Carlo simulation\label{sec:PositronRangeModelling}}
\sectionauthor{Chong Li, Joaquin L. Herraiz, J\"urgen Scheins, Christoph Lerche}

\subsection{State of the art\label{sec:posisota}}

\begin{figure}[!t]
    \centering
    \includegraphics[width=0.75\linewidth]{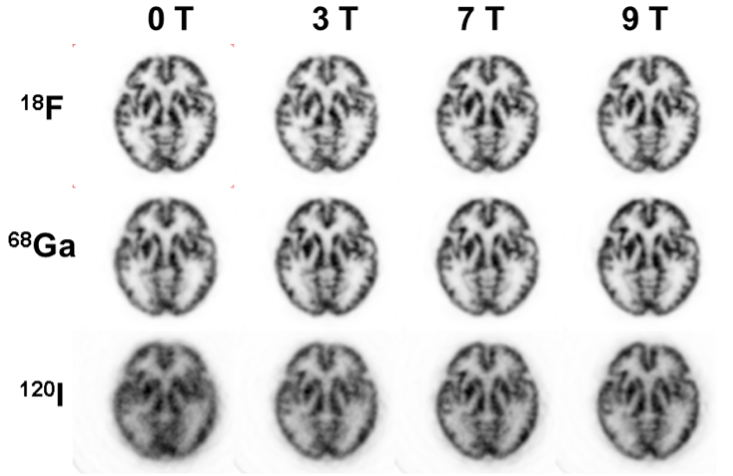}
    \caption{Transaxial \ac{pet}-images of a fillable anthropomorphic brain phantom.
    Cortex compartment filled with one of three positron emitters at magnetic field strength 0~T, 3~T, 7~T and 9.4~T.
    It demonstrates the effect of positron range on the reconstructed \ac{pet} image for different field strengths and
    nuclides.
    Image adapted from \cite{Shah2014EffectsMR-BrainPET}.}
    \label{fig:posRangeMagField}
\end{figure}

The two 511\,keV annihilation gamma-ray photons, which define the \acp{lor} used to localise the origin of the detected
events, are, in most cases, emitted after the positron has been thermalised.
This thermalisation process involves the positron losing its initial kinetic energy through inelastic collisions with
atomic electrons, elastic scattering off nuclei within the surrounding matter, and radiative losses (bremsstrahlung).
This energy loss is a stochastic process, as are the energy and direction of the emitted positron.
Therefore, the finite distance a positron travels before annihilating, known as the positron range, introduces a
stochastic uncertainty in the annihilation event's localisation relative to the radionuclide's actual decay site. 
The positron range depends on the density and atomic composition of the traversed tissues or materials and the emitted
positron's initial kinetic energy.
Positrons are emitted with a continuous energy spectrum, ranging from zero up to a maximum endpoint energy ($\epmax$),
which is characteristic of the specific radionuclide and its decay branch.
For instance, \isotope[18]{F} and \isotope[64]{Cu} have relatively low positron energies, with $\epmax$ of 0.634~MeV and
0.653~MeV, respectively.
In contrast, other \ac{pet} nuclides, such as \isotope[15]{O} and \isotope[68]{Ga}, possess significantly higher
positron energies, with $\epmax$ values of 1.73~MeV and 1.90~MeV, respectively \parencite{Conti2016}.
In homogeneous media, the positron annihilation point distribution is characterised by a sharply peaked central maximum
with extended, non-Gaussian tails.
The 1D histograms of the 3D distributions exhibit \ac{fwtm} values that can be approximately 5 to 10 times larger than
their corresponding \ac{fwhm} values, e.g.~in water, the corresponding \ac{fwhm}/\ac{fwtm} values were found by
simulation to be 0.1\,mm / 1.0\,mm for \isotope[18]{F} and around 0.5\,mm / 4.1\,mm for \isotope[15]{O}
\parencite{Levin1999CalculationResolution}.
It is important to note that the related quantities $R_\mathrm{mean}$ (mean positron range) and $R_\mathrm{max}$
(maximum positron range) refer to the mean (resp. max) distance between the emission and annihilation point, while in
\ac{pet}-imaging the combined effect including also the projection of the three-dimensional positron annihilation
distribution through the \acp{lor} onto the detectors leads to the image blurring caused by positron range
\parencite{Lehnert2011AnalyticalSimulation}.
As the positron range also depends on the density and composition of the matter or tissue it traverses, the
nuclide-specific positron range distribution acts as an object-dependent blurring component within the system's overall
\ac{psf}, and not only leads to blurring but also to \emph{shine-through} caused by air-cavities such as the trachea and
by highly heterogeneous, low-density tissue such as the lung which can lead to artefacts and quantification errors even
in standard clinical imaging \parencite{Emond2019EffectLungs,Abdul-Fatah2009IdentificationPET/CT,kolb2015shine}.
These effects limit the achievable spatial resolution and quantification accuracy, as the reconstructed image represents
a convolution of the true tracer biodistribution with the system \ac{psf}, motivating the development of \ac{prc}
techniques. 
The importance of \ac{prc}  varies significantly depending on the intrinsic spatial resolution capabilities of the
\ac{pet} system.
\Ac{prc} is highly beneficial for small-animal \ac{pet} scanners with resolutions often $\lesssim$\,1\,mm.
For dedicated organ imaging systems (resolution $\approx$\,1.5 to $\approx$\,2.5\,mm), it remains important, while for
standard clinical \ac{pet} scanners (resolution $\gtrsim$\,3.5\,mm), its impact is relatively less critical compared to
other resolution-degrading factors, especially for lower-energy positron emitters such as \isotope[18]{F}. 
Simple \ac{prc} was proposed as early as in the 1970s \parencite{Derenzo1979PrecisionEmitters} using a deconvolution
with a uniform \ac{psf}, however, several factors may affect the typically assumed isotropic nature of the positron
range distribution, such as tissue inhomogeneities and organ boundaries (e.g.~lung-tissue interfaces), and the presence
of strong magnetic fields, as encountered in integrated \ac{mr}/\ac{pet} imaging systems
\parencite{kolb2015shine,Li2017CompressiveGeant4} (see \cref{fig:posRangeMagField}).
In those cases, an accurate \ac{prc} requires a more sophisticated, position- and direction-variant modelling of the
positron range, with the methods summarised below.

Various methodologies for \ac{prc} have been developed over time, each differing in terms of accuracy,
computational requirements, and clinical applicability.
These methodologies can be broadly categorised into analytical or parametrised models, \ac{mc} simulations,
\ac{mal}-based approaches, and combinations of these. 
\Ac{mc} simulations have emerged as the gold standard for positron range correction; however, full \ac{mc} simulations
are computationally very demanding and still too slow for routine clinical application on the fly.
This is impractical even for moderately sized kernels and few different tissue types
\parencite{Kraus2012SimulationSystem}.
\Ac{gpu}-accelerated \ac{mc} implementations have enabled a substantial increase in speed, addressing a key limitation
for clinical use.
\textcite{Ma_Scatter_TMI2020} proposed a \ac{gpu}-accelerated \ac{mc} approach for scatter correction, including
positron propagation. 
They achieved 0.9 million positron propagations per second, including the subsequent tracing of the annihilation
photons, in a voxelised brain phantom study.
Although this was still too slow for on-the-fly \ac{prc} within an iterative correction algorithm,  advances in \ac{gpu}
capabilities are expected to make it a viable option in the future. 
Hybrid methods that combine \ac{mc} simulations with analytical models represent a promising balance between accuracy
and computational efficiency.
These can involve:
\begin{itemize}
    \item \textbf{\Ac{mc}-derived kernels:} Pre-computing libraries of positron range kernels using \ac{mc} simulations
    for different tissue types (identified from \ac{ct} or \ac{mr} data).
    These kernels (often anisotropic and spatially variant) are then incorporated into the image reconstruction
    algorithm, typically within the \ac{psf} or as a post-reconstruction deconvolution step.
    This approach is computationally efficient during the actual patient reconstruction.
    Cal-González et al., for example, used the PeneloPET framework to generate tissue-dependent and spatially variant
    \acp{psf} (incorporating positron range) from \ac{mc} simulations, demonstrating minimal computational overhead
    during reconstruction due to the use of these pre-computed kernels \parencite{Cal-Gonzalez2015Tissue-DependentPET}.
    Recent work from \textcite{Kertesz2022ImplementationImaging} has demonstrated its application in clinical scanners,
    obtaining significant
    improvements with \isotope[82]{Rb} \parencite{Lassen2025PositronPET/CT}. 
    \item \textbf{\Ac{mc}-informed models:} Using \ac{mc} data to derive, calibrate, or refine parameters within
    analytical or simplified physical models.
    \textcite{Li2023FastPET} proposed a method that approximates positron trajectories using steps smaller than voxel
    sizes, informed
    by \ac{mc}-derived data for adjustments, to generate medium-dependent anisotropic 3D kernels.
    Their approach reportedly links positron stopping power considerations with material properties, with the aim of
    generating accurate kernels.
\end{itemize}

These hybrid strategies effectively combine the precision of \ac{mc} simulations for characterising positron range with
greater computational tractability, enabling the implementation of accurate and efficient \ac{prc} in current \ac{pet}
systems.
The goal is to achieve a robust correction that accounts for tissue-dependent variations in positron range, thereby
improving the quantitative accuracy and spatial resolution of the final reconstructed \ac{pet} images.

The synergy between \ac{mc} simulations and \ac{mal} models is proving highly beneficial for \ac{prc}.
By training on extensive, high-precision data generated via \ac{mc} simulations, \ac{mal} approaches can learn complex,
non-linear relationships to approximate \ac{mc}-level accuracy in correcting for positron range effects, while achieving
significantly faster execution times during inference.
Within this framework, \ac{mal} models, especially deep learning architectures, have emerged as an innovative and
powerful strategy for \ac{prc}.
For instance, \textcite{Herraiz2020Deep-LearningImages} developed a deep learning model, referred to as Deep-\ac{prc}.
Trained on \ac{mc}-simulated data, this model aims to restore \ac{pet} images by mitigating the blurring of the positron
range.
The authors reported high fidelity in image restoration, with a quantified accuracy of up to 95\% compared to the ground
truth in their study metrics, and minimal noise amplification, offering a rapid correction that is particularly
advantageous for high-energy positron emitters like \isotope[68]{Ga}.
\textcite{encinabaranda2026tissue} further incorporated tissue-dependent anatomical information into this approach,
training 3D
residual encoder-decoder \acp{cnn} on simulated data with an attenuation-image-dependent loss function and using both
the initial \ac{pet} and \ac{ct} images as inputs.
In simulated and preclinical \isotope[68]{Ga} mouse studies, the Two-Channel model outperformed Richardson-Lucy
deconvolution, recovering lung activity with 97\% agreement to the ground truth compared to 77\%, without increasing
noise.
Similarly, \textcite{Yang2021CompensatingStudy} investigated the application of various \acp{cnn} for \ac{prc}, also
using \ac{mc} simulations
for training.
Their findings indicated that \ac{cnn}-based correction can significantly sharpen image boundaries and improve
quantitative
metrics, such as recovery coefficients, in \isotope[68]{Ga} \ac{pet} images without undue noise increase.
These approaches use the accuracy of Monte Carlo simulations in conjunction with the short inference time of deep
learning models, facilitating real-time \ac{prc} and demonstrating the effectiveness of \acp{cnn} in preclinical and
clinical positron range correction. 

Rather than restoring an already reconstructed image, deep learning can also be used to model the \ac{pr} within the
forward model of iterative image reconstruction.
In this setting, the network predicts the \ac{pr} blurring of the activity distribution, which is then followed by the
conventional forward projection, which preserves consistency with the measured data and reduces the risk of hallucinated
structures.
For instance, \textcite{mellak2026dual} proposed a Dual-Input Dynamic Convolution network that uses the activity image
together
with the attenuation map to generate spatially variant, tissue-dependent positron range kernels.
The network is trained on \ac{mc}-simulated data and embedded within iterative reconstruction.
Such forward-modelling approaches combine the accuracy of \ac{mc} simulations with the short inference time of deep
learning,
while retaining the robustness of model-based reconstruction. 

\subsection{Challenges and needs\label{sec:poscan}}

Integrating \ac{pr} models into on-the-fly \ac{psf} computations requires their inclusion in both the forward and
backward projection operations of iterative reconstruction algorithms.
The use of unmatched projectors (i.e., a more accurate physics model in the forward projector than in the back
projector, see also the discussion on photon transport and its adjoint in \cref{sec:rte,sec:WhereDoesMCfit}, in
particular \cref{eq:adjointRTEpropagator}) has been explored by several groups (e.g., \cite{Fu2010ARange} and references
therein), to lower the computational burden of \ac{mc} simulations.
In this context, \textcite{Fu2010ARange} used a residual error correction implemented into a \ac{map} reconstruction to
address the high
computational burden of \ac{mc} range simulation. 
Despite the active development of positron range correction methods, achieving Monte Carlo-level accuracy while
substantially reducing processing time in clinical settings remains a significant challenge.
For example, a straightforward single-thread \ac{cpu} \ac{mc} computation of a \isotope[120]{I} positron range kernel (1
million events) for one image voxel (water) requires about 5~min using the \ac{geant4} library.
Although efficient parallelisation may yield computation speed improvements of 3 to 4 orders of magnitude,
parallelisation alone will not be sufficient in the near future to enable pure \ac{mc} on-the-fly positron range kernel
computation. 
The main challenge is that the positron range is spatially variant due to tissue borders and/or magnetic fields, and
thus the kernel needs to be computed for all voxels with activity, although avoiding repetitions by  
precomputing a dictionary of all classes of tissue interfaces and neighbourhoods could potentially lower the
computational burden while maintaining the required accuracy. 
A state-of-the-art \ac{pet} image can have up to 100 million voxels. 
Even though the positron range kernels, unlike for instance the scatter estimate, do not need to be updated during the
reconstruction (the positron range does not depend on the biodistribution of the tracer and is part of those
\ac{psf} components that
can be precomputed, see \ac{mc} realisation of the forward model in \cref{fig:where_mc_fits}), the combination with
algorithmic optimisations, such as factorisation, enhanced analytical precomputations, \ac{mc} variance reduction
methods, and machine learning models into hybrid methods will still be required in the next few years.
Given their potential, hybrid \ac{mc}-based methods that aim to find a balance between accuracy and computational
efficiency can serve as an effective solution until sufficient computational power is available for pure on-the-fly
\ac{mc} simulations. 

\subsection{Final remarks\label{sec:posfr}}

Finally, it is important to remember that the positron range component is an integral part of the overall system's
\ac{psf}.
As demonstrated by researchers like \textcite{Rahmim2013ResolutionPitfalls} and
\textcite{Nuyts2014UnconstrainedSolution}, incorporating resolution modelling (including \ac{prc})
into unconstrained iterative reconstruction algorithms can lead to issues.
These include the potential for non-unique solutions in the presence of noise and the increase of Gibbs artefacts
(ringing) near sharp discontinuities in activity (see the discussion of ill-posedness in \cref{sec:DeterministicRecon}).
This may be particularly important when attempting to perform on-the-fly positron range computations, as this often
involves compromising accuracy.
Such compromises include, for example, stopping early or using less precise approximations in favour of speed.
In this case, the kernel may not accurately model the real situation, potentially causing artefacts, as in the case of
blind deconvolution \parencite{Prasath2009RingingMethods}.

\newsectionpage
\section{Joint estimation of activity and attenuation in PET with Monte Carlo simulation\label{sec:JAAE}}
\sectionauthor{Georg Schramm}

This section discusses how fast and accurate \ac{mc} simulation can benefit the \acf{jeaa} in
\ac{pet}, for which an accurate scatter model has long been a limiting factor.
In \ac{pet}, the primary goal is a quantitatively correct activity image, and errors of the attenuation estimate
matter only as far as they affect the activity.
With \ac{tof}, the emission data determine the attenuation factors up to a global scale, which translates into a
global bias of the activity.
The scatter is both a problem and a potential solution here.
Errors of the scatter model bias the joint estimate, but the absolute scatter level carries the information that
fixes the scale, provided that it is modelled accurately and that its dependence on the attenuation enters the
estimation.
Fast \ac{mc} simulation makes such absolute scatter estimates possible, including multiple scatter, 
and provides the reference against which faster, analytic, differentiable scatter models can be validated.
\Ac{jeaa} has also been studied in \ac{spect}, partly exploiting scattered
photons~\cite{cade2013,Courdurier2015,bousse2016,rahman2020,li2026}.
Since \ac{spect} provides no \ac{tof} information, the problem is
more ill-posed than in \ac{tof} \ac{pet}.
It may nevertheless be worth revisiting with accurate, possibly differentiable, models of single and higher-order
scatter, analytic or \ac{mc}-based.
\Cref{sec:jeaabg} recalls attenuation modelling in \ac{pet}.
\Cref{sec:jeaapet} summarises the status of \ac{jeaa}, the information on the attenuation that scattered
coincidences carry beyond the trues, and the role of fast \ac{mc} scatter estimation.
\Cref{sec:jeaachall} discusses the current and future challenges, in particular the non-uniqueness of the
\ac{jeaa} problem and the required accuracy of the scatter model.
\Cref{sec:jeaanew} outlines the developments needed to resolve the non-uniqueness by a joint estimation from
true and scattered coincidences.
\Cref{sec:jeaaconcl} concludes the section.

\subsection{Background\label{sec:jeaabg}}

Quantitative \ac{pet} requires an accurate attenuation image $\attnvec$.
It enters the forward model through the attenuation factors of the \ac{srm} (\cref{sec:ModelPipeLine}) and
through the scatter term (\cref{sec:ScatterModel}).
In clinical \ac{pet}/\ac{ct} imaging, $\attnvec$ is derived from a \ac{ct} scan by converting Hounsfield units to linear
attenuation coefficients at 511\,keV.
Where a \ac{ct} is not available, as in \ac{pet}/\ac{mri}~\cite{krokos_2023}, for ultra-low-dose \ac{ct}-less imaging
with \ac{lafov} scanners, or in organ-dedicated standalone \ac{pet} scanners, $\attnvec$ has to be obtained otherwise.
Options include \ac{mri}-based methods~\cite{krokos_2023}, the \isotope[176]{Lu} background radiation of the
scintillator~\cite{Teimoorisichani2022}, deep learning~\cite{Lee2021_DLAC} and the emission data themselves via
\ac{jeaa}, or a combination of these methods.
If $\attnvec$ is precisely known, \acl*{am} poses no major obstacle, and this section is concerned with the case
where it is not.

As in \cref{eq:fwd_model1}, the expected prompt counts in \ac{pet} are the sum of true (unscattered), scattered
and random coincidences,
\begin{equation}
  \bar{\promptsvec}(\actvtyvec,\attnvec) =
  \underbrace{\srm^{(0)}(\attnvec)\,\actvtyvec}_{\bar{\truesvec}}
  + \underbrace{\overbrace{\srm^{(1)}(\attnvec)\,\actvtyvec}^{\text{single scatter}}
    + \overbrace{\textstyle\sum_{k\geq2}\srm^{(k)}(\attnvec)\,\actvtyvec}^{\text{higher-order
    scatter}}}_{\bar{\scattervec}}
  + \bar{\randomsvec}(\actvtyvec,\attnvec)\,,
\label{eq:fwd_model}
\end{equation}
where $\srm^{(k)}(\attnvec)$ is the discretised contribution of coincidences with $k$ scatter events in total
(\cref{eq:scatterterms}).
In double scatter ($k=2$), both photons scatter once or one photon scatters twice.
The trues and the scatter are linear in $\actvtyvec$; the trues depend on $\attnvec$ only through the attenuation
factors of the \acp{lor}, the scatter through the whole attenuation image.
In this section we write the \ac{tof} bin index $t$ explicitly, so that $\bar{\prompts}_{\datind t}$ denotes the
element of $\bar{\promptsvec}$ for a geometrical \ac{lor} $\datind$ and \ac{tof} bin $t$, whereas
\cref{sec:reconbasics} includes $t$ in the data index $\datind$.
Given a voxelised activity distribution $\actvtyvec$ and attenuation map $\attnvec$, the elements of
$\srm^{(0)}(\attnvec)\,\actvtyvec$, the expected trues, are typically modelled as the discretised version of the
attenuated Radon transform in \cref{eq:AttenuatedRadonTransform}
\begin{equation}
  \bar{\trues}_{\datind t}(\actvtyvec,\attnvec)
    = n_\datind\;\exp\!\Bigl(-\sum_\pixind p_{\datind\pixind}\,\attn_\pixind\Bigr)\;
      \Bigl(\sum_\pixind p_{\datind t\pixind}\,\actvty_\pixind\Bigr)
    = n_\datind\,a_\datind(\attnvec)\,\sum_\pixind p_{\datind t\pixind}\,\actvty_\pixind\,,
  \label{eq:trues_jeaa}
\end{equation}
where $p_{\datind t\pixind}$ and $p_{\datind\pixind}$ denote the elements of a \ac{tof} and a non-\ac{tof}
forward projector.
This is the factorisation in \cref{eq:factorised_srm} with the \ac{pet} propagator in \cref{eq:factorised_srm_pet}.
The normalisation factors $n_\datind$ are the diagonal of $\srm_{\rm norm}$, the attenuation factors
$a_\datind(\attnvec)$ are the diagonal of $\srm_{\rm att}$, and $p_{\datind t\pixind}$ are the elements of
$\srm_{\rm proj}$ (resolution modelling omitted for brevity).
Note that $a_\datind$ denotes an attenuation factor of \ac{lor} $\datind$, not an element $\srmel_{\datind\pixind}$
of the \ac{srm}.

The expected randoms $\bar{\randomsvec}(\actvtyvec,\attnvec)$ depend on the activity and attenuation through the
singles rates.
Their dependence on $\actvtyvec$ is non-linear, approximately quadratic, since they scale with the product of the
singles rates of the two detectors.
They are commonly pre-estimated from the measured singles rates or a delayed coincidence window and then held fixed
during reconstruction.
In principle, the singles, and hence the randoms, carry information on the scale, since the attenuation of a single
photon depends on the depth of its emission point; combined with \ac{tof} data, single events were shown to determine
the scale in a 2D setting~\cite{Feng2018}.
Modelling the singles and randoms accurately for clinical systems is, however, probably challenging and
computationally very expensive.

Most reconstruction algorithms (e.g.\ \ac{mlem}, \ac{osem}) maximise the Poisson log-likelihood in
\cref{eq:loglikelihood} with respect to $\actvtyvec$ for an attenuation image $\attnvec$ that is fixed, e.g.\ derived
from a \ac{ct}.
Since no simple closed-form expression exists for the scatter, its computation is expensive, and the scatter is smooth,
current implementations usually neglect the dependence of $\bar{\scattervec}$ on $\actvtyvec$ and treat it as a
nuisance term, pre-computed with a simplified analytical model~\cite{ollinger_model-based_1996, watson_single_1996,
watson_evaluation_1997, watson_double_2020} (\cref{sec:WhereDoesMCfit}).

Unfortunately, in practice, the attenuation image $\attnvec$ is rarely perfect, e.g.\ due to patient motion, errors in
the conversion from \ac{ct} or \ac{mri} to 511\,keV attenuation coefficients, or \ac{ct} artefacts caused by truncation
or metal implants.
These mismatches introduce systematic biases into both true and scattered forward models,
which in turn produce local, spatially variant errors in the reconstructed activity distribution $\actvtyvec$.

\subsection{Status\label{sec:jeaapet}}

\subsubsection{Joint estimation of activity and attenuation \label{sec:jeaa_status}}

\Ac{jeaa} estimates $\attnvec$ from the emission data themselves, so that the attenuation is consistent with the
\ac{pet} data, e.g.\ in the presence of respiratory or patient motion.
\textcite{defrise_time--flight_2012} showed that \ac{tof} \ac{pet} data determine the attenuation sinogram up to a
constant for all
\acp{lor} that intersect the activity when the scatter expectation is known and 
treated as a nuisance, which makes \ac{jeaa} in principle feasible with \ac{tof}.
Moreover, improving \ac{tof} resolution enhances the precision of the attenuation and activity estimates and reduces
cross-talk between both estimates, rendering \ac{jeaa} particularly attractive for modern and future \ac{pet} systems
with sub-200\,ps \ac{tof} resolution.

A variety of \ac{jeaa} algorithms have been proposed, such as \ac{mlaa}, \acl*{mlacf} and
\acl*{mlrr}~\cite{nuyts_simultaneous_1999, rezaei_simultaneous_2012, Hamill2012, rezaei_ml-reconstruction_2014,
rezaei_joint_2018, Feng2018, Rezaei2019, Li2020}.
However, they all require accurate models of scattered (and random) coincidences, which, like the trues, have to be
modelled in every \ac{tof} bin.

As noted above, the scatter expectation itself depends on the unknown attenuation image $\attnvec$, creating a
``chicken-and-egg'' problem in \ac{jeaa}.
A common strategy~\cite{Rezaei2019} starts with \ac{jeaa} updates that do not model the scatter, and then alternates
between a re-estimation of the scatter from the current $\actvtyvec$ and $\attnvec$ and \ac{jeaa} updates of
$\actvtyvec$ and $\attnvec$ with this scatter estimate held fixed.
The scatter is usually re-estimated by an analytical single-scatter simulation followed by tail fitting
(\cref{sec:ScatterEstimation}).
This is the step where fast \ac{mc} simulation can enter directly (\cref{sec:jeaaben}).

Once the scatter estimate is frozen, the constant left undetermined by the \ac{tof} data makes the \ac{jeaa} problem
non-unique (\cref{sec:jeaanonuniq}).
Moreover, \ac{jeaa} methods exhibit a greater sensitivity to systematic errors in scatter modelling and other
calibrations than standard reconstructions that estimate only $\actvtyvec$ with a fixed
$\attnvec$~\cite{Nuyts2018, Rezaei2019}.
This ``decreased robustness'', combined with the non-uniqueness, has hindered the widespread clinical adoption of
\ac{jeaa} so far.

\subsubsection{Fixing the non-uniqueness with prior information\label{sec:jeaascale}}

The non-uniqueness is commonly resolved by prior information on $\attnvec$.
The scale can be fixed by a region of known attenuation, either in an attenuation image reconstructed from the
estimated attenuation factors~\cite{rezaei_ml-reconstruction_2014} or, without repeated transmission reconstructions,
by means of a unit attenuation medium~\cite{Li2020}.
Other prior information includes known tissue-class values, the body outline, a low-dose \ac{ct}, and the sources of
attenuation information listed in \cref{sec:jeaabg}, such as \isotope[176]{Lu} transmission
data~\cite{Teimoorisichani2022}
or deep learning~\cite{Lee2021_DLAC}.

\subsubsection[Information on the attenuation in scattered coincidences]{Information on the attenuation in scattered
coincidences\label{sec:jeaainfo}}

\begin{figure}[t]
  \centering
  \resizebox{\linewidth}{!}{\input{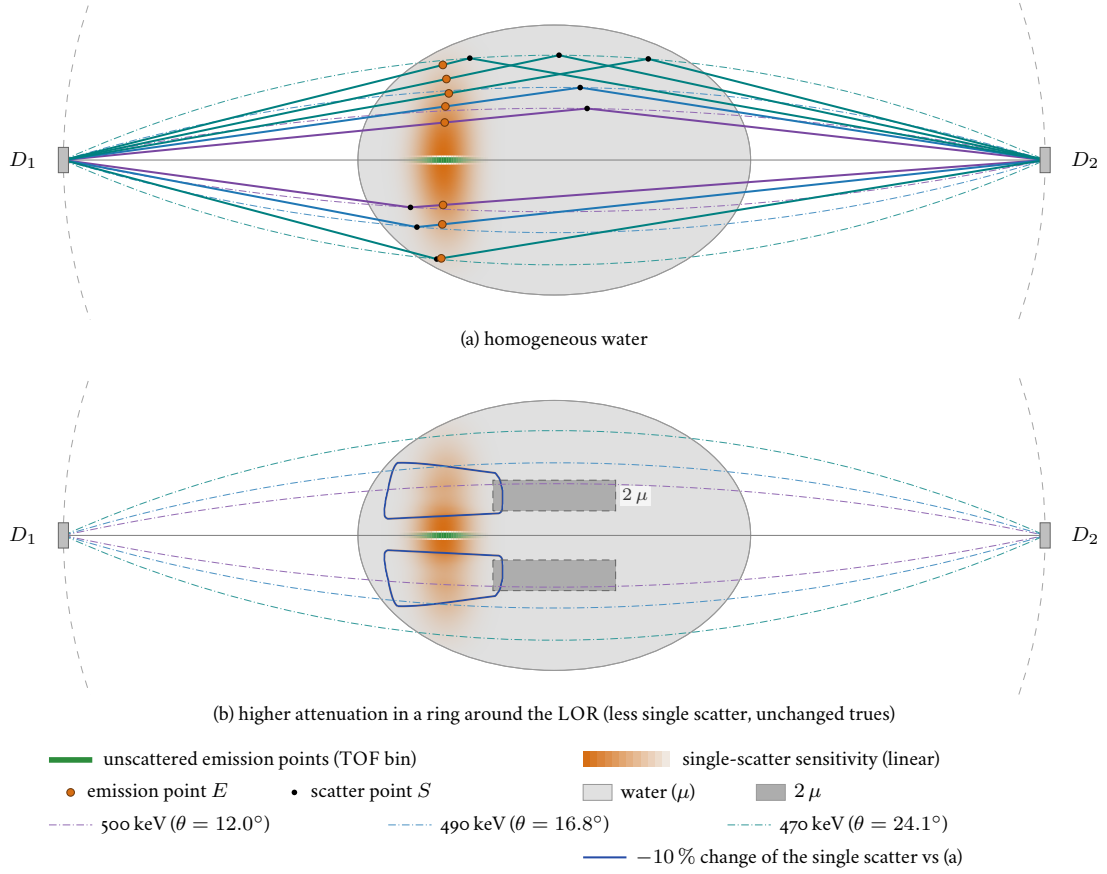}}
  \caption{Single-scatter coincidences carry information about the attenuation that the trues do not.
  Shown is one \ac{lor} ($D_1$, $D_2$) of a \ac{tof}-\ac{pet} scanner (200\,ps) and one off-centre \ac{tof} bin.
  Unscattered emission points of this bin lie on the green segment.
  The orange map shows the emission points from which single-scatter coincidences are detected in the same \ac{lor}
  and \ac{tof} bin, within a 490--600\,keV energy window (11\,\% energy resolution).
  All scatter points lie on equal-angle arcs (dash-dotted), and the coloured paths show example emission and scatter
  points.
  The map includes attenuation, the Klein--Nishina cross-section, the energy window, the \ac{tof} kernel and solid
  angle, for uniform activity in a water body.
  (a)~Homogeneous water.
  (b)~Twice the attenuation in a ring around the \ac{lor}, rotationally symmetric about it, so that the \ac{lor} and
  hence its trues are unchanged.
  The single scatter in this \ac{tof} bin decreases, since the reduced photon survival outweighs the additional
  Compton interactions in the ring.}
  \label{fig:jeaa_tof_scatter}
\end{figure}

The trues in a \ac{tof} bin of a \ac{lor} depend on $\attnvec$ only through the attenuation factor $a_\datind$ of the
\ac{lor}, i.e.\ through the line integral of $\attnvec$ along the whole \ac{lor}, which is the same for all its
\ac{tof} bins (\cref{eq:trues_jeaa}).
The single-scatter coincidences recorded in the same \ac{lor} and \ac{tof} bin, in contrast, originate from a 
whole region around the \ac{lor} (\cref{fig:jeaa_tof_scatter}).
For higher-order scatter, this region becomes even bigger.
For a given scattered energy, the scatter points lie on an arc of equal scattering angle through the two detectors,
and the \ac{tof} difference fixes the emission point on the path of the unscattered photon.
Because of the finite energy resolution, a single energy window accepts a range of scattering angles.
The possible emission points therefore form an extended region around the \ac{tof} position of the trues, and
several combinations of emission and scatter point lead to the same measured event.

As a consequence, the scatter expectation of a \ac{lor} changes when $\attnvec$ changes away from the \ac{lor}, even
if its trues remain unchanged.
\Cref{fig:jeaa_tof_scatter}(b) illustrates this for a ring of higher attenuation around the \ac{lor}.
A local change of $\attnvec$ has two competing effects.
It changes the local probability of Compton scattering, which is proportional to $\attnvec$
(assuming that Compton scattering is the dominant interaction).
It also changes the survival probability of all photons that pass through the region, before and after the scatter.
For a patient-sized object, the second effect dominates in most of the region that contributes to a given
\ac{lor} and \ac{tof} bin, so that lower attenuation there increases, and higher attenuation decreases, the scatter,
just as lower attenuation along a \ac{lor} increases its trues.
Only close to the outer edge of the region accepted by the energy window can the change of the local Compton
probability dominate, and there the effect on the scatter is small.
Integrated over the whole object, the survival effect dominates unless the in-tissue path lengths are of the order of
the mean free path ($\attn L \lesssim 1$), as in small objects.

Two consequences follow for \ac{jeaa}.
First, the scatter in a single photopeak energy window already carries information on $\attnvec$ that the trues do
not provide.
Second, a pair of activity and attenuation estimates that reproduces the trues but has the wrong scale does not
reproduce the scatter, so that the scatter can be used to resolve the non-uniqueness (\cref{sec:jeaadirect}).
Intuitively, much of this information should come from \acp{lor} close to the object boundary, where the scatter
fraction is highest, and from \acp{lor} just outside the object, which record scatter but no trues.
The scatter information is, however, coarse.
A local change of $\attnvec$ near a \ac{lor} changes the scatter in all its \ac{tof} bins in a similar way, since the
photon paths of every \ac{tof} bin extend over the whole object along the \ac{lor}.
The scatter therefore constrains $\attnvec$ in the region around the \ac{lor} as a whole, but hardly localises it
along the \ac{lor}.
It is better suited to fix a few global parameters, such as the scale, than to reconstruct $\attnvec$ in detail.

The role of the scatter differs between non-\ac{tof} and \ac{tof} data.
Without \ac{tof}, the joint estimation suffers from severe cross-talk between $\actvtyvec$ and $\attnvec$, and the
additional information from scattered coincidences in multiple energy windows is most valuable.
With \ac{tof}, the cross-talk is much less severe~\cite{defrise_time--flight_2012, rezaei_simultaneous_2012}, and a
single energy window is likely sufficient.

\textcite{berker_scattered_2014,berker_scattered_2014-1,Berker2019} showed that attenuation maps can be reconstructed
from scattered coincidences, and proposed joint
reconstruction algorithms from scattered and unscattered data.
Their evaluation used 2D phantoms without \ac{tof}, assumed a perfect energy resolution and, like all work discussed
here, considered single scatter only.
\textcite{Brusaferri2020} extended this to a 3D maximum-likelihood framework for non-\ac{tof} \ac{pet} with a photopeak
and lower energy windows, including the finite energy resolution.
They derived the Jacobian of their single-scatter model with respect to $\attnvec$ (and $\actvtyvec$), so that the
scattered coincidences of the lower energy windows enter the gradient of the joint log-likelihood directly, while the
photopeak scatter is re-estimated one step late.
Using multiple energy windows in this way substantially reduced the cross-talk of non-\ac{tof} \ac{mlaa}.

\subsubsection[Fast MC scatter estimation for JEAA]{Fast \acs{mc} scatter estimation for \acs{jeaa}\label{sec:jeaaben}}

Fast and accurate \ac{mc} simulation can reduce the scatter-modelling errors to which \ac{jeaa} is particularly
sensitive (\cref{sec:jeaa_status}).
In the iterative scheme above, it can replace the analytical single-scatter simulation, with several advantages for
\ac{jeaa}.
\begin{itemize}
  \item It includes multiple scatter, which single-scatter simulation neglects or approximates.
  \item It provides an absolutely quantitative scatter estimate, without the rescaling to the scatter tails that
        single-scatter simulation commonly relies on, and thus retains the information on the scale, which a joint
        estimation can exploit (\cref{sec:jeaanonuniq,sec:jeaadirect}).
  \item It can simulate the energy deposited in the detector, including incomplete deposition and inter-crystal
        scatter, which, blurred with the measured energy resolution, determines the scatter accepted by a given
        energy window, in particular in lower-energy windows (\cref{sec:jeaanonuniq,sec:jeaaacc}).
  \item Low-noise \ac{mc} simulations can serve as a reference for faster analytical scatter models, including
        differentiable ones whose gradient is used in a joint estimation (\cref{sec:jeaadirect}).
\end{itemize}
An absolute scatter level requires that \ac{mc} simulations, as well as analytical scatter simulators, are
cross-calibrated against a few calibration measurements, so that their absolute scatter level matches that of the
scanner.\footnote{Neither \ac{mc} nor analytical simulators reproduce the detection chain exactly, e.g.\ the light
collection, the electronic thresholds and the coincidence processing, which all affect the absolute detection
efficiency.
Since these effects depend on the scanner and its settings rather than on the patient, a few phantom measurements
suffice to calibrate them.}
The detection efficiency of scattered coincidences differs from that of the trues, so that their normalisation is not
the same as for the trues \cite{noe2026}.
Since the scatter has to be re-estimated whenever $\actvtyvec$ and $\attnvec$ change, the runtime matters.
Current fast \ac{mc} simulators provide accurate low-noise \ac{pet} scatter estimates within minutes
(\cref{sec:scatterpet}), which makes an update in almost every outer iteration of \ac{jeaa} feasible if needed.

\subsection{Current and future challenges\label{sec:jeaachall}}

\subsubsection{Non-uniqueness and its resolution by the scatter\label{sec:jeaanonuniq}}

\begin{figure}[t]
  \centering
  \resizebox{\linewidth}{!}{
\definecolor{mcblue}{RGB}{31,119,180}
\definecolor{mcorange}{RGB}{214,110,20}
\begin{tikzpicture}
\def\datadir{images/data/jeaa_scale}
\pgfplotsset{
  every axis/.style={width=5.9cm, height=5.0cm, xmin=-18, xmax=18, xtick={-10,0,10},
    tick label style={font=\footnotesize}, label style={font=\small}, title style={font=\small},
    axis line style={black!60}, legend style={font=\footnotesize, draw=none, fill=none}, legend cell align=left,
    xlabel={position (cm)}},
  mu axis/.style={ymin=-0.01, ymax=0.16, ytick={0,0.05,0.1,0.15}, yticklabel style={/pgf/number format/fixed},
    extra y ticks={0}, extra y tick style={grid=major, grid style={black!30}, tick label style={opacity=0}}}}
\tikzset{support/.style={black!45, densely dotted, thin}} 
\begin{axis}[name=a, ymin=0, ymax=1.6, ytick={0,0.8,1,1.25}, ylabel={activity (a.u.)},
  title={(a) activity $\lambda$}, legend style={at={(axis cs:0,0.4)}, anchor=center}]
  \addplot[mcorange, thick, const plot] coordinates {(-18,0)(-10,1.25)(10,0)(18,0)};
  \addplot[black, thick, const plot] coordinates {(-18,0)(-10,1)(10,0)(18,0)};
  \addplot[mcblue, thick, const plot] coordinates {(-18,0)(-10,0.8)(10,0)(18,0)};
  \legend{$1.25\,\lambda$, $\lambda$, $0.8\,\lambda$}
\end{axis}
\begin{axis}[name=b, at={(a.east)}, anchor=west, xshift=1.5cm, mu axis, ylabel={$\mu$ (cm$^{-1}$)},
  title={(b) $\mu$, same support as $\lambda$}]
  \draw[support] (axis cs:-10,-0.01) -- (axis cs:-10,0.16) (axis cs:10,-0.01) -- (axis cs:10,0.16);
  \addplot[mcorange, thick] table {\datadir/same_up.dat};
  \addplot[black, thick] table {\datadir/same_ref.dat};
  \addplot[mcblue, thick] table {\datadir/same_down.dat};
\end{axis}
\begin{axis}[name=c, at={(b.east)}, anchor=west, xshift=0.5cm, mu axis, yticklabels={},
  title={(c) $\mu$, wider support than $\lambda$}]
  \draw[support] (axis cs:-10,-0.01) -- (axis cs:-10,0.16) (axis cs:10,-0.01) -- (axis cs:10,0.16);
  \addplot[mcorange, thick] table {\datadir/wide_up.dat};
  \addplot[black, thick] table {\datadir/wide_ref.dat};
  \addplot[mcblue, thick] table {\datadir/wide_down.dat};
  \addplot[mcorange, thick, dashed] table {\datadir/wide2_up.dat};
  \addplot[mcblue, thick, dashed] table {\datadir/wide2_down.dat};
\end{axis}
\end{tikzpicture}}
  \caption{Scale non-uniqueness of the trues, shown for central profiles through a disk (2D) or sphere (3D) with uniform
  activity within a radius of 10\,cm (dotted lines in (c)); the profiles are the same in 2D and 3D, since the line
  integrals depend only on the distance of the line from the centre.
  (a)~Activity scaled by 1.25 and 0.8.
  (b), (c)~Attenuation images that compensate the scaling, i.e.\ change the line integral of $\attnvec$ by the same
  amount, $\pm\ln 1.25$, along every \ac{lor} through the activity, so that all activity--attenuation pairs in a panel
  yield identical expected trues.
  A constant offset of $\attnvec$ cannot achieve this, since its line integrals grow with the chord length.
  (b)~If $\attnvec$ has the same support as $\actvtyvec$, the radially symmetric compensating change is unique and
  diverges at the boundary; for $0.8\,\actvtyvec$ it becomes negative and thus non-physical.
  (c)~If $\attnvec$ can extend beyond the activity (support radius 15\,cm), bounded compensating changes exist;
  two of infinitely many are shown (solid and dashed), which differ mainly outside the activity.}
  \label{fig:jeaa_scale}
\end{figure}
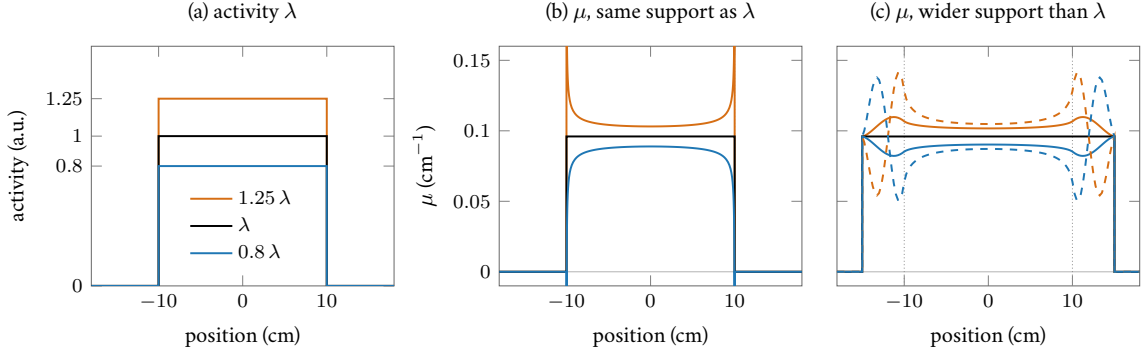

If an activity image $\hat{\actvtyvec}$ and a sinogram of attenuation factors $\hat{\bm a}=(\hat a_\datind)$
(\cref{eq:trues_jeaa}) reproduce the expectation of the trues, so do $\hat{\actvtyvec}/\alpha$ and $\alpha\hat{\bm a}$
for any $\alpha>0$.
We discuss the problem for the attenuation image rather than the attenuation factors, since the scatter depends on
$\attnvec$ itself.
In terms of $\attnvec$, this corresponds to adding the constant $-\ln\alpha$ to the line integrals of all \acp{lor}
that intersect the activity~\cite{defrise_time--flight_2012}.
Since this offset is the same for all these \acp{lor}, whatever their length inside the object, the corresponding
change of $\attnvec$ is not a constant but an image with equal line integrals along them (\cref{fig:jeaa_scale}).
The line integrals of \acp{lor} that do not intersect the activity are not constrained by the trues at all.
Even if the supports of the activity and the attenuation are known and coincide, the required change of $\attnvec$
is singular at the boundary, but for a higher activity bounded attenuation images can approximate it arbitrarily
well, so that knowing the object outline does not fix the scale (\cref{fig:jeaa_scale}(b)).
If attenuating material extends beyond the activity, e.g.\ tissue with negligible uptake, the patient bed or
\ac{mr} coils, its attenuation is in addition left undetermined, and bounded compensating changes of $\attnvec$ exist
(\cref{fig:jeaa_scale}(c)).
Prior information (\cref{sec:jeaascale}) removes these ambiguities if it is available and accurate~\cite{jurew2026},
whereas the scatter is needed most when the supports and the attenuation outside the activity are unknown.

Within this ambiguity, e.g.\ between the pairs in \cref{fig:jeaa_scale}, the scatter changes in two ways, in its
absolute level and, because it depends non-linearly
on $\attnvec$, in its distribution over the \acp{lor} and \ac{tof} bins.
A fixed scatter estimate uses neither, even if it is the true expectation, since it does not change with the scale.
A scatter estimate that is recomputed with the current $\attnvec$ but rescaled to the scatter tails retains the
information in the shape of the distribution, but loses that in the absolute level.
The full information is therefore only available if the absolute scatter level is modelled.
It is, moreover, only used if the dependence of the scatter on $\attnvec$ enters the estimation itself, through the
gradient of the log-likelihood (\cref{sec:jeaadirect}) or a search for the scale.
An alternating scheme that holds the scatter fixed during the updates of $\actvtyvec$ and $\attnvec$
(\cref{sec:jeaa_status}) does not in general select the scale, even if the scatter is recomputed between the updates.

Additional lower-energy windows can add scattered coincidences from larger scattering angles and thus further
information
on the distribution of the scatter~\cite{Brusaferri2020}.
However, they also contain more higher-order scatter, so that they require fast and accurate models of higher-order
scatter, for which fast \ac{mc} simulation is probably the most practical option.
Whether the practical gain is substantial for data with good \ac{tof} resolution, where the trues already constrain
the problem much more, is an open question.

\subsubsection{Accuracy of the absolute scatter model\label{sec:jeaaacc}}

Errors of the scatter model affect \ac{jeaa} in two ways.
As in any \ac{jeaa} that treats the scatter as a nuisance term, they are absorbed by the model of the trues and bias
$\attnvec$ and $\actvtyvec$ locally~\cite{Nuyts2018}.
If the scatter is used to resolve the non-uniqueness, they also bias the global scale of the activity
(\cref{sec:jeaanonuniq}).
When the activity is scaled up, the scatter increases with the activity but decreases with the accompanying additional
attenuation (orange curves in \cref{fig:jeaa_scale}), which reduces the survival of the photons.
Since both effects partly cancel, the scatter changes less than proportionally with the scale, and a relative error of
the scatter expectation translates into a larger relative error of the activity.
An accurate activity therefore requires a scatter model with a high absolute accuracy.
For this purpose, higher-order scatter should not be neglected even in a narrow energy window around the photopeak,
since it is not negligible compared to the single scatter in patient-sized objects.
Double scatter probably accounts for most of it, so that modelling single and double scatter may be sufficient.
Accurate energy calibration is equally crucial, because the scatter accepted by the window depends strongly on the
position of the energy threshold relative to the photopeak.
We expect this to improve with better detector hardware and per-scan calibration procedures.
Besides the cross-calibration of the scatter model (\cref{sec:jeaaben}), scatter from activity outside the axial
\ac{fov} should be included.
If that activity has never been inside the \ac{fov}, e.g.\ with a scanner whose axial \ac{fov} covers only part of
the patient, it is unknown, and its scatter can only be approximated, e.g.\ by extrapolation.
Finally, the randoms have to be estimated accurately in the \acp{lor} outside the object, where errors of the randoms
cannot be distinguished from scatter.

\subsection{Development needed to meet the challenges\label{sec:jeaanew}}

Prior information fixes the scale only if it is available, accurate and correctly registered (\cref{sec:jeaascale}).
The alternative is to let the scatter resolve the non-uniqueness in a joint estimation from true and scattered
coincidences.

\subsubsection{Joint estimation from true and scattered coincidences\label{sec:jeaadirect}}

The most general formulation embeds the full forward model of \cref{eq:fwd_model}, including multiple scatter, in a
joint maximum-likelihood problem,
\begin{equation}
  (\actvtyvec^{\mathrm{ML}}, \attnvec^{\mathrm{ML}}) = \argmax_{\actvtyvec, \attnvec \geq 0}\,
  \sum_{\datind, t} \Bigl[\prompts_{\datind t} \log \bar{\prompts}_{\datind t}(\actvtyvec,\attnvec)
  - \bar{\prompts}_{\datind t}(\actvtyvec,\attnvec)\Bigr]\,,
  \label{eq:jeaa_joint_ml}
\end{equation}
summed over several energy windows if these are used.
It resolves the non-uniqueness without prior information, since the scatter then determines the scale together with
the attenuation in the regions that the trues do not constrain; a search for the scale (\cref{sec:jeaanonuniq}) is a
restricted form of the same problem.
Its key ingredient is the gradient of the log-likelihood with respect to $\attnvec$, including the dependence of the
scatter expectation on $\attnvec$.
In principle, this gradient can be derived from \ac{mc} simulation, e.g.\ via \ac{mc} gradient
estimators~\cite{li2018, Hochuli2016}, path replay as in inverse rendering~\cite{yao2018,NimierDavid2020}, or the
coupling of \ac{mc} simulators with automatic differentiation~\cite{Aehle2024,Aehle2025}, but this is non-trivial
and noisy.
The general form of this gradient is given in \cref{eq:LogLikelihoodGradMCattn} in \cref{sec:InvProb}.
A more practical route uses the different roles of single and higher-order scatter.
The single scatter carries most of the information on $\attnvec$, and its gradient can be derived
analytically~\cite{Brusaferri2020}, so that a differentiable single-scatter simulation is probably the most
efficient approach.
Evaluating it and its gradient for all \acp{lor} and \ac{tof} bins in every iteration is nevertheless demanding at
clinical scale, and will likely require coarse grids of scatter points and \acp{lor} and efficient \ac{gpu}
implementations.
Higher-order scatter probably does not have to be differentiated, since it adds comparatively little information on
$\attnvec$.
It can enter the expectation as an additive term that is recomputed from the current estimates every few iterations,
e.g.\ by an analytical double-scatter model or by fast \ac{mc} simulation.
An \ac{mc} estimate of this term has to have low noise, since noise biases the gradient through the ratio
$\promptsvec/\bar{\promptsvec}$ (\cref{sec:StochasticReconstruction}), which current simulators achieve with suitable
post-processing (\cref{sec:scatterpet}).
It remains to be investigated whether neglecting the gradient of the higher-order scatter biases the scale,
to which the scatter level is critical (\cref{sec:jeaaacc}),
and how the resulting objective can be optimised efficiently along its weakly determined directions.

Fast \ac{mc} simulation is essential for this route even if it does not provide the gradient.
Fast analytical scatter models take short cuts, e.g.\ a finite number of scatter points, and fast \ac{mc} simulations
with very low noise are needed to verify that both their expectation and their gradient are correct.
\Ac{mc} simulation also provides the contributions that are hard to model analytically, such as scatter of third and
higher order and detector effects.

The choice of the energy window offers a further degree of freedom.
A narrow window around the photopeak reduces the higher-order scatter that has to be modelled.
With list-mode data that include the photon energies, the joint estimation can use such a narrow window, and the final
activity image can be reconstructed from a wider window with the estimated $\attnvec$, where errors of the scatter
estimate no longer affect the scale.  This approach will need to balance the accuracy of the scatter estimate, the extra
information at lower energy and the decreased sensitivity of a narrow window.

\subsection{Concluding remarks\label{sec:jeaaconcl}}

Scatter has long been a weak point of \ac{jeaa}, since \ac{jeaa} is very sensitive to inaccuracies of the scatter
model.
Fast \ac{mc} simulation addresses this in several ways.
It provides absolutely quantitative scatter estimates that include multiple scatter, it serves as a low-noise reference
to validate faster, possibly differentiable, analytical scatter models
and their gradients, and, in the longer term, differentiable \ac{mc} simulation may provide the gradient of the
scatter itself.
Beyond being a source of error, the scatter, provided that its absolute level is modelled accurately, can probably
resolve the non-uniqueness of the \ac{jeaa} problem, which so far has required prior information on the attenuation.
Research is needed on how to do this in a way that is fast enough and robust to small residual modelling errors such
that it can be integrated into clinical use.

\newsectionpage
\section{Solving inverse problems with Monte Carlo models\label{sec:InvProb}}
\sectionauthor{\v{Z}eljko Kereta, Simon Arridge, Kris Thielemans}

\subsection{Motivation for Monte Carlo models in reconstruction}

In this section we discuss the use of \ac{mc} physics engines in \ac{pet} and \ac{spect} reconstruction, as defined in
\cref{sec:WhereDoesMCfit}. 
\Ac{mc}-based reconstruction does not represent a single method. 
Instead, several regimes should be distinguished.

\begin{itemize}
    \item \textbf{\Ac{mcs} as a Numerical Tool:} Use \ac{mc} to compute and replace a deterministic  \ac{srm}, followed
    by classical reconstruction tools.
    This does not need a separate adjoint \ac{mc}.   
    \item \textbf{Fully \ac{mc} reconstruction:} Use \ac{mc} forward and adjoint operators inside a stochastic
    reconstruction loop.
    The \Ac{rte} provides a framework for defining the corresponding transport operators.
    \item \textbf{Hybrid approaches:} Use \ac{mc} only for selected parts of the pipeline (e.g. consider the
    factorisation in \cref{eq:factorised_srm}), and deterministic components elsewhere.
\end{itemize}
A reconstruction framework must state exactly where \ac{mc} enters the model, what is simulated once, what is
patient-specific, and whether the forward and adjoint operations are physically and statistically consistent.
\subsection{Model differentiation}

Optimisation-based techniques, which are the most prevalent in current image reconstruction approaches, require a
suitable notion of \emph{differentiation}, either to obtain the gradient of the log-likelihood $\loglik$ for
use in first-order methods, or to obtain the model derivative (i.e. the Jacobian) for use in second-order methods. 
For reconstruction of a (discretised) activity $\actvtyvec$, assuming a fixed attenuation $\attnvec$, the linear model
$\srm(\attnvec)$ can be pre-computed with \ac{mc}, in which case the backprojection is simply the transpose
$\srm\transp$.
Alternatively, it can be evaluated on-the-fly, in which case backprojection requires the adjoint \ac{mc} model, also
computed on-the-fly. 
In practice, however, a  deterministic analytic backprojection operator is often used in place of a true adjoint of the
\ac{mc} forward model.
Note that this adjoint mismatch is different from the one induced by a limited \ac{mc} budget.
Namely, while the latter arises from stochastic estimations, and can in principle be reduced by increasing the \ac{mc}
budget, the former can be systematic, persisting even in the infinite budget limit, and may cause the reconstruction
updates to deviate from the true gradient.
If we also reconstruct $\attnvec$, we need to evaluate $\dbyd{\loglik}{\attnvec}$, which by the chain rule might require
 $\dbyd{\srm}{\attnvec}$. 
The direct evaluation of $\dbyd{\loglik}{\attnvec}$ is described in detail in \cref{sec:AdjointMCrole}. 
For evaluation of  $\dbyd{\srm}{\attnvec}$ we mention the ``photon replay'' concept~\cite{yao2018}. 
The idea is to keep the seed or state of the random number generator needed to reproduce each simulated photon history.
One can then ``replay'' only those photons that reached a given detector to determine the sensitivity map for that
source-detector pair. 
This allows constructing $\dbyd{\srm}{\attnvec}$ \emph{without} requiring the adjoint \ac{mc} simulation. 
Photon replay has proved highly influential in diffuse optical tomography, but it remains an open question whether it
can be
applied in \ac{spect} and \ac{pet}. 

Automatic differentiation suggests an alternative for reconstruction and for addressing the adjoint mismatch.
In particular, instead of treating \ac{mc} simulation as a black box, one could make parts of the transport and
detection model differentiable with respect to object and system dependent parameters, such as activity, attenuation,
detector parameters, calibration or other acquisition parameters.

One means of achieving this is through a deterministic surrogate of \ac{mc} transport.
The bolder and more challenging approach is differentiable \ac{mc} itself: estimating not only the expected projection
data, but also derivatives of those expectations with respect to physical parameters. 
In principle, this would allow computing gradients of a reconstruction objective using the same underlying transport
model that generates the predictions. 
This could be an attractive direction for joint estimation, calibration, motion correction, uncertainty propagation, and
task-driven scanner design.

However, the main challenges are substantial. 
First, photon transport contains discrete events such as interaction type, scattering order, detection, absorption,
rejection, and bin assignment. 
Naively differentiating through sampled histories might provide neither a useful nor an unbiased gradient estimator. 
Instead, reparameterisation methods, pathwise derivatives (already studied for optical tomography), and other
probabilistic tools would be needed. 
These estimators can have high variance, especially for rare detection events or deep scatter paths. 
Discontinuous detector thresholds and energy-window binning further complicate the problem.

Differentiable \ac{mc} is promising, with an underdeveloped but growing interest and body of work in statistical
modelling, machine learning, and partial differential equation theory.
However, as discussed in \cref{sec:mcFinalRemarks}, it is not a straightforward application of automatic differentiation
to an existing transport code, and its scalability and utility for practical, large scale reconstruction problems are
unresolved. 
Key objectives would be to design gradient estimators that are physically correct, computationally feasible, compatible with reliable and stable reconstruction algorithms, and low in variance.

\subsection{Requirements for an adjoint \acl{mc} model\label{sec:AdjointMCrole}}

\begin{figure}[t]
    \centering
    \resizebox{0.9\linewidth}{!}{
\definecolor{mcblue}{RGB}{31,119,180}
\definecolor{mcorange}{RGB}{214,110,20}
\begin{tikzpicture}[
  font=\small,
  fwd/.style={-{Latex[length=2mm]}, thick, mcblue},
  adj/.style={-{Latex[length=2mm]}, thick, mcorange, dashed},
  vtx/.style={circle, fill=black, inner sep=1.1pt},
  src/.style={circle, fill=mcblue!25, draw=mcblue, inner sep=1.6pt},
  det/.style={rectangle, fill=black!15, draw=black!60, minimum width=2mm, minimum height=4mm, inner sep=0pt},
  lab/.style={font=\footnotesize},
]
\begin{scope}[shift={(0,0)}]
  \filldraw[black!6, draw=black!35] (1.1,0.8) ellipse[x radius=1.75cm, y radius=1.3cm];
  \node[font=\scriptsize\itshape, black!55] at (1.0,1.72) {patient, $\attn(\pos)$};
  \node[src] (a0) at (0,0) {};
  \node[det, rotate=-35] (a1) at (3.4,1.95) {};
  \draw[adj] (a1) -- (a0);
  \node[lab, above left] at (1.7,0.98) {$\propop^{\mathrm{LS}\ast}(\attn)$};
  \node[lab, below] at (a0.south) {$\pos_0$};
  \node[lab, right=1pt] at (a1.east) {$\pos_1$, $\adjsrc_{\datind}$};
  \node[font=\footnotesize, align=center] at (1.7,-1.0) {(a) gradient w.r.t.\ activity};
\end{scope}
\begin{scope}[shift={(6.0,0)}]
  \filldraw[black!6, draw=black!35] (1.1,0.8) ellipse[x radius=1.75cm, y radius=1.3cm];
  \node[src] (b0) at (0,0) {};
  \node[vtx] (bk) at (2.1,0.45) {};
  \node[det, rotate=-35] (b1) at (3.4,1.95) {};
  \draw[fwd] (b0) -- (bk);
  \draw[adj] (b1) -- (bk);
  \node[lab, above] at (1.05,0.27) {$\propLS(\attn)$};
  \node[lab, right] at (2.85,1.0) {$\propop^{\mathrm{LS}\ast}(\attn)$};
  \node[lab, below] at (b0.south) {$\pos_0$};
  \node[lab, below right] at (bk) {$\pos_k$};
  \node[lab, below=7pt] at (bk) {$\scatop_{k}$};
  \node[lab, right=1pt] at (b1.east) {$\pos_1$, $\adjsrc_{\datind}$};
  \node[font=\footnotesize, align=center] at (1.7,-1.0) {(b) gradient w.r.t.\ attenuation};
\end{scope}
\draw[fwd] (10.9,1.7) -- ++(0.8,0) node[right, black, lab] {forward};
\draw[adj] (10.9,1.2) -- ++(0.8,0) node[right, black, lab] {adjoint};
\end{tikzpicture}}
    \caption{Adjoint propagators in the gradient of the negative log-likelihood.
    (a) For the activity $\actvty$, the adjoint source $\adjsrcvec = \promptsvec/\bar{\promptsvec}-\onevec$ is
    back-propagated from the detector by $\propop^{\mathrm{LS}\ast}(\attn)$ (\cref{eq:LogLikelihoodGradMCactivity}).
    (b) For the attenuation $\attn$ at $\pos_k$, the back-propagated adjoint source is combined with the forward
    radiance $\propLS(\attn)\actvty$ at that point (\cref{eq:LogLikelihoodGradMCattn}).}
    \label{fig:AdjointBornSeries1}
\end{figure}

Let us revisit the variational approach for reconstruction to include the concept of an \emph{adjoint operator}.
First, we recall the compositional structure from \cref{eq:FwdOpLippmannSchwinger}
(omitting the term $\bar{\randomsvec}$)
\begin{equation}
    \cdsrm(\actvty,\attn) = \measop \circ \propLS(\attn) \circ \srcop(\actvty)\,.
\end{equation}
We make some simplifying assumptions for clarity of exposition, some of which  can be relaxed.
\paragraph{Adjoint of the measurement operator}    
    Assume that the measurement operator $\measop$ is linear, i.e. the detected counts are proportional to the arriving number of
    photons and do not exhibit saturation or dead-time effects.
    Then the expected counts at $\datind$ can be represented as a (possibly weighted) sum over the arriving radiance
    $\phi$ as 
    \[
\bar\prompts_{\datind} = \measop_{\datind}\phi := \left< \meas_\datind, \phi \right>_{Y},   
    \]
    where the measurement kernel is expressed as
    \[
    \meas_{\datind}(\pos,\angdir) \qquad \pos \in \partial \Omega_\datind, \, \angdir \cdot \hat {\bf n}_\datind > 0,
    \]
    and $\partial \Omega_\datind $ is the spatial extent of detector/detector pairs $\datind$ over the boundary
    $\partial
    \Omega$ and $ \hat {\bf n}_\datind $ the corresponding outward directed normal vector  (or vector pair) at the detectors. 
    Then $\measop' = \measop$ and $\measop^{\ast}$ is an operator taking a single scalar as input and generating a
    spatial and angular distribution 
    \[
    \measop_{\datind}^{\ast} h_\datind \rightarrow  h_\datind \meas_\datind (\pos,\angdir^\ast)\qquad \pos \in \partial
    \Omega_\datind, \, \angdir^{\ast} \cdot \hat {\bf n}_\datind < 0,
    \]
    with $\angdir^\ast$ the reverse direction to $\angdir$.

\paragraph{Adjoint of the propagation step}
The premise of using \ac{mcs} as a ``physics engine'' is to assume that the adjoint propagation step plays the role of a
stochastic proxy for ${\propop^{\mathrm{LS}}}^\ast$ (given in \cref{eq:FwdOpLippmannSchwinger}). 
Even though the adjoint equation~\eqref{eq:adjointRTEpropagator} is well-defined, constructing a comparatively
consistent \ac{mc} adjoint for \ac{mc}-based forward models remains an active research question, see for example recent
work in relation to rendering of scattering volumes in computer graphics~\cite{NimierDavid2020}.
If we are modelling \ac{tof} we may choose to record ``reverse-time'' rather than forward time. 
For scattering, energy change will be upward not downward, with the upward change $\Delta\Energy$ being selected from
the interval $[0,\Energy^{\rm max}-\Energy^{\rm current}]$ which in turn determines the scattering angle change from the
Klein--Nishina formula in \cref{eq:KleinNishina}.
Note that here $\Energy^{\rm max}$ refers to photon energy rather than positron energy as in
\cref{sec:PositronRangeModelling}.

\paragraph{Adjoint of the source model}  
Assume that the source model is linear, i.e. $\srcop(\actvty)= \srcop\actvty$.
The form of the adjoint source model
is different for the \ac{spect} and \ac{pet} cases. 
For \ac{spect} we should count an adjoint photon as contributing to the activity reconstruction if it arrives at voxel
$\pixind$ with the energy of the emitting source.
For \ac{pet} we have a more complex model.
We should launch \emph{two} adjoint photons, one from each detector in the pair corresponding to bin $\datind$.
These are launched at independent times that are within the width of the detector coincidence window.
At each voxel we determine if both photons arrive within a suitable time interval (for \ac{tof}) and with the  energy of
the emitting source (as in \ac{spect}).
If so we add it to the activity count at voxel $\pixind$.\footnote{There may be a role for an adjoint equivalent to
Forced Detection here, see \cref{sec:varred}.}

\paragraph{Adjoint calculation of the gradient}
Under standard methods for differentiation of operators, we can state the gradient of the Poisson log-likelihood (ignoring the background term $\backgrvec$) as
\begin{align}
\nabla_{\actvty}\loglik(\promptsvec,\cdsrm(\attn), \actvty)
&= \sum_{\datind} \cdsrm^{\ast}_{\datind} \left(\frac{\prompts_{\datind}}{\cdsrm_{\datind}(\attn)\actvty}-{1}\right)  \,,
\label{eq:NPLLgradActvty} \\
\nabla_{\attn}\loglik(\promptsvec,\cdsrm(\attn) ,\actvty)
&=  \sum_{\datind}\left[\dbyd{\cdsrm_{\datind}}{\attn}\right]^{\ast}  \underbrace{
\left(\frac{\prompts_{\datind}}{\cdsrm_{\datind}(\attn)\actvty}- {1} \right)
}_{\adjsrc_{\datind}}
   \label{eq:NPLLgradAttn}\,,
\end{align}
where $\dbyd{\cdsrm}{\attn}$ 
is the Fr\'{e}chet derivative of $\cdsrm$ and   $\adjsrcvec =\left( \frac{\promptsvec}{\bar{\promptsvec}} -
\onevec\right) $ is an \emph{adjoint source}.
The expected counts at
bin $\datind$ will then be represented as 
\begin{equation}
    \bar\prompts_{\datind} = \left< \meas_\datind, \propLS(\attn) \srcop  \actvty \right>_{Y},   
\end{equation}
where $\meas_{\datind}$ is the kernel of the detector (detector pairs in \ac{pet}), $\propLS$ is the Lippmann--Schwinger
operator (see \cref{eq:FwdOpLippmannSchwinger}), and $Y$ is the space of data. 
It follows that we can define the derivative of the forward operator for a single detector bin with respect to $\actvty_{\pixind}$, with basis
element $\imbss_j$, as 
\begin{equation}
    \dbyd{\bar{\prompts}_{\datind}}{\actvty_{\pixind}} = \left< \meas_\datind, \underbrace{\propLS(\attn) \srcop \imbss_{\pixind}}_{\phi_\pixind}
    \right>_{Y} = \left< {\srcop^\ast 
    \underbrace{ 
    {\propop^{\mathrm{LS}}}^{\ast}(\attn)  \meas_\datind
    }_{\phi^\ast_\datind},\imbss_{\pixind}} \right>_{X}. 
\end{equation}
The gradient of the log-likelihood is then given as 
\begin{equation}
    \dbyd{\loglik}{\actvty_{\pixind}} = \sum_{\datind}\left<\meas_{\datind} \adjsrc_\datind, \propLS(\attn) \srcop
    \imbss_{\pixind} \right>_{Y} = \sum_{\datind}\left< {\srcop^\ast \propop^{\mathrm{LS}}}^{\ast}(\attn)  \meas_{\datind}
    \adjsrc_\datind, \imbss_{\pixind} \right>_{X}. 
        \label{eq:LogLikelihoodGradMCactivity}
\end{equation}
Similarly for attenuation,
\begin{equation}
    \dbyd{\bar{\prompts}_{\datind}}{\attn_{\pixind}} = -\left< \meas_\datind, \propLS(\attn) \mathcal{V}_{\pixind}
    \propLS(\attn) \srcop \actvty \right>_{Y} 
    = -\left< {\propop^{\mathrm{LS}}}^{\ast}(\attn)  \meas_\datind,  \mathcal{V}_{\pixind} \propLS(\attn)
    \srcop\actvty\right>_{X},
\end{equation}
and the gradient of the log-likelihood
\begin{equation}
    \dbyd{\loglik}{\attn_{\pixind}} = -\sum_{\datind}\left< \meas_{\datind} \adjsrc_\datind, \propLS(\attn)
    \mathcal{V}_{\pixind} \propLS(\attn) \srcop\actvty \right>_{Y} 
    = -\sum_{\datind}\left< {\propop^{\mathrm{LS}}}^\ast(\attn)  \meas_{\datind} \adjsrc_\datind, \mathcal{V}_{\pixind}
    \propLS(\attn)\srcop \actvty \right>_{X},
    \label{eq:LogLikelihoodGradMCattn}
\end{equation}
where $\mathcal{V}_{\pixind}=\frac{\partial(\transop_{\attn} - \attn \scatop)}{\partial \attn_{\pixind}}$ is the
``potential'' operator.

\paragraph{Sampling from the data residual}
Since the adjoint vector $\adjsrcvec$ may have negative entries, it cannot directly define a sampling distribution.
Thus, we need
to define an everywhere non-negative distribution from which to draw adjoint sources, such that it is supported on all
detectors. 
Although the adjoint source may contain negative entries, the linearity of the forward model with respect to activity
allows the adjoint operation to be decomposed into two terms with non-negative sources.
In current implementations in optical tomography this is achieved by using a cumulative probability distribution for the
positive and negative parts of $\adjsrcvec$ using a lookup-table to infer the source position from a uniform sample of
the cumulative distribution~\cite{macdonald2020}.
For the Poisson log-likelihood gradient \cref{eq:gradientloglikelihood}, it is natural to use the backprojection of the
ratio of data-to-prediction $\cdsrm^*(\promptsvec/\bar\promptsvec)$ and the sensitivity image $\cdsrm^* \onevec$.
The latter can be precomputed to high accuracy for a fixed system model. 
What remains is to construct an efficient adjoint \ac{mc} estimator and to analyse the resulting errors.
Each adjoint photon is initiated in a direction sampled from the acceptance distribution of the detector/detector-pairs
(ignoring scatter). 
In \ac{spect} the acceptance distribution is determined by the collimator details and is the adjoint of the measurement
operator $\measop^{\ast}$. 
In \ac{pet} each initiated adjoint event must consist of two photons, one from each of the detectors in the  pair
corresponding to the randomly selected histogram bin. 
This can still be cast as an adjoint measurement operator but the distribution of the initial paths is more nuanced if
we want to emphasise the unscattered photons.
This can be considered an open problem.

\subsection{Reconstruction with \acs{mcs} engines\label{sec:StochasticReconstruction}}

The central mathematical issue when using \ac{mc} models inside a reconstruction loop 
is that the resulting \ac{mc} forward model is not a static object but rather a constantly changing random estimate of
the true forward operator that depends on the photon budget (\ac{mc} accuracy). 
Consequently, the reconstructed image depends both on the measured data and on the \ac{mc}  histories that are generated
constantly and are used by the reconstruction engine to update the activity estimate. 
This turns reconstruction into a stochastic approximation problem 
in which standard deterministic algorithms no longer apply directly. 
However, a fully-\ac{mc} approach allows numerical and computational benefits, such as allowing for an efficient
allocation of the simulation budget throughout the reconstruction process, along with other benefits (in terms of
physical accuracy).
In particular, a possible simulation-efficient strategy, with respect to reconstruction quality, would be to start with
a low
\ac{mc} simulation budget, when the reconstruction is still of a lower quality, 
and to increase the \ac{mc} budget as needed through the iterations as the reconstruction quality improves. 

A generic pseudo-code for a gradient-based fully \ac{mc} reconstruction is provided below.
Let $\itrind>0$ be the iteration index and $\actvtyvec^{\itrind}$ the corresponding activity estimate. 
\begin{itemize}
    \item Allocate a budget of $N_{\itrind}$ samples for \ac{mc} simulations in iteration $\itrind$;
    \item Draw $N_{\itrind}$ samples $\mcsmpl_{\itrind}$;
    \item Compute the \ac{mc} update direction ${\bm g}_{\itrind}=\text{MC-Gradient}(\actvtyvec^{\itrind};
    \,\mcsmpl_{\itrind})$;
    \item Update the activity $\actvtyvec^{\itrind+1}$ using ${\bm g}_{\itrind}$, $\actvtyvec^{\itrind}$, and additional
    terms as required by the optimiser.  
\end{itemize}
We will now briefly discuss each component.
The exact form of the sample set $\mcsmpl_{\itrind}$ depends on the used \ac{mc} method, for example the entire \ac{mc}
budget $N_{\itrind}$
can be distributed to only the forward simulation, or it can be split into forward and adjoint simulations. 
Let $\bar{\promptsvec}_{\mcsmpl}$ be the \ac{mc} generated array of detector-bin counts, which here corresponds to an
\ac{mc} estimate of the entire physics, $\bar{\promptsvec}_{\mcsmpl}(\actvty)=\mcop_{\mcsmpl}(\actvty;\attn)$, where in
 a standard construction, the \ac{mc} estimator is unbiased, i.e.
$\cdsrm(\attn)\actvty=\mathbb{E}_\mcsmpl[\mcop_{\mcsmpl}(\actvty;\attn)]$.
In the following, we assume that $\bar{\promptsvec}$ includes the background term $\backgrvec$, which is non-zero everywhere to simplify the discussion.
In typical terms, the data-fidelity component of the objective is the log-likelihood, whose derivative can be written as
$D\bar{\promptsvec}(\actvty)^\ast\left(\frac{\promptsvec}{\bar{\promptsvec}(\actvty)}-\onevec\right)$.
The exact form of this derivative, and how it is evaluated, depends on the meaning of each term inside.
In particular, in the case of classical deterministic algorithms with discretised operators, the derivative of the log-likelihood is computed as
$\srm\transp\left(\frac{\promptsvec}{\srm\actvtyvec+\backgrvec} -\onevec\right)$, see \cref{eq:gradientloglikelihood}.
However, when using \ac{mc} estimators, one or both terms do not use the (static) \ac{srm} $\srm$.
The critical question is thus the computation of the derivative, required to obtain the \ac{mc} update direction ${\bm
g}_{\itrind}$, for which there are several options.
\begin{itemize}
    \item Option 1: Keep $D\bar{\promptsvec}(\actvty)^\ast$ deterministic, and use \ac{mc} to estimate
    $\bar{\promptsvec}(\actvty)$.
    \item Option 2: Compute both  $D\bar{\promptsvec}(\actvty)^\ast$ and $\bar{\promptsvec}(\actvty)$ using (separate)
    \ac{mc} estimators.
    \item Option 3: Compute an \ac{mc} estimator of the gradient directly.
\end{itemize}
To discuss their impact, consider gradient-based update equations for activity, which after omitting regularisation and
other terms for the sake of simplicity,  are typically of the form
\[ 
\actvtyvec^{\itrind+1} = \projection_{\ge 0} \left( \actvtyvec^{\itrind} + \gamma_{\itrind} {\bm g}_{\itrind}\right),
\]
where $\projection_{\ge 0}$ is the non-negativity projection and $\gamma_{\itrind}$ is a (possibly vector-valued)
non-negative step size.
The update direction ${\bm g}_{\itrind}$ is an estimator of the true gradient ${\bm g}(\actvtyvec^{\itrind})$, and can
be decomposed as 
\[{\bm g}_{\itrind} = {\bm g}(\actvtyvec^{\itrind}) + {\bm \beta}_{\itrind}+{\bm \epsilon}_{\itrind},\]
where ${\bm \beta}_{\itrind} = \mathbb{E}[{\bm g}_{\itrind}\vert \actvtyvec^{\itrind}] - {\bm g}(\actvtyvec^{\itrind})$
is the bias of the gradient estimator and ${\bm \epsilon}_{\itrind} = {\bm g}_{\itrind} - \mathbb{E}[{\bm
g}_{\itrind}\vert \actvtyvec^{\itrind}]$ denotes the variation of the gradient estimator around the mean.
Under standard regularity and finite-variance assumptions on \ac{mc} estimators we have $\|{\bm \epsilon}_{\itrind}\| = \mathcal{O}_p(N_{\itrind}^{-1/2})$ and by
definition $\mathbb{E}[{\bm \epsilon}_{\itrind}\vert\actvtyvec^{\itrind}]=0$.
However, the gradient estimator ${\bm g}_{\itrind}$ is typically biased, i.e. ${\bm
\beta}_{\itrind}\neq0$.
For instance, consider the first option for computing the gradient, and moreover, assume that we have access to the
true adjoint $D\bar{\promptsvec}(\actvty)^\ast$ and that we have an unbiased \ac{mc} estimator of the forward model,
$\mathbb{E}_\mcsmpl[\bar{\promptsvec}_\mcsmpl(\actvty)]=\bar{\promptsvec}(\actvty)$.
The bias of the gradient thus satisfies
\begin{equation}
 {\bm \beta}_{\itrind} = D\bar{\promptsvec}(\actvtyvec^{\itrind})^\ast
 \left(\promptsvec\odot\left(\mathbb{E}_{\mcsmpl_{\itrind}}\left[\frac{1}{\bar{\promptsvec}_{\mcsmpl_{\itrind}}(\actvtyvec^{\itrind})}
   \Big\vert\actvtyvec^{\itrind}\right]-\frac{1}{\bar{\promptsvec}(\actvtyvec^{\itrind})}
   \right)\right).
     \label{eq:MCgradientBias}
\end{equation}
However, the term inside the parentheses is zero only for degenerate distributions, due to Jensen's inequality applied component-wise on positive values.
In this case, standard \ac{mc} bounds would provide only $\|{\bm \beta}_{\itrind}\|=\mathcal{O}(N_{\itrind}^{-1})$ in the
asymptotic regime with respect to $N_{\itrind}$.
Thus, classical results on the convergence of supermartingales (such as Robbins--Siegmund or its variants
\cite{robbins1971convergence})
would suggest 
\[ \sum_{\itrind} \gamma_{\itrind} \|{\bm \beta}_{\itrind}\|<\infty, \,\, \sum_{\itrind}
\gamma_{\itrind}^2\,\mathbb{E}[\|{\bm \epsilon}_{\itrind}\|^2\vert \actvtyvec^{\itrind}] <\infty, \text{ and }
\sum_{\itrind} \gamma_{\itrind} =\infty,\] would suffice to ensure convergence to a maximiser.
Assuming $\|{\bm \beta}_{\itrind}\|=\mathcal{O}(N_{\itrind}^{-1})$ and $\|{\bm \epsilon}_{\itrind}\| =
\mathcal{O}(N_{\itrind}^{-1/2})$, this leads to 
\[ \sum_{\itrind} \frac{\gamma_{\itrind}}{N_{\itrind}}<\infty, \,\, \sum_{\itrind}
\frac{\gamma_{\itrind}^2}{N_{\itrind}} <\infty, \text{ and } \sum_{\itrind} \gamma_{\itrind} =\infty.\]
Therefore, for the estimator considered here, the \ac{mc} budget would generally need to be increased over iterations for theoretical convergence, with the
exact regime balanced out with step sizes, unless suitable bias reduction mechanisms are used.
It is important to note that biased gradient estimators can still allow constant \ac{mc} budgets; however, this requires
additional theoretical assumptions and gradients of a particular form, cf. \cite[Theorem~4]{atchade2017perturbed}.
On the other hand,  if we have access only to an approximate rather than an exact adjoint, then a persistent structural bias is
imposed, which cannot be resolved by increasing the \ac{mc} budget with the iterations.

Similar issues to the ones discussed above need to be addressed for other gradient constructions using \ac{mc} estimators.
There are many open research directions and questions regarding algorithmic,  theoretical, and practical choices,  including constructing an adequate estimator of the gradient, establishing its properties in terms of bias, reliance on \ac{mc}
and its adjoint, designing photon budget strategies and step size rules.
Moreover, the discussion above relates to using the Kullback--Leibler divergence in the data fidelity, which results in
the issues with bias of the gradient estimator. 
Using least squares (or weighted least squares) as the data fidelity term would help resolve some of the issues and the
mathematical analysis would be more straightforward.

\subsection{Concluding Remarks}

The most promising future methods will probably be hybrid rather than fully \ac{mc}-based. 
Deterministic models should be retained where the physics is simple, stable, and well calibrated. 
At this first stage, \ac{mc} should be used where object-dependent transport, scatter, detector physics, or
radionuclide-specific effects materially affect clinical practice or some quantitative endpoint.

Important open problems include convergence theory for stochastic \acl*{em} and \ac{osem} variants, principled treatment
of \ac{mc} noise, adjoint \ac{mc} methods, adaptive allocation of simulation budgets, uncertainty propagation,
differentiable or surrogate \ac{mc} models, and robust validation against measured scanner data.
A common challenge underlying these developments is determining how the fidelity of the \ac{mc} forward model, the
corresponding computational costs, and the reconstruction accuracy are related.
In particular, increased simulation fidelity does not necessarily translate into improved image reconstruction,
especially when statistical noise and model uncertainties are taken into account.

\newsectionpage
\section{Computing hardware for fast Monte Carlo simulation\label{sec:comparch}}
\sectionauthor{Jens Henrik G\"obbert, J\"urgen Scheins, Christoph Lerche}

\subsection{Fast-growing \acs{pet} system-matrix size}

\begin{figure}
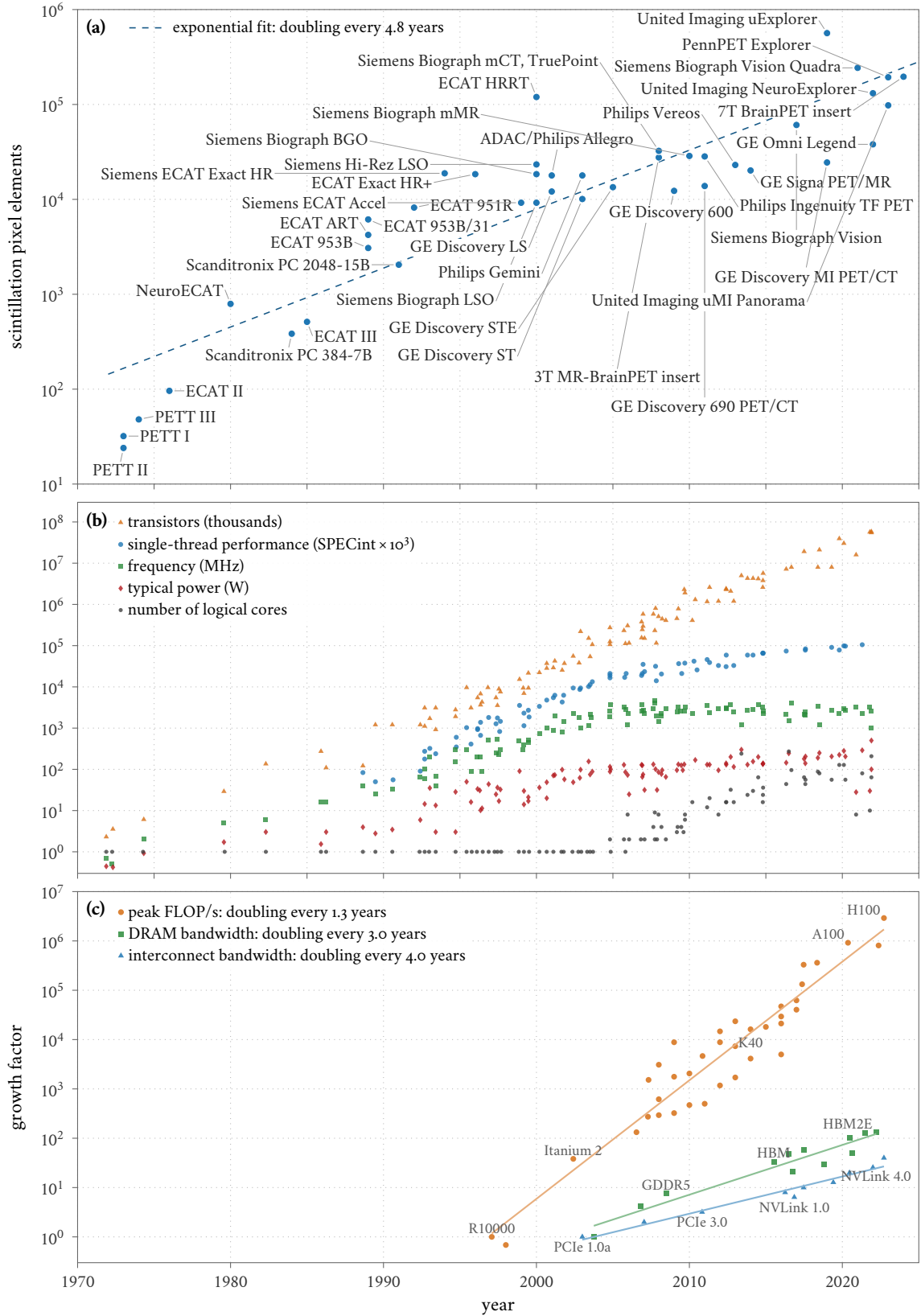

    \centering
    \ifjournalversion\scalebox{0.9}{
\begin{tikzpicture}
\input{images/fig91_hw_trends_a}
\begin{axis}[roadmap, ymode=log, roadmap log y, name=b,
  at={(a.south west)}, anchor=north west, yshift=-0.3cm,
  width=13.83cm, height=6.1cm,
  xmin=1970, xmax=2025, ymin=0.3, ymax=3e8, xticklabels={},
  ytick={1,1e1,1e2,1e3,1e4,1e5,1e6,1e7,1e8},
  legend style={at={(rel axis cs:0.045,0.985)}, anchor=north west},
  every axis plot/.append style={only marks, mark size=1.15pt, mark options={fill opacity=0.8, draw opacity=0}},
]
  \def\datadir{images/data/microprocessor_trends}
  \addplot[plotorange, mark=triangle*, mark size=1.45pt] table[comment chars=\#] {\datadir/transistors.dat};
  \addlegendentry{transistors (thousands)}
  \addplot[plotblue,   mark=*]                           table[comment chars=\#] {\datadir/specint.dat};
  \addlegendentry{single-thread performance (SPECint\,\texttimes\,10\textsuperscript{3})}
  \addplot[plotgreen,  mark=square*, mark size=1.0pt]   table[comment chars=\#] {\datadir/frequency.dat};
  \addlegendentry{frequency (MHz)}
  \addplot[plotred,    mark=diamond*, mark size=1.45pt] table[comment chars=\#] {\datadir/watts.dat};
  \addlegendentry{typical power (W)}
  \addplot[black!75,   mark=*, mark size=0.95pt]        table[comment chars=\#] {\datadir/cores.dat};
  \addlegendentry{number of logical cores}
  \node[panel] at (rel axis cs:0,1) {(b)};
\end{axis}
\begin{axis}[roadmap, ymode=log, roadmap log y, name=c,
  at={(b.south west)}, anchor=north west, yshift=-0.3cm,
  width=13.83cm, height=6.1cm,
  xmin=1970, xmax=2025, ymin=0.3, ymax=1e7,
  xtick={1970,1980,1990,2000,2010,2020},
  ytick={1,1e1,1e2,1e3,1e4,1e5,1e6,1e7},
  xlabel={year}, ylabel={growth factor},
  legend style={at={(rel axis cs:0.045,0.985)}, anchor=north west},
  every axis plot/.append style={only marks, mark size=1.35pt, mark options={fill opacity=0.85, draw opacity=0}},
]
  \def\datadir{images/data/memory_wall}
  \addplot[plotorange, mark=*]                           table[comment chars=\#] {\datadir/peak_flops.dat};
  \addlegendentry{peak FLOP/s: doubling every 1.3 years}
  \addplot[plotgreen,  mark=square*, mark size=1.15pt]   table[comment chars=\#] {\datadir/dram_bandwidth.dat};
  \addlegendentry{DRAM bandwidth: doubling every 3.0 years}
  \addplot[plotblue,   mark=triangle*, mark size=1.6pt] table[comment chars=\#] {\datadir/interconnect_bandwidth.dat};
  \addlegendentry{interconnect bandwidth: doubling every 4.0 years}
  \addplot[forget plot, sharp plot, no marks, line width=0.8pt, plotorange!55] coordinates {(1997.08,1.173) (2022.72,1.711e6)};
  \addplot[forget plot, sharp plot, no marks, line width=0.8pt, plotgreen!55]  coordinates {(2003.75,1.677) (2022.22,122.3)};
  \addplot[forget plot, sharp plot, no marks, line width=0.8pt, plotblue!55]   coordinates {(2003.0,0.8761) (2022.72,26.82)};
  \begin{scope}[every node/.style={font=\scriptsize\plotnums, text=black!65, inner sep=1.5pt}]
    \node[above]                at (axis cs:1997.08,1)      {R10000};
    \node[above]                at (axis cs:2002.41,38.2)   {Itanium~2};
    \node[anchor=north, yshift=-2pt] at (axis cs:2014.0,1.62e4) {K40};
    \node[above left]           at (axis cs:2020.37,9.18e5) {A100};
    \node[above left]           at (axis cs:2022.72,2.91e6) {H100};
    \node[above]                at (axis cs:2008.48,7.55)   {GDDR5};
    \node[above]                at (axis cs:2015.52,33.6)   {HBM};
    \node[above left]           at (axis cs:2022.22,133.7)  {HBM2E};
    \node[below]                at (axis cs:2003.0,0.95)    {PCIe~1.0a};
    \node[below]                at (axis cs:2010.83,3.0)    {PCIe~3.0};
    \node[below]                at (axis cs:2016.86,6.0)    {NVLink~1.0};
    \node[anchor=north, yshift=-3pt] at (axis cs:2022.2,36) {NVLink~4.0};
  \end{scope}
  \node[panel] at (rel axis cs:0,1) {(c)};
\end{axis}
\end{tikzpicture}}\else
\begin{tikzpicture}
\input{images/fig91_hw_trends_a}
\begin{axis}[roadmap, ymode=log, roadmap log y, name=b,
  at={(a.south west)}, anchor=north west, yshift=-0.3cm,
  width=13.83cm, height=6.1cm,
  xmin=1970, xmax=2025, ymin=0.3, ymax=3e8, xticklabels={},
  ytick={1,1e1,1e2,1e3,1e4,1e5,1e6,1e7,1e8},
  legend style={at={(rel axis cs:0.045,0.985)}, anchor=north west},
  every axis plot/.append style={only marks, mark size=1.15pt, mark options={fill opacity=0.8, draw opacity=0}},
]
  \def\datadir{images/data/microprocessor_trends}
  \addplot[plotorange, mark=triangle*, mark size=1.45pt] table[comment chars=\#] {\datadir/transistors.dat};
  \addlegendentry{transistors (thousands)}
  \addplot[plotblue,   mark=*]                           table[comment chars=\#] {\datadir/specint.dat};
  \addlegendentry{single-thread performance (SPECint\,\texttimes\,10\textsuperscript{3})}
  \addplot[plotgreen,  mark=square*, mark size=1.0pt]   table[comment chars=\#] {\datadir/frequency.dat};
  \addlegendentry{frequency (MHz)}
  \addplot[plotred,    mark=diamond*, mark size=1.45pt] table[comment chars=\#] {\datadir/watts.dat};
  \addlegendentry{typical power (W)}
  \addplot[black!75,   mark=*, mark size=0.95pt]        table[comment chars=\#] {\datadir/cores.dat};
  \addlegendentry{number of logical cores}
  \node[panel] at (rel axis cs:0,1) {(b)};
\end{axis}
\begin{axis}[roadmap, ymode=log, roadmap log y, name=c,
  at={(b.south west)}, anchor=north west, yshift=-0.3cm,
  width=13.83cm, height=6.1cm,
  xmin=1970, xmax=2025, ymin=0.3, ymax=1e7,
  xtick={1970,1980,1990,2000,2010,2020},
  ytick={1,1e1,1e2,1e3,1e4,1e5,1e6,1e7},
  xlabel={year}, ylabel={growth factor},
  legend style={at={(rel axis cs:0.045,0.985)}, anchor=north west},
  every axis plot/.append style={only marks, mark size=1.35pt, mark options={fill opacity=0.85, draw opacity=0}},
]
  \def\datadir{images/data/memory_wall}
  \addplot[plotorange, mark=*]                           table[comment chars=\#] {\datadir/peak_flops.dat};
  \addlegendentry{peak FLOP/s: doubling every 1.3 years}
  \addplot[plotgreen,  mark=square*, mark size=1.15pt]   table[comment chars=\#] {\datadir/dram_bandwidth.dat};
  \addlegendentry{DRAM bandwidth: doubling every 3.0 years}
  \addplot[plotblue,   mark=triangle*, mark size=1.6pt] table[comment chars=\#] {\datadir/interconnect_bandwidth.dat};
  \addlegendentry{interconnect bandwidth: doubling every 4.0 years}
  \addplot[forget plot, sharp plot, no marks, line width=0.8pt, plotorange!55] coordinates {(1997.08,1.173) (2022.72,1.711e6)};
  \addplot[forget plot, sharp plot, no marks, line width=0.8pt, plotgreen!55]  coordinates {(2003.75,1.677) (2022.22,122.3)};
  \addplot[forget plot, sharp plot, no marks, line width=0.8pt, plotblue!55]   coordinates {(2003.0,0.8761) (2022.72,26.82)};
  \begin{scope}[every node/.style={font=\scriptsize\plotnums, text=black!65, inner sep=1.5pt}]
    \node[above]                at (axis cs:1997.08,1)      {R10000};
    \node[above]                at (axis cs:2002.41,38.2)   {Itanium~2};
    \node[anchor=north, yshift=-2pt] at (axis cs:2014.0,1.62e4) {K40};
    \node[above left]           at (axis cs:2020.37,9.18e5) {A100};
    \node[above left]           at (axis cs:2022.72,2.91e6) {H100};
    \node[above]                at (axis cs:2008.48,7.55)   {GDDR5};
    \node[above]                at (axis cs:2015.52,33.6)   {HBM};
    \node[above left]           at (axis cs:2022.22,133.7)  {HBM2E};
    \node[below]                at (axis cs:2003.0,0.95)    {PCIe~1.0a};
    \node[below]                at (axis cs:2010.83,3.0)    {PCIe~3.0};
    \node[below]                at (axis cs:2016.86,6.0)    {NVLink~1.0};
    \node[anchor=north, yshift=-3pt] at (axis cs:2022.2,36) {NVLink~4.0};
  \end{scope}
  \node[panel] at (rel axis cs:0,1) {(c)};
\end{axis}
\end{tikzpicture}\fi
    \caption{Growth of \ac{pet} scanners and of computing hardware over five decades.
    (a) Number of scintillation pixel elements of \ac{pet} scanners; the dashed line is a least-squares fit of an
    exponential growth (doubling every 4.8 years).
    The selection is representative, not complete.
    (b) Microprocessor trend data, showing the plateauing of clock frequency, typical power and single-thread
    performance alongside the continued growth of transistor counts and the rise of multicore processors; replotted
    from the data of \textcite{Rupp2022} (CC~BY~4.0), data up to 2010 collected by M.~Horowitz, F.~Labonte,
    O.~Shacham, K.~Olukotun, L.~Hammond and C.~Batten.
    (c) Peak hardware FLOP/s, \ac{dram} bandwidth and interconnect bandwidth, shown as growth factor relative to the
    first
    data point of each series; the widening gap between compute and memory bandwidth is the ``memory wall''.
    Lines are least-squares fits on a logarithmic scale.
    Data from \textcite{Gholami2024}.}
    \label{fig:hw_trends}
\end{figure}

Since Gordon L. Brownell developed the first rectilinear scanner in 1952, the number of scintillation pixels in \ac{pet}
systems has grown constantly, as higher scintillation detector granularity is linked to improved spatial resolution.
At the same time, the axial \ac{fov} has steadily increased to improve sensitivity and patient coverage for faster
scans.
Although increasing the scintillator volume and cutting scintillators into ever smaller pixels contributes considerably
to the system's cost, the growth in numbers is exponential.
Starting with fewer than 100 pixels in the 1970s, this figure has doubled roughly every five years, reaching over half a
million with the first total-body scanner in 2019 (see \cref{fig:hw_trends}(a)). 
However, the growth of the system matrix is even steeper, almost doubling every 1.5 years.
This is because the number of voxels scales with the axial \ac{fov} and the voxel size must be reduced for
reconstruction to benefit from higher detector granularity.
With the advent of \ac{tof}-assisted \ac{pet} reconstructions, a further dimension that increases the \ac{srm} comes
into
play, as the \ac{lor} bins are divided into \ac{tof} bins, the number of which is proportional to
$\sqrt{D^2+L^2}/\sigma_\mathrm{TOF}$.
For example, a scanner with a \ac{fov} diameter of $D = \text{50\,cm}$, an \ac{fov} length of $L = \text{50\,cm}$ and a
time resolution of $\sigma_\mathrm{TOF} = \text{200\,ps}$ will require approximately 20 \ac{tof} bins along the longest
\ac{lor}.
An increasing number of scanners are also being built with \ac{doi}-capable scintillation detectors.
However, this does not necessarily result in an increase in the \ac{srm}, as the \ac{doi} information can be converted
into pixel information using a virtual detector layer, see \cite{Scheins2011}, for example.
The rapidly increasing number of pixels and voxels poses fewer problems for \ac{spect} imaging because the number of
scintillator pixels only contributes linearly to the matrix size.
In \ac{pet} imaging, however, it contributes to the power of two and leads to a growing demand for the computational
power of systems used for image reconstruction and matrix computation.
This section provides an overview of the latest developments and makes short- and long-term projections.

\subsection{Past developments: Moore’s law and Dennard scaling}

The development of computer systems over the past decades has been strongly shaped by Moore’s Law, a principle that
evolved from an empirical observation into a de facto industrial roadmap guiding semiconductor development.
It states that the number of transistors on a microchip doubles approximately every two years while the cost per
transistor falls.
This observation was first formulated by Gordon E. Moore in 1965 \cite{Moore1965} and later refined in 1975
\cite{Moore1975}, and it accurately predicted the exponential growth of integrated circuit complexity for more than five
decades.

The success of Moore’s Law during the late 20th century was largely underpinned by Dennard Scaling, also known as
metal--oxide--semiconductor
field-effect transistor (MOSFET) scaling \cite{Dennard1974}.
This principle observed that as transistors shrink, their power density remains constant because voltage and current
scale downward in proportion to their physical dimensions.
This synergy allowed the industry to pack more transistors onto a chip while simultaneously increasing clock frequencies
without exceeding thermal limits.
As visualised in historical trend data in \cref{fig:hw_trends}(b), this era saw transistor counts, clock speeds, and
typical power consumption rise in lockstep, providing a ``free performance'' dividend that enabled the transition from
massive mainframes to the personal computing revolution.

However, as the industry reached the mid-2000s, this ``virtuous cycle'' encountered physical barriers.
As transistors approached atomic scales, leakage current became a dominant factor, leading to the breakdown of Dennard
Scaling.
This resulted in the ``Power Wall'', where increasing clock frequencies further would lead to unsustainable heat
dissipation.
While transistor density continued to climb into the billions, the divergence in historical trends became evident:
single-thread performance improvements slowed significantly and became increasingly dependent on microarchitectural
techniques rather than higher clock speeds.
The era of near-automatic single-core performance gains driven primarily by frequency scaling had effectively come to an
end.

\subsection{Present developments: multicore architectures and heterogeneous computing}

As traditional transistor scaling approaches physical limits, progress in computer systems is increasingly driven by
architectural and system-level innovations rather than simple miniaturisation.
When single-core clock speeds plateaued around 2005, the industry’s first major response was the shift towards multicore
architectures (see \cref{fig:hw_trends}(b)).
Instead of striving for higher frequencies, manufacturers began integrating multiple processing units onto a single die,
allowing for increased throughput via parallel execution.

However, as the number of cores increased, the industry eventually encountered thermal and power density constraints.
Simply adding more identical cores led to diminishing returns because of thermal constraints and the challenge of
parallelising software.
This has led to the current era of heterogeneous computing, where a diverse mix of \acp{cpu}, \acp{gpu}, and specialised
\ac{ai}
accelerators (such as \acp{tpu} or neural processing units) work together.
By assigning specific tasks to the most efficient hardware—such as offloading matrix multiplication to a tensor core
while leaving general logic to the \ac{cpu}—modern systems achieve performance gains that general-purpose processors can
no
longer provide \cite{Hennessy2019}.

To support this heterogeneity, advanced chip packaging techniques such as chiplets and 3D integration have emerged.
These allow manufacturers to bypass the yield issues and physical constraints of massive monolithic chips by ``tiling''
smaller, specialised components together.
This strategy is part of a broader vision known as ``More than Moore''.
Originally conceptualised by the European Technology Platform for Nanoelectronics (ENIAC) \cite{Zhang2006}, the term
describes a shift from pure digital scaling towards integrating non-digital functionalities—such as sensors and power
management—onto the same platform.
This vision was subsequently adopted by the International Technology Roadmap for Semiconductors \cite{Arden2010},
becoming the global standard for sustaining system-level performance as traditional transistor scaling slows.

Furthermore, the end of Dennard Scaling means that modern chips are now thermally limited.
This has ushered in the era of Dark Silicon, where large portions of a chip must remain powered off to avoid
overheating, reinforcing the shift towards specialised, energy-efficient accelerators rather than general-purpose
processors \cite{Esmaeilzadeh2011}.

\subsection{Bottlenecks: compute- vs.\ memory-bound workloads}

While transistor density continues to increase, system performance is increasingly limited by the Von Neumann
Bottleneck—the physical and logical separation between the processor and memory.
Because data transfer speeds have not kept pace with the massive increases in processing power, computers have hit the
``Memory Wall''---a term coined by Wulf and McKee to describe the exponentially growing performance gap between the
\ac{cpu}
and main memory \cite{Wulf1995, Gholami2024} (see \cref{fig:hw_trends}(c)).
This gap has resulted in a state where high-speed compute units remain idle for significant periods while waiting for
data to arrive from relatively slow memory.

This divergence is best quantified through the metric of arithmetic intensity (also called operational intensity),
expressed as the ratio of FLOP/byte
(floating-point operations per byte of data transferred).
This ratio defines a fundamental boundary in system architecture: compute-bound applications possess a high arithmetic
intensity, performing many calculations on a small amount of data, and are thus limited by the raw speed of the
processor.
Conversely, memory-bound applications, such as big data analytics, modern \ac{ai} inference or \ac{pet} imaging
simulations,
exhibit a low FLOP/byte ratio.
For these workloads, increasing the raw clock speed or the number of cores provides at some point negligible benefits,
as the processor remains stalled by the inadequate data transfer rate (bandwidth) or the time required to initiate a
transfer (latency).

\begin{figure}
    \centering
\begin{tikzpicture}[
  roofnew/.style={line width=1.3pt, plotblue},
  roofold/.style={line width=1.1pt, plotblue!40},
  arr/.style={-{Latex[length=2mm]}, line width=0.6pt},
  lab/.style={font=\footnotesize},
]
\begin{axis}[roadmap,
  width=14.19cm, height=5.6cm,
  xmin=0, xmax=10, ymin=0, ymax=6.6,
  xtick=\empty, ytick=\empty, xmajorgrids=false, ymajorgrids=false,
  xlabel={operational intensity (FLOP/byte, log scale)},
  ylabel={attainable performance\\(FLOP/s, log scale)}, ylabel style={align=center},
  clip=false, axis on top,
]
  \fill[plotorange!8] (axis cs:0,0) rectangle (axis cs:5.5,6.6);
  \fill[plotblue!6] (axis cs:5.5,0) rectangle (axis cs:10,6.6);
  \node[lab, plotorange!80!black, font=\footnotesize\bfseries] at (axis cs:2.75,6.15) {memory-bound};
  \node[lab, plotblue!80!black, font=\footnotesize\bfseries] at (axis cs:7.75,6.15) {compute-bound};
  \draw[roofold] (axis cs:0.5,0) -- (axis cs:4,3.5) -- (axis cs:9.8,3.5);
  \node[lab, plotblue!55, above] at (axis cs:8.0,3.5) {peak compute, older generation};
  \draw[roofnew] (axis cs:0,0) -- node[lab, sloped, above, pos=0.69, text=plotblue!90!black] {peak memory bandwidth}
    (axis cs:5.5,5.5) -- (axis cs:9.8,5.5);
  \node[lab, plotblue!90!black, above] at (axis cs:8.0,5.5) {peak compute, newer generation};
  \draw[densely dotted, plotblue!40, thick] (axis cs:4,0) -- (axis cs:4,3.5);
  \draw[densely dotted, plotblue, thick] (axis cs:5.5,0) -- (axis cs:5.5,5.5);
  \fill[plotblue!40] (axis cs:4,3.5) circle (2.3pt);
  \fill[plotblue] (axis cs:5.5,5.5) circle (2.3pt);
  \node[lab, anchor=north west] at (axis cs:5.6,5.4) {ridge point};
  \draw[arr, black!60] (axis cs:4.07,0.3) -- node[above, lab, black!70, inner sep=1pt] {next generation} (axis cs:5.43,0.3);
  \begin{scope}[box/.style={lab, fill=plotorange!8, inner sep=1pt, text=plotorange!80!black}]
    \fill[plotorange] (axis cs:2.6,2.32) circle (2.6pt);
    \node[box, anchor=south east] at (axis cs:2.42,2.5) {Monte Carlo simulation};
    \draw[arr, plotorange!80!black] (axis cs:2.6,2.5) -- (axis cs:2.6,3.45);
    \node[box, anchor=south east] at (axis cs:2.62,3.5) {higher memory bandwidth};
    \draw[arr, plotorange!80!black] (axis cs:2.75,2.32) -- (axis cs:3.6,2.32);
    \node[box, anchor=west, align=left] at (axis cs:3.68,2.32) {higher FLOP/byte\\(algorithm)};
  \end{scope}
\end{axis}
\end{tikzpicture}
    \caption{Schematic roofline model~\cite{Williams2009}.
    The attainable performance is bounded by the peak memory
    bandwidth (slanted roof) and the peak compute performance (flat roof), which meet at the ridge point.
    Since peak compute grows faster than memory bandwidth, the ridge point moves to higher operational intensity with
    each hardware generation.
    Monte Carlo \ac{pet} simulations are typically memory-bound, so they gain from higher
    memory bandwidth or from algorithms with more operations per byte, but not from more compute alone.}
    \label{fig:roofline}
\end{figure}
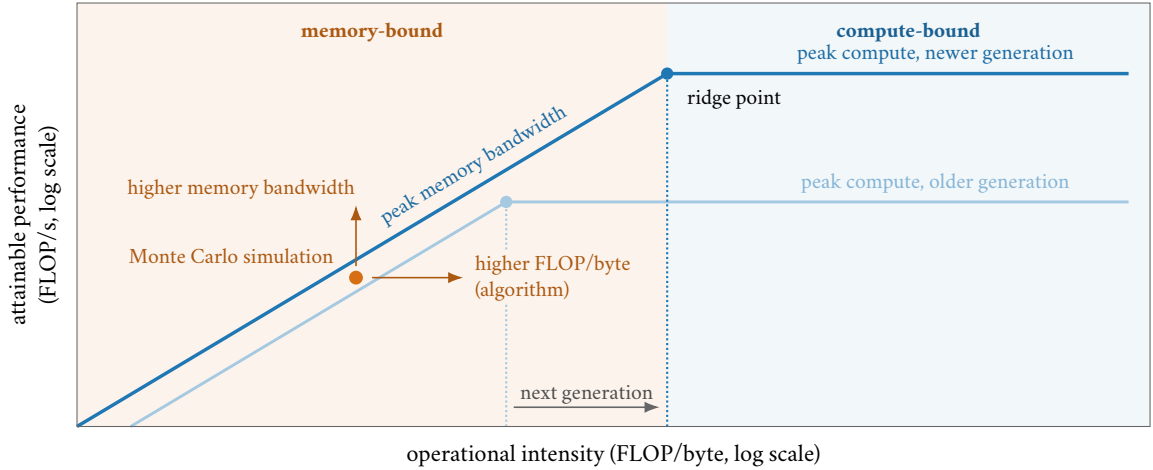

The relationship between these hardware and software constraints is formalised through the Roofline Model
\cite{Williams2009}, which provides a visual bound and bottleneck analysis by plotting attainable performance against
operational intensity (operations per byte) on a log-log scale (see \cref{fig:roofline}).
The model defines an upper bound on performance using two intersecting trajectories: a slanted ``roof'' representing the
peak memory bandwidth and a flat horizontal line representing the peak computational performance.
The critical intersection of these lines, known as the ridge point, identifies the minimum operational intensity
required to achieve the machine's maximum performance.

Applications with an operational intensity lower than this point—such as many Monte Carlo-based \ac{pet} simulations—hit
the
slanted portion of the roof, indicating they are fundamentally memory-bound.
For these workloads, the Roofline Model demonstrates that increasing raw compute power is ineffective; instead,
performance can mostly be improved by increasing memory bandwidth or by optimising the underlying algorithm to increase
the FLOP/byte ratio, shifting the application to the right, towards the ridge point.
However, because peak compute performance has historically grown faster than memory bandwidth, the ridge point shifts to
higher operational intensities with each new hardware generation.

\subsection{Escaping the memory wall}

Historically, the primary architectural response to the Memory Wall has been the implementation of a sophisticated
hierarchical cache structure (L1, L2, and L3) \cite{Jacob2010}.
In the context of the Roofline Model, caches effectively ``push the roof higher'' by providing a significantly higher
local bandwidth than the main memory bus could ever sustain.

However, the effectiveness of traditional caching is severely diminished when applied to \ac{mc} simulations for
\ac{pet}
imaging.
These simulations are characterised by irregular memory access patterns, as the stochastic interaction of a photon
involves paths that do not follow a linear or predictable spatial route.
This randomness results in frequent ``cache misses'', forcing a costly round-trip to the main memory
\cite{Bielajew2001}.
Consequently, \ac{mc}-based \ac{pet} workloads spend a disproportionate amount of time in a stalled state as the cache
hierarchy
fails to hide the underlying latency of the Von Neumann architecture \cite{Tramm2024}.

To address these limitations, recent hardware advancements focus on the physical consolidation of memory and logic
through 3D-stacked architectures and \ac{hbm}.
Rather than placing \ac{dram} chips side-by-side on a motherboard, \ac{hbm} stacks memory dies vertically and connects
them to the
processor via a silicon interposer using thousands of through-silicon vias \cite{Loh2008}.
A specialised implementation of this concept is 3D V-Cache, where additional layers of static random-access memory are
bonded directly onto the
\ac{cpu} die, providing a hardware-level mitigation of main memory bandwidth limitations \cite{Domke2023}.

A more disruptive solution that fundamentally breaks with the Von Neumann architecture is \ac{pim}.
While \ac{hbm} and 3D-stacking bring memory closer to the \ac{cpu}, \ac{pim} significantly reduces off-chip data
movement by executing
selected operations within the memory array itself \cite{Mutlu2019}.
In a \ac{pim}-enabled system, the memory is no longer a passive storage bin but an active participant in the simulation.
This paradigm shift addresses the massive energy and performance ``tax'' incurred by data movement \cite{Ghose2019}.
Recent hardware prototypes have demonstrated the potential for specialised in-memory operators to achieve execution
speeds significantly faster than traditional general-purpose processors \cite{Martinez2026, Kim2020}.

\subsection{Scaling via distributed computing}

A straightforward method for increasing available memory bandwidth and computational throughput is the expansion of the
system into distributed computing nodes.
Because each additional node contributes not only its own processors but also its own local memory controllers and
memory bus, the aggregate memory bandwidth of the cluster increases linearly with the number of nodes.
For \ac{mc} simulations, which are ``embarrassingly parallel'' by nature, each photon history can be computed
independently,
allowing for massive performance gains across distributed partitions.
However, global synchronisation and interconnect bandwidth eventually limit strong-scaling efficiency, a phenomenon
interpreted through Amdahl’s Law \cite{sarrut2014a}.
In addition, care must be taken to avoid the excessive and inefficient creation of independent threads or processes.
For example, variance reduction techniques such as \emph{photon splitting} can have this effect.
This technique considers all potential photon trajectories, including physically extinct paths, to provide an
effective computational method for simulating a large number of trajectories.
Limiting thread and process creation to the history of primary photons only, or implementing pre-configurable variance
reduction multiplicity limits \cite{scheins2021}, can effectively avoid this problem. 
\subsection{Future outlook: projected performance gains for Monte Carlo simulators}
Looking forward, the performance of \ac{mc} simulations for imaging will be defined by the growing divergence between
compute capability and data delivery, constrained by a strict power-per-node ceiling.
Empirical analysis of the last 20 years shows that while peak compute (FLOP/s) has doubled about every 1.3 years,
\ac{dram}
bandwidth has only doubled about every 3 years (see \cref{fig:hw_trends}(c), \cite{Gholami2024}).
This widening gap means that without architectural intervention, \ac{mc} simulations—which are inherently
memory-bound—will
spend an increasing percentage of their power budget on ``stalled'' cycles.

While physical bus bandwidth is projected to increase by approximately 10.5$\times$ over the next decade ($1.6^5$), this
scaling remains capped by the ``Von Neumann bottleneck''.
The path to a 50$\times$ speedup lies in treating domain-specific \ac{pim} \cite{Mutlu2022}, which
integrates computation directly into \ac{dram} arrays, as the primary compute substrate.
By embedding \ac{mc}-specific logic directly into the \ac{dram} dies, the simulation gains access to the massive
internal
bank-level parallelism of the memory arrays—bandwidth that is often two orders of magnitude higher than what the
external bus can support.
The memory then acts as a compute fabric tailored for the
high-entropy, irregular access patterns inherent in Monte Carlo tracking for \ac{srm} computation.

Within a fixed power budget, these complementary avenues offer a path to a total 10--50$\times$ performance gain per
node over the next decade.
Since parallel \ac{mc} workloads scale almost linearly across nodes, these per-node gains multiply further with the
number of nodes.
This paradigm shift—from ``brute force'' scaling to architectural and energy-driven efficiency—will be crucial in moving
high-fidelity \ac{srm} simulations into the clinic, making real-time, high-accuracy dosimetry a sustainable reality.

\newsectionpage
\section{Synthesis and outlook\label{sec:outlook}}
This section returns to the questions raised in the introduction, identifies the research directions that recur
across the preceding sections, and concludes the paper.

\subsection{Back to the four questions\label{sec:outlookquestions}}

\paragraph{Where on the spectrum should a given application sit?}
The preceding sections suggest a pragmatic answer.
\Ac{mc} is most valuable where analytic models fail or become inefficient, e.g.\ for the system response of
\ac{spect} with high-energy emitters (\cref{sec:FastSysResp}), for multiple scatter
(\cref{sec:ScatterEstimation}) and for positron range in heterogeneous tissue
(\cref{sec:PositronRangeModelling}).
Where analytic models are accurate and computationally feasible, 
there is little reason to replace them.
Most practical solutions will therefore be hybrid.
It is still largely open how the decomposition into deterministic and stochastic components affects the bias,
variance and convergence of the reconstruction.
If \ac{mcs} provides only an additive nuisance term, 
it enters the forward model but not the ``back-projection'' step.
If it also provides parts of the system matrix, its noise, and a possible mismatch with a deterministic
back-projection, affect every update (\cref{sec:InvProb}).

\paragraph{How precise does the simulation have to be?}
The required precision need not be the same throughout the reconstruction.
At early iterations far from the solution, a rough estimate of the gradient direction, e.g.\ from a small number of
simulated photons, may suffice,
whereas close to the solution the precision has to increase, similar to stochastic optimisation with growing
batch sizes or variance reduction (\cref{sec:StochasticReconstruction}).
Whether the \ac{mc} noise has to be driven below the noise of the measured data, or only kept low enough for the
quantitative task at hand, and how to assess this trade-off, remains to be investigated.
Since the \ac{mc} noise enters the gradient non-linearly through the ratio of measured and expected data, it causes
a bias that grows when the expected counts per detector bin are low.
For sparse data, as in \ac{tof} and long axial \ac{fov} \ac{pet}, the requirement can therefore be much stricter
than a comparison with the noise of the measured data suggests.

\paragraph{Is in-patient scatter a nuisance or a signal?}
In the near term, one of the most concrete benefits of fast \ac{mc} is a more accurate scatter estimate,
including multiple
scatter and an absolute scatter level, and with it improved quantification in single- and multi-energy-window
\ac{pet} and \ac{spect} (\cref{sec:ScatterEstimation}).
This is most relevant where analytic models struggle, e.g.\ for \ac{spect} of $\alpha$-emitters, with low
signal-to-noise ratios, high scatter-to-primary ratios and scattering in the camera head.
Scatter can also be a signal, mainly about the attenuation rather than the activity.
In \ac{tof} \ac{pet}, it can probably resolve the non-uniqueness of the joint estimation of activity and attenuation
(\cref{sec:JAAE}), provided that the scatter model is accurate (also in absolute terms) and its gradient
with respect to activity and attenuation is available.

\paragraph{How much of the physics has to be simulated?}
The answer depends on what is considered as the data, i.e.\ counts with energy and timing information, or the
scintillation light, photodetector response and digitised waveforms from which they are derived.
For new detector concepts, e.g.\ based on Cherenkov light, \ac{mcs} may be the only practical way to obtain the
required response models.
It also depends on what is to be reconstructed and on the task at hand, which is often less obvious than assumed.
Faster simulators allow more detailed physics within the same time budget, but each additional level of detail has
to improve the reconstructed image enough to justify its cost.
Simulators that include more of the physics in the same time can also help to identify which steps of the
data-generation chain are well described by deterministic models and which are not.

\subsection{Cross-cutting research directions\label{sec:outlookdirections}}

The following directions recur across the preceding sections and are listed in the order of their priority.

\paragraph{Validation, calibration and benchmarks.}
The most immediate value of fast \ac{mcs} may lie in the validation of more detailed analytic models and of their
gradients.
Such models usually take short cuts, e.g.\ a finite number of scatter points, and their validation against low-noise
\ac{mc} simulations is currently often skipped, or done only for a few scanning scenarios and objects, because the
simulations take too long.
Analytic and \ac{mc} models alike have to be cross-calibrated against measurements, so that their absolute
predictions match the scanner.
Shared benchmarks would make different approaches comparable, including simplified physics models, e.g.\ reduced
transport models, in which the effects of \ac{mc} noise, model mismatch and reconstruction strategy can be isolated
and studied systematically.

\paragraph{Expectations instead of realisations.}
Reconstruction requires the expectation of the measurement, whereas \ac{mcs} provides noisy
realisations.
Simulators and reconstruction algorithms have to be developed together to close this gap, e.g.\ by variance
reduction that targets the expectation in the detector bins that matter, by post-processing of \ac{mc} estimates
with controlled bias, and by reconstruction algorithms that remain stable with stochastic forward models
(\cref{sec:InvProb}).

\paragraph{Gradients and adjoints.}
The gradient of the Poisson log-likelihood consists of a back-projection of the ratio of measured and expected data,
which only requires the detector bins that recorded counts, and of the sensitivity image $\srm\transp\onevec$, which
requires all bins.
If the sensitivity image can be precomputed and stored, simulators should support the first term, the back-projection
of the ratio, directly, e.g.\ by forced detection or photon replay for the detected bins.
Whether differentiable \ac{mc} is needed, or whether an analytic gradient of the dominant component combined with an
\ac{mc}-based estimate of the remaining contributions suffices, as suggested for the joint estimation of activity
and attenuation (\cref{sec:JAAE}), is an open question.
The bias introduced by combining an \ac{mc} forward model with a mismatched deterministic back-projection is
discussed in \cref{sec:InvProb}.

\paragraph{Allocation of the computational budget.}
For a fixed budget, it is unclear how the \ac{mc} samples are best distributed over source voxels, detector bins,
energy windows and iterations.
The estimation of the sensitivity image by subsampling the detector bins is an example
(\cref{sec:WhereDoesMCfit}).
The idea is not new, but how to subsample with the least damage to the final image is open, since the estimate
enters every update as a divisor.
Splitting the simulation into a patient and a detector part, which is precomputed once, could reduce the cost
further, but currently requires approximations, and generative models that learn the detector response are a
possible alternative.
Since \ac{mc} simulation is typically memory-bound, these choices also depend on the hardware
(\cref{sec:comparch}).

\paragraph{Uncertainty quantification.}
In the longer term, fast forward models make it conceivable to quantify and propagate the statistical uncertainty
that arises from the measured counts and the finite number of simulated histories, e.g.\ by posterior sampling or
conditional means.
The uncertainty of the models themselves, e.g.\ of physics approximations, cross sections and calibrations, affects
\ac{mc} and analytic models alike, and can at best be explored by perturbing the model parameters.

\subsection{Opportunities beyond reconstruction\label{sec:outlookbeyond}}

Fast \ac{mcs} also benefits tasks outside the reconstruction itself.
It provides realistic data sets with known ground truth for training and testing deep-learning methods.
It allows new system concepts to be evaluated by simulation before expensive prototypes are built, in the sense of
digital twins.
Both require appropriate figures of merit, and evaluating resolution and noise separately 
can be of limited use, whereas
task-based measures, e.g.\ the Fisher information for a given task, could be more meaningful.

\subsection{Concluding remarks\label{sec:outlookconcl}}

This roadmap asked what it would take to make \ac{mc} simulation part of image reconstruction in \ac{pet} and
\ac{spect}.
Faster simulators are necessary but not sufficient.
It also takes simulators that deliver what reconstruction needs, i.e.\ expectations in the relevant detector bins
and gradient information, reconstruction algorithms that tolerate stochastic forward models, validation and
calibration against measurements, and implementations that match the hardware.
Some steps are within reach, such as \ac{mc}-based system response models in \ac{spect}, absolute quantitative
scatter estimates that are updated in every outer iteration, and the validation of fast analytic models against
\ac{mc}.
Others, such as stochastic reconstruction with controlled \ac{mc} noise, the use of scatter as a signal, 
and the simulation of the complete acquisition, from the activity and material distribution in the patient to the
detected data, inside the reconstruction,
certainly require more research in the coming decade.

\newsectionpage
\section*{Acknowledgements}
\markright{Acknowledgements}

\paragraph{AI disclosure statement}
During the preparation of this manuscript, the following generative AI tools
were used by individual authors.
In sections~1 and~10, GS used Claude Opus~5.5 (Anthropic) to edit text written
by the authors and to generate text.
In section~3, DS used Gemini (Google) to edit and restructure text.
In section~4, PG used Claude Opus~5.5 (Anthropic) to support the literature
search, and MR used deepseek-v4-flash and gpt-oss-120b to edit and
restructure text.
In section~7, GS used Claude Opus~5.5 (Anthropic) to edit text written by the
authors, to generate text, and to create the explanatory figure 7.1.
In section~9, JHG used Qwen3.5-122B-A10B-FP8 to edit and restructure text.
All AI-generated or AI-edited text was critically 
reviewed and revised by the respective author. 
The AI-generated illustrations are schematic and do not
represent research data. 
All references, including those identified with AI
assistance, were checked against the original publications. 
The authors take
full responsibility for the content of the manuscript.

\paragraph{Funding}
KT acknowledges support from the Computational Collaborative Project in
Synergistic Reconstruction for Biomedical Imaging (CCP SyneRBI), funded by the
UKRI Science and Technology Facilities Council (STFC) through the UKRI Digital
Research Infrastructure programme.
PG acknowledges support from grant JDC2023-051754-I, funded by
MCIU/AEI/10.13039/501100011033 and by ESF+.
DS acknowledges support from grants
SIRIC LYriCAN INCa-DGOS-INSERM-ITMO cancer\_18003 and ANR-11-LABX-0063.
JLH acknowledges support from grant TEC-2024/TEC-43 LUNABRAIN-CM, funded by the Comunidad de Madrid (Spain) through the R\&D activities program in technologies, granted by Order 5696/2024, and research grant ``Implementation of Novel Reconstructions for Siemens Scanners'' at UCM funded by Siemens Healthineers.
ZK acknowledges support from the UK Engineering and Physical Sciences Research
Council (EPSRC) under grant EP/X010740/1.
JG acknowledges support from the Mrs Berta Kamprad Foundation
(FBKS-2025-21-689) and the Swedish Cancer Society (243785Pj01H).
GS acknowledges support from the Research Foundation -- Flanders (FWO) under
grant G016626N.
\paragraph{Competing interests}
BA received travel reimbursement for speaking engagements from Spectrum Dynamics Medical and in-kind research support from GE HealthCare. 
JG acts as a consultant for Spago Nanomedical AB. 
The remaining authors declare no competing interests.

\section*{Data availability statement}

No new data were created or analysed in this study.

\section*{List of abbreviations}
\markright{List of abbreviations}
\addcontentsline{toc}{section}{List of abbreviations}
\setlength{\columnsep}{1.5em}\begin{multicols}{2}\small 
\begin{acronym}[UMC-PET]
\acro{ai}[AI]{artificial intelligence}
\acro{am}[AM]{attenuation modelling}
\acro{ann}[ANN]{artificial neural network}
\acro{arf}[ARF]{angular response function}
\acro{cfd}[CFD]{convolution forced detection}
\acro{cnn}[CNN]{convolutional neural network}
\acro{cpu}[CPU]{central processing unit}
\acro{ct}[CT]{X-ray computed tomography}
\acro{dew}[DEW]{dual-energy window}
\acro{diffmc}[diffMC]{differentiable Monte Carlo}
\acro{doi}[DOI]{depth of interaction}
\acro{dram}[DRAM]{dynamic random-access memory}
\acro{egsnrc}[EGSnrc]{Electron Gamma Shower nrc}
\acro{em}[EM]{expectation maximisation}
\acro{esse}[ESSE]{effective scatter source estimation}
\acro{fd}[FD]{forced detection}
\acro{fov}[FOV]{field of view}
\acro{fwhm}[FWHM]{full width at half maximum}
\acro{fwtm}[FWTM]{full width at tenth maximum}
\acro{gan}[GAN]{generative adversarial network}
\acro{gate}[GATE]{Geant4 Application for Tomographic Emission}
\acro{geant4}[Geant4]{Geometry and Tracking Toolkit}
\acro{ggems}[GGEMS]{GPU Geant4-based Monte Carlo Simulations}
\acro{gpet}[gPET]{GPU-accelerated Monte Carlo for PET}
\acro{gpu}[GPU]{graphics processing unit}
\acro{hbm}[HBM]{high-bandwidth memory}
\acro{hpc}[HPC]{high performance computing}
\acro{jeaa}[JEAA]{joint estimation of activity and attenuation}
\acro{lafov}[LAFOV]{long axial field of view}
\acro{lor}[LOR]{line of response}
\acroplural{lor}[LORs]{lines of response}
\acro{lut}[LUT]{look-up table}
\acro{mal}[MAL]{machine learning}
\acro{map}[MAP]{maximum a posteriori}
\acro{mc}[MC]{Monte Carlo}
\acro{mcgpupet}[MCGPU-PET]{Monte Carlo GPU for PET}
\acro{mcs}[MCS]{Monte Carlo simulation}
\acro{mcss}[MCSS]{Monte Carlo scatter simulation}
\acro{ml}[ML]{maximum likelihood}
\acro{mlaa}[MLAA]{maximum likelihood activity and attenuation estimation}
\acro{mlacf}[MLACF]{maximum likelihood activity and attenuation correction factor estimation}
\acro{mlem}[MLEM]{maximum likelihood expectation maximisation}
\acro{mlrr}[MLRR]{maximum likelihood reconstruction of activity and registration of the attenuation image}
\acro{mr}[MR]{magnetic resonance}
\acro{mri}[MRI]{magnetic resonance imaging}
\acro{osem}[OSEM]{ordered subsets expectation maximisation}
\acro{pca}[PCA]{principal component analysis}
\acro{pet}[PET]{positron emission tomography}
\acro{pim}[PIM]{processing-in-memory}
\acro{pps}[PPS]{PET Physics Simulator}
\acro{pr}[PR]{positron range}
\acro{prc}[PRC]{positron range correction}
\acro{psf}[PSF]{point spread function}
\acro{rm}[RM]{resolution model}
\acro{rte}[RTE]{radiative transfer equation}
\acro{saga}[SAGA]{unbiased stochastic average gradient}
\acro{simd}[SIMD]{single instruction multiple data}
\acro{simind}[SIMIND]{SIMulation of Imaging Nuclear Detectors}
\acro{simset}[SimSET]{Simulation System for Emission Tomography}
\acro{snr}[SNR]{signal-to-noise ratio}
\acro{spect}[SPECT]{single photon emission computed tomography}
\acro{srm}[SRM]{system response matrix}
\acro{sss}[SSS]{single scatter simulation}
\acro{stir}[STIR]{Software for Tomographic Image Reconstruction}
\acro{svrg}[SVRG]{stochastic variance reduced gradient}
\acro{tew}[TEW]{triple-energy window}
\acro{tof}[TOF]{time-of-flight}
\acro{tor}[TOR]{tube of response}
\acroplural{tor}[TORs]{tubes of response}
\acro{tpu}[TPU]{tensor processing unit}
\acro{umcpet}[UMC-PET]{Ultra-fast Monte Carlo for PET}
\acro{vor}[VOR]{volume of response}
\acroplural{vor}[VORs]{volumes of response}
\acro{vrt}[VRT]{variance reduction technique}
\acro{wls}[WLS]{weighted least squares}
\end{acronym}

\end{multicols}

\AtNextBibliography{\footnotesize}
\printbibliography[heading=bibintoc]

\iffinetypo
  \vspace*{\fill}
  \begin{center}
    \footnotesize
    \textsc{\textls[80]{colophon}}\\[0.6ex]
    \ifcrimson
      Set in Crimson, with Libre Franklin for titles and headings and Computer Modern for mathematics.\\
    \else
      Set in Computer Modern.\\
    \fi
    Typeset with \LaTeX{} and pdf\TeX{} using microtype; figures drawn with TikZ and pgfplots;
    references processed with biblatex and Biber.
  \end{center}
\fi

\end{document}